\documentclass[useAMS,usenatbib]{mn2e}
\usepackage{epsfig,lscape}

\usepackage{graphicx}
\usepackage{times}

\DeclareGraphicsExtensions{.eps,.eps.gz,.epsi}

\newcommand\apj{ApJ}% 
\newcommand\aaps{A\&AS}% 
\newcommand\Te{$T_{\rm e}$}%
\newcommand\Ne{$N_{\rm e}$}%
\newcommand\oii{O~{\sc ii}}%
\newcommand\oiii{O~{\sc iii}}%

\title[Elemental abundances of GBPNe from ORLs]{Elemental abundances of 
Galactic bulge planetary nebulae from optical recombination lines}
\author[W. Wang \& X.-W. Liu]{W. Wang$^1$
and X.-W. Liu$^{1,2}$\thanks{E-mail: liuxw@bac.pku.edu.cn}\\
$^1$Department of Astronomy, Peking University, Beijing 100871, P. R. China\\
$^2$Kavli Institute for Astronomy and Astrophysics, Peking University,
Beijing 100871, P. R. China}

\begin{document}
\date{Received:}

\pagerange{\pageref{firstpage}--\pageref{lastpage}} \pubyear{2007}

\maketitle

\label{firstpage}

\begin{abstract} {Deep long-slit optical spectrophotometric observations are
presented for 25 Galactic bulge planetary nebulae (GBPNe) and 6 Galactic disk
planetary nebulae (GDPNe). The spectra, combined with archival ultraviolet
spectra obtained with the {\it International Ultraviolet Explorer}\ ({\it IUE})
and infrared spectra obtained with the {\it Infrared Space Observatory}\ ({\it
ISO}), have been used to carry out a detailed plasma diagnostic and element
abundance analysis utilizing both collisional excited lines (CELs) and optical
recombination lines (ORLs).

Comparisons of plasma diagnostic and abundance analysis results obtained from
CELs and from ORLs reproduce many of the patterns previously found for GDPNe.
In particular we show that the large discrepancies between electron
temperatures (\Te's) derived from CELs and from ORLs appear to be mainly caused
by abnormally low values yielded by recombination lines and/or continua.
Similarly, the large discrepancies between heavy element abundances deduced
from ORLs and from CELs are largely caused by abnormally high values obtained
from ORLs, up to tens of solar in extreme cases. It appears that whatever
mechanisms are causing the ubiquitous dichotomy between CELs and ORLs, their
main effects are to enhance the emission of ORLs, but hardly affect that of
CELs. It seems that heavy element abundances deduced from ORLs may not reflect
the bulk composition of the nebula. Rather, our analysis suggests that ORLs of
heavy element ions mainly originate from a previously unseen component of 
plasma of \Te's of just a few hundred Kelvin, which is too cool to excite any
optical and UV CELs.

We find that GBPNe are on the average 0.1--0.2\,dex more metal-rich than GDPNe
but have a mean C/O ratio that is approximately 0.2\,dex lower.  By comparing
the observed relative abundances of heavy elements with recent theoretical
predictions, we show that GBPNe probably evolved from a relatively metal-rich
environment of initial $Z \sim 0.013$, compared to an initial $Z \la 0.008$ for
GDPNe. In addition, we find that GBPNe tend to have more massive progenitor
stars than GDPNe. GBPNe are found to have an average magnesium abundance about
0.13~dex higher than GDPNe. The latter have a mean magnesium abundance almost
identical to the solar value. The enhancement of magnesium in GBPNe and the
large [$\alpha$/Fe] ratios of bulge giants suggest that the primary enrichment
process in the bulge was Type\,II SNe. PN observations yield a Ne/O abundance
ratio much higher than the solar value, suggesting that the solar neon
abundance may have been underestimated by 0.2~dex.

} 
\end{abstract}


\section{Introduction}
\label{intr} 
 
Knowledge of the chemical composition of the Galactic bulge is of paramount
importance for the understanding of the history of formation and evolution of
the Galaxy (e.g. \citealt{bergh1996}; \citealt{william1997}).  While early work
suggested that bulge stars are super metal-rich (eg. \citealt{WR1983};
\citealt{rich1988}; \citealt{GF1992}), more recent high resolution
spectroscopic studies showed that the bulge is actually slightly iron-poor
compared to the solar neighborhood (\citealt{MR1994}), a result supported by
the analysis of low resolution spectroscopic and photometric observations for a
large sample of bulge K and M giants by \cite{IG1995}.  In addition, a careful
metallicity analysis of low resolution spectra of integrated light from the
Galactic bulge in Baade's window by \cite{idiart1996} suggested that the bulge
is enhanced in $\alpha$-elements (represented by magnesium). 

All the aforementioned studies were based on analysis of light from bulge stars
and yielded essentially no information of abundances of light elements such as
carbon and oxygen. In this context, abundance analyses using GBPNe as probes
can play an important role.  Planetary nebulae (PNe) are products of low- and
intermediate-mass stars (LIMS). Ionizing stars of PNe have typical luminosities
of several hundreds to thousands solar luminosity. Much of the energy is
emitted in a few strong narrow emission lines, making them easily observable to
large distances. In the optical and infrared, many PNe in the Galactic bulge
can be readily observed even though the bulge is heavily obscured by
intervening dust grains in the Galactic disk. For most elements, PNe are
believed to preserve much of the original composition of the interstellar
medium (ISM) at the time of their formation. Elemental abundances derived from
PN observations can therefore be compared and related to results from stellar
abundance analyses. Possible exceptions are light elements such as He, N and
C, whose abundances can potentially be modified by various dredge-up processes,
depending on mass and initial metallicity of the progenitor star.  Measurements
of abundances of those elements in PNe thus shed light on the nucleosynthesis
and dredge processes that occurred during the late evolutionary stages of their
progenitor stars.

Hitherto, spectroscopic abundance determinations have been carried out for
approximately 250 GBPNe. The analyses were almost invariably based on
observations of relatively strong CELs (\citealt{ratag1992,ratag1997};
\citealt{escudero04}; \citealt{exter2004}). The results show that, albeit with
large scatter, GBPNe have average abundances and abundance distributions
similar to GDPNe for elements unaffected by evolution of their progenitor
stars, such as oxygen and other heavier elements. For helium and nitrogen,
whose abundances may have been altered by the various dredge-up processes,
studies so far have yielded different results. Part of the discrepancies may
be due to differences in PN populations sampled by individual studies.

The heavy extinction towards the Galactic bulge prohibits extensive ultraviolet
(UV) spectroscopy of GBPNe. Only a very restrictive number of UV spectra,
obtained mainly with the {\it IUE}, are available for GBPNe, mostly located
within Baade's window. This poses a serious limitation for the study of some
key elements, such as carbon for which the only observable strong CELs, the
C~{\sc iv} $\lambda\lambda$1548,1550 and C~{\sc iii}] $\lambda\lambda$1907,1909
lines, fall in the ultraviolet (the infrared [C~{\sc ii}] 158$\mu$m line arises
mainly from the photodissociation region surrounding the ionized region, rather
than from the latter itself). As a consequence, very little is known at the
moment about the abundance and its distribution of this key element for the
bulge population. Even with suitable UV observations, the analysis is
complicated by the uncertain extinction law towards the Galactic bulge, which
is known to deviate from that of the general ISM of the Galactic disk (Walton,
Barlow \& Clegg 1993; Liu et al. 2001). 

Advances in the past decade in observational techniques and in calculations of
the basic atomic data, such as effective recombination coefficients for
non-hydrogenic ions with multiple valence electrons, have made it possible to
determine nebular abundances using faint optical recombination lines
from heavy element ions (e.g. Liu et al. 1995, 2000, 2001). Ionic abundances
deduced from ORLs have the advantage that they are almost independent of the
physical conditions, such as \Te\, and electron density (\Ne), of the nebula
under study. Dominant ionic species of carbon typically found in photoionized
gaseous nebulae possess a number of relatively bright ORLs, such as C~{\sc ii}
$\lambda$4267, C~{\sc iii} $\lambda$4187 and $\lambda$4650 and C~{\sc iv} 4658
(blended with the [Fe~{\sc iii}] $\lambda$4658 line in some objects), allowing
measurements of ionic abundances of C$^{2+}$, C$^{3+}$ and C$^{4+}$. These
lines are routinely detected in spectra of reasonable quality for bright
Galactic PNe (e.g. Liu 1998).

Deep, medium resolution optical spectroscopy, allowing determinations from ORLs
of abundances of carbon, nitrogen, oxygen and neon, have been carried out for
several dozen GDPNe (Tsamis et al. 2003, 2004; Liu et al. 2004a,b; Wesson, Liu
\& Barlow 2005). It is found that ionic abundances derived from ORLs are
ubiquitously higher than values derived from CELs, i.e. abundance discrepancy
factor (adf) $\ga 1$ for all the nebulae analyzed so far. The adf is found to vary
from nebula to nebula and typically falls in the range 1.6--3.0, but with a
significant tail extending to much higher values, reaching 70 in the most
extreme case found so far (c.f. Liu 2003, 2006b for recent reviews). On the
other hand, analyses also show that for a given nebula, adf's derived for all
abundant second-row elements, C, N, O and Ne, are of similar magnitude; in
other words, heavy element abundance ratios, such as C/O, N/O and Ne/O are
unaffected, regardless of the actual value of the adf (Liu et al. 1995, 2000,
2001; Mathis et al. 1998; Luo et al. 2001; Tsamis et al. 2004; Liu et al.
2004b). In stark contrast to those abundant second-row elements, ORL
abundance determinations for magnesium, the only third-row element that has
been studied using an ORL, yield nearly a constant Mg/H abundance ratio almost 
identical to the solar value (\citealt{BPL2003,liu2006b}). 

Detailed plasma diagnostics using ORLs indicate that ORLs may originate in
regions of \Te\, and \Ne\, quite different from those where CELs are
emitted. While new observations have confirmed the earlier results of
\cite{peimbert1971} and \cite{liu1993} that \Te's, derived
from the Balmer discontinuity of the hydrogen recombination spectrum are
systematically lower than values deduced from the collisionally excited [O~{\sc
iii}] nebular to auroral forbidden line ratio, careful analyses of the relative
intensities of He~{\sc i} ORLs suggest that He~{\sc i} ORLs may arise from
plasmas of \Te's even lower than indicated by the H~{\sc i} Balmer
discontinuity.  Much more importantly, careful comparisons of high quality
measurements of observed relative intensities of O~{\sc ii} ORLs in spectra of
PNe exhibiting particularly large adf's ($\ga 5$) clearly show that the lines
arise from plasmas of \Te~$\sim 1000$\,K, about an order of
magnitude lower than values deduced from the [O~{\sc iii}] forbidden line
ratios. The observations thus point to the presence of another component of
cold plasma, previously unknown, embedded in the diffuse nebula. The gas has
much higher metallicity and, because of the much enhanced cooling, a much lower
\Te. In addition, because of the very low 
\Te, much lower than the typical excitation energies of optical and UV
CELs, it emits essentially no optical and UV CELs, and is thus invisible via
those lines. The existence of such a cold, H-deficient component of plasma thus
provides a natural explanation for the systematic higher heavy elemental
abundances and lower \Te's deduced from ORLs compared to those
derived from CELs (Liu et al.  2000, 2006). The nature and origin of the
high-metallicity gas remain elusive.  Accurate determinations of the prevailing
physical conditions (\Te\, and \Ne) and elemental abundances
will be essential.

In this paper, we present deep, high-quality optical spectra for a sample of 25
GBPNe plus 6 GDPNe. The data are combined with archival IR and UV spectra to
study the nebular physical conditions and elemental abundances using both CELs
and ORLs. The sample, together with those published previously, mainly for
GDPNe, bring the total number of PNe for which ORL elemental abundances have
been determined to nearly a hundred. The purposes of the study are two folds:
1) to determine and characterize the distribution of adf's for a significant
sample of GBPNe and compare it to that for GDPNe; 2) to determine carbon
abundances and C/O ratios for the first time for a significant sample of GBPNe
which are poorly known for the bulge population.  The paper is organized as
follows. In Section 2, we describe briefly our new optical observations and
archival IR and UV data.  Plasma diagnostics, ionic and elemental abundance
determinations using CELs and ORLs are presented respectively in Sections 3 and
4. In Section 5, we compare our results with those in the literature. In
Section 6, we present a statistical analysis of adf's and other nebular
physical properties. In Section 7, we contrast the abundance patterns deduced
for GDPNe and for GBPNe and compare them to theoretical models. We conclude by
summarizing the main results in the last Section.

\begin{table*}
\caption{Basic data. Angular diameters are optical measurements except 
as specified.}
\label{obj}
\begin{tabular}{l  c c l l r r c c}
\hline
\noalign{\smallskip}
  Name &     PNG     &Diam &\multicolumn{2}{c}{$-$log$F(\rm H\beta)$}     &$F$(5\,GHz)&RV~~&B&V\\                                                   
       &         & ($\arcsec$) &\multicolumn{2}{c}{(erg cm$^{-2}$ s$^{-1}$)} & (mJy~~~~) & (km s$^{-1}$) & (mag) & (mag) \\ 
       &             &A92  &CKS92         & WL07                        &        &D98~~&\multicolumn{2}{c}{T91,T89}  \\
\noalign{\smallskip}
\hline
\noalign{\smallskip}
\multicolumn{9}{l}{Bulge PNe}\\
Cn 1-5   &002.2$-$09.4 & 7.0 &11.25     &11.40 &44.0 & $-$29&15.50&15.20 \\
Cn 2-1   &356.2$-$04.4 & 2.4 &11.64     &11.63 &49.0 &$-$271&    *&    * \\
H 1-41   &356.7$-$04.8 & 9.6 &11.7$^a$ &11.90 &12.0 &  76&16.30&16.20 \\
H 1-42   &357.2$-$04.5 & 5.8 &11.7$^a$ &11.68 &40.0 & $-$79&17.30&    * \\
H 1-50   &358.7$-$05.2 & 1.4$^b$ &11.68     &11.70 &31.0 &  28&    *&17.10 \\
H 1-54   &002.1$-$04.2 & 4.8 &11.89     &11.87 &31.0 &$-$116&15.70&15.40 \\
IC 4699  &348.0-13.8 & 5.0 &11.69     &11.70 &20.0 &$-$123&14.82&15.10 \\
M 1-20   &006.1+08.3 & 1.9$^c$ &11.93     &11.94 &51.0 &  91&17.70&17.10 \\
M 2-4   &349.8+04.4 & 5.0 &11.84     &11.94 &32.0 &$-$184&17.60&17.00 \\
M 2-6   &353.3+06.3 & 8.0 &12.16     &12.23 &17.0 & $-$88&16.67&16.40 \\
M 2-23   &002.2$-$02.7 & 8.5 &11.58     &11.58 &41.0 & 224&16.70&    * \\
M 2-27   &359.9$-$04.5 & 4.8 &12.21     &12.23 &50.0 & 170&    *&    * \\
M 2-31   &006.0$-$03.6 & 5.1 &12.11$^a$&12.10 &51.0 & 157&    *&    * \\
M 2-33   &002.0$-$06.2 & 5.8 &11.6$^a$&11.85 &12.0 &$-$112&14.40&14.40 \\             
M 2-39   &008.1$-$04.7 & 3.2 &12.13     &12.07 & 8.0 &  71&16.20&15.80 \\
M 2-42   &008.2$-$04.8 & 3.8 &12.12     &12.10 &14.0 & 157&18.20&    * \\
M 3-7   &357.1+03.6 & 5.8 &12.32     &12.38 &28.0 &$-$191&17.30&16.40 \\
M 3-21   &355.1$-$06.9 & 5.0$^d$ &11.42     &11.39 &30.0 & $-$68&16.20&15.30 \\
M 3-29   &004.0-11.1 & 8.2 &11.7$^a$&11.78 &12.0 &  50&15.42&15.50 \\
M 3-32   &009.4$-$09.8 & 6.0 &11.9$^a$&11.85 &12.0 &  46&17.40&17.10 \\
M 3-33   &009.6-10.6 & 5.0 &12.00     &11.93 & 7.5 & 173&15.70&15.90 \\
NGC 6439 &011.0+05.8 & 5.0 &11.71     &11.73 &55.0 & $-$93&20.20&    * \\
NGC 6565 &003.5$-$04.6 &13.6 &11.22     &11.25 &38.2 & $-$20&    *&18.50 \\
NGC 6620 &005.8$-$06.1 & 8.0 &11.74     &11.73 & 3.5 &  72&    *&19.60 \\
VY 2-1  &007.0$-$06.8 & 7.0$^d$ &11.50     &11.57 &37.0 & 115&16.60&    * \\
\noalign{\smallskip}
\multicolumn{9}{l}{Disk PNe}\\
H 1-35   &355.7$-$03.5 & 2.0 &11.52     &11.50 &72.0 & 160&15.70&15.40 \\
M 1-29   &359.1$-$01.7 & 7.6 &12.2$^a$ &12.41 &97.0 & -62&    *&    * \\
M 1-61   &019.4$-$05.3 & 1.8$^c$ &11.43     &11.46 &97.0 &  40&17.10&    * \\
NGC 6567 &011.7$-$00.6 & 7.6 &10.95     &10.94 &76.0 & 119&14.42&14.36 \\
He 2-118 &327.5+13.3 & 5.0 &11.70     &11.67 &10.0$^d$ &$-$164&18.20&18.70 \\  
IC 4846  &027.6$-$09.6 & 2.0 &11.34     &11.30 &43.0 & 151&15.19&15.19 \\ 
\hline                                
\end{tabular}                         
\begin{list}{}{}
\item[References:] A92 - \cite{acker1992}; CKS92 - \cite{CKS92}; 
D98 - \cite{durand1998}; T89 - \cite{tylenda1989}; 
T91 - \cite{tylenda1991}; WL07 - from long-slit observations of the
current work
\item[$^{a}$] From \cite{acker1992}.
\item[$^{b}$] Radio measurement from \cite{ZPB89}.
\item[$^{c}$] Radio measurement from \cite{aaquist1990}.
\item[$^{d}$] Upper limit.
\end{list}
\end{table*}

\section{Targets and Observations}
\label{obs} 

In total 31 PNe in the direction of the Galactic center were observed. The
basic parameters of the targets, including observed absolute H$\beta$ flux,
radio flux density at 5.0\,GHz, angular diameter (value measured in the optical
is adopted if available, otherwise radio measurement is used), heliocentric
radial velocity and magnitude of the central star are given in Table~\ref{obj}.
We assume that a PN probably belongs to the bulge population if it falls within
20 degrees of the Galactic center, has an angular diameter less than 20\arcsec,
a 5\,GHz radio flux density of less than 50\,mJy and a relatively large
heliocentric radial velocity \citep{ratag1997}.  25 of our targets satisfy
these criteria. For the remaining 6 nebulae, H~1-35, M~1-29, M~1-61 and
NGC~6567 have a radio flux density in excess of 70\,mJy, and IC~4846 and
He~2-118 fall nearly 30 degrees or more from the Galactic center. They are thus
more likely belonging to the disk rather than the bulge population. We include
these six PNe in our sample of GDPNe (c.f. Section~\ref{discussion}).   
 
\subsection{Optical observations}
\label{obs:opt} 

\begin{table}
\caption{Journal of ESO 1.52-m observations.}
\label{obsj}
\begin{tabular}{l c c c r}
\hline
\noalign{\smallskip}
  Name & Date &$\lambda$-range&FWHM &Exp. Time \\                                                   
\noalign{\smallskip}
       &(UT)  & (\AA)         &(\AA)&(sec)    \\ 
\noalign{\smallskip}
\hline
\noalign{\smallskip}
Cn~1-5  &07/1996 &3520--7420&4.5 &30,300\\
         &07/1995 &3994--4983&1.5 &2$\times$1200\\
Cn~2-1  &07/1995 &3520--7420&4.5 &15,300 \\
         &07/1995 &3994--4983&1.0 &2$\times$1800 \\
	 &06/2001 &3500--4805&1.5 &1200 \\ 
H~1-35   &07/1995 &3520--7420&4.5 &30,300 \\
         &07/1996 &3994--4983&1.0 &2$\times$1800\\
         &06/2001 &3500--4805&1.5 &1200 \\
         &06/2001 &4660--7260&3.0 &900  \\
H~1-41   &07/1995 &3520--7420&4.5 &30,2$\times$300\\
         &07/1996 &3994--4983&1.5 &2$\times$1200\\
H~1-42   &07/1995 &3520--7420&4.5 &30,300 \\
         &07/1996 &3994--4983&1.0 &2$\times$1800\\
H~1-50   &07/1995 &3520--7420&4.5 &30,300   \\
         &07/1995 &3994--4983&1.0 &2$\times$1800\\
	 &06/2001 &3500--4805&1.5 &1200 \\
H~1-54   &07/1996 &3527--7431&6.0 &60,300   \\
         &07/1996 &3994--4983&1.5 &2$\times$1800\\
         &06/2001 &4660--7260&1.5 &900 \\
He~2-118 &07/1996 &3520--7420&4.5 &30,300\\  
         &07/1995 &3994--4983&1.5 &2$\times$1200\\
	 &06/2001 &3500--4805&1.5 &1200 \\
IC~4699  &07/1995 &3520--7420&4.5 &90,300       \\
         &07/1995 &3993--4979&0.9 &2$\times$1800\\
IC~4846  &07/1995 &3520--7420&4.5 &30,300  \\ 
         &07/1995 &3993--4980&0.9 &2$\times$1800\\
M~1-20   &07/1996 &3527--7431&6.0 &60,300   \\
         &07/1996 &3994--4984&1.5 &2$\times$1800\\
	 &06/2001 &3500--4805&1.5 &1800 \\
M~1-29   &07/1996 &3527--7431&6.0 &60,600 \\
M~1-61   &07/1995 &3520--7420&4.5 &20,300 \\
         &07/1995 &3993--4980&0.9 &300,2$\times$1800\\
M~2-4   &07/1995 &3520--7420&4.5 &30,300   \\
         &07/1996 &3994--4984&1.5 &2$\times$1800\\
	 &06/2001 &3500--4805&1.5 &1800\\
M~2-6   &07/1996 &3520--7420&6.0 &30,300   \\
         &07/1996 &3994--4984&1.5 &1800 \\
M 2-23   &07/1995 &3520--7420&4.5 &15,300    \\
         &07/1996 &3994--4984&1.5 &2$\times$1500\\
M~2-27   &07/1996 &3527--7431&6.0 &60,600 \\
M~2-31   &07/1996 &3527--7431&6.0 &60,300\\
M~2-33   &07/1996 &3527--7431&6.0 &60,300 \\
         &07/1996 &3994--4984&1.5 &2$\times$1800\\
M~2-39   &07/1995 &3520--7420&4.5 &120,300  \\
         &07/1996 &3994--4984&1.5 &2$\times$1800\\
M~2-42   &07/1996 &3520--7431&6.0 &60,300  \\
M~3-7   &07/1996 &3527--7431&6.0 &60,300   \\
M~3-21   &07/1995 &3520--7420&4.5 &10,300  \\
         &07/1996 &3994--4984&0.9 &2$\times$1800\\
         &06/2001 &3500--4805&1.5 &2$\times$1200\\
M~3-29   &07/1996 &3520--7431&6.0 &60,300 \\
M~3-32   &07/1995 &3530--7430&4.5 &60,2$\times$300\\
         &07/1996 &3030--4050&1.5 &2$\times$1800\\
	 &07/1996 &3990--5000&1.5 &2$\times$1800\\
M~3-33   &07/1995 &3520--7420&4.5 &60,300 \\
         &07/1996 &3994--4984&1.5 &2$\times$1800\\
NGC~6439 &07/1995 &3520--7420&4.5 &30,300   \\
         &07/1996 &3994--4983&1.5 &2$\times$1200 \\
	 &06/2001 &3500--4805&1.5 &1200,2$\times$1800 \\
NGC~6565 &07/1995 &3520--7420&4.5 &90,300     \\
         &07/1995 &3993--4979&0.9 &2$\times$1800\\
\noalign{\smallskip}
\hline
\end{tabular}
\end{table}

\setcounter{table}{1}
\begin{table}
\caption{{ \it --continued} }
\begin{tabular}{l  c c c r}
\hline
\noalign{\smallskip}
  Name & Date &$\lambda$-range&FWHM &Exp. Time \\                                                   
\noalign{\smallskip}
       &(UT)  & (\AA)         &(\AA)&(sec)    \\ 
\noalign{\smallskip}
\hline
NGC~6567 &07/1995 &3520--7420&4.5 &20,300  \\
         &07/1995 &3900--4980&0.9 &600,2$\times$1800\\
         &06/2001 &3500--4805&1.5 &1200 \\
NGC~6620 &07/1995 &3520--7420&4.5 &60,300\\
         &07/1995 &4000--4987&0.9 &2$\times$1800\\
	 &06/2001 &3500--4805&1.5 &3$\times$1800\\
VY~2-1  &07/1995 &3520--7420&4.5 &30,300 \\
         &07/1996 &3994--4984&1.5 &2$\times$1800 \\
\noalign{\smallskip}	 
\hline           
\end{tabular}    
\end{table}

The spectra were obtained in 1995, 1996 and 2001 with the European Southern
Observatory (ESO) 1.52\,m telescope using the Boller \& Chivens (B\&C)
long-slit spectrograph. A journal of observations is presented in
Table~\ref{obsj}.  In 1995, the spectrograph was equipped with a Ford
2048$\times$2048 $15\,\mu{\rm m}\times 15\,\mu{\rm m}$ CCD, which was superseded in 1996 by
a UV-enhanced Loral $2048\times2048$ $15\,\mu{\rm m}\times 15\,\mu{\rm m}$ chip
and in 2001 by a Loral $2688\times 2688$ $15\,\mu{\rm m}\times 15\,\mu{\rm m}$
chip. 

Two grating setups were used. The low resolution setting yielded a FWHM of
$\sim 4.5$ or 6.0~{\AA} and covered the wavelength range from 3520~{\AA} to
7420~{\AA}. The high resolution setting gave a full width half maximum (FWHM)
of $\sim 0.9$ or 1.5~{\AA} and covered the wavelength range 3994 -- 4983 or
3500 -- 4800~{\AA}.  Typical exposure times were about 5\,min for the low
resolution setting.  For the high spectral resolution, integration times of 20
to 30\,min were used, and in most cases two or more frames were taken to
increase the signal-to-noise (S/N) ratio and to facilitate removal of cosmic
rays. The observations were aimed to detect at least the strongest ORLs of
O~{\sc ii}, i.e. the $\lambda$4649 line from Multiplet V\,1.  In order to
obtain strengths of the very strong emission lines, such as [O~{\sc iii}]
$\lambda\lambda$4959, 5007 and H$\alpha$ which saturated on long exposures,
short exposures of duration of 10 to 60 seconds were also obtained.  A slit
width of 2\arcsec\ was used for all nebular observations. However, in order to
obtain the total flux of H$\beta$, a short exposure was taken using an
8\arcsec\ wide slit for each nebula for the low spectral resolution setup. 

All nebulae were observed with a fixed position long-slit, normally passing
through the nebular center and sampling the brightest parts of the nebula. The
only exception was Cn\,1-5 for which scanned spectra were obtained by uniformly
driving a long slit across the nebular surface, thus yielding average spectra
for the entire nebula. 

All the two-dimensional long-slit spectra were reduced using the {\sc long92}
package in {\sc midas}\footnote{{\sc midas} is developed and distributed by the
European Southern Observatory.} following the standard procedure. Spectra were
bias-subtracted, flat-fielded and cosmic-rays removed, and then wavelength
calibrated using exposures of a calibration lamp. Flux-calibration was carried
out using the {\sc IRAF}\footnote{{\sc IRAF} is developed and distributed by
the National Optical Astronomy Observatories.} package by using wide-slit
spectroscopic observations of {\it HST} standard stars.

The 3.5~arcmin long-slit was long enough to cover the entire nebula for all
sample objects while still leaving ample clear area to sample the sky background.
The sky spectrum was generated by choosing sky windows on both sides of the
nebular emission while avoiding regions of background/foreground stars. The sky
spectrum was then subtracted from the two-dimensional spectrum.  After
sky-subtraction, the two-dimensional spectra were integrated along the slit
direction over the nebular surface.  For a few nebulae with a bright central
star, such as NGC~6567 and M~2-33, a few CCD rows centered on the star were
excluded in the summation to avoid strong contamination of the stellar light.

\begin{figure*}
\centering
\epsfig{file=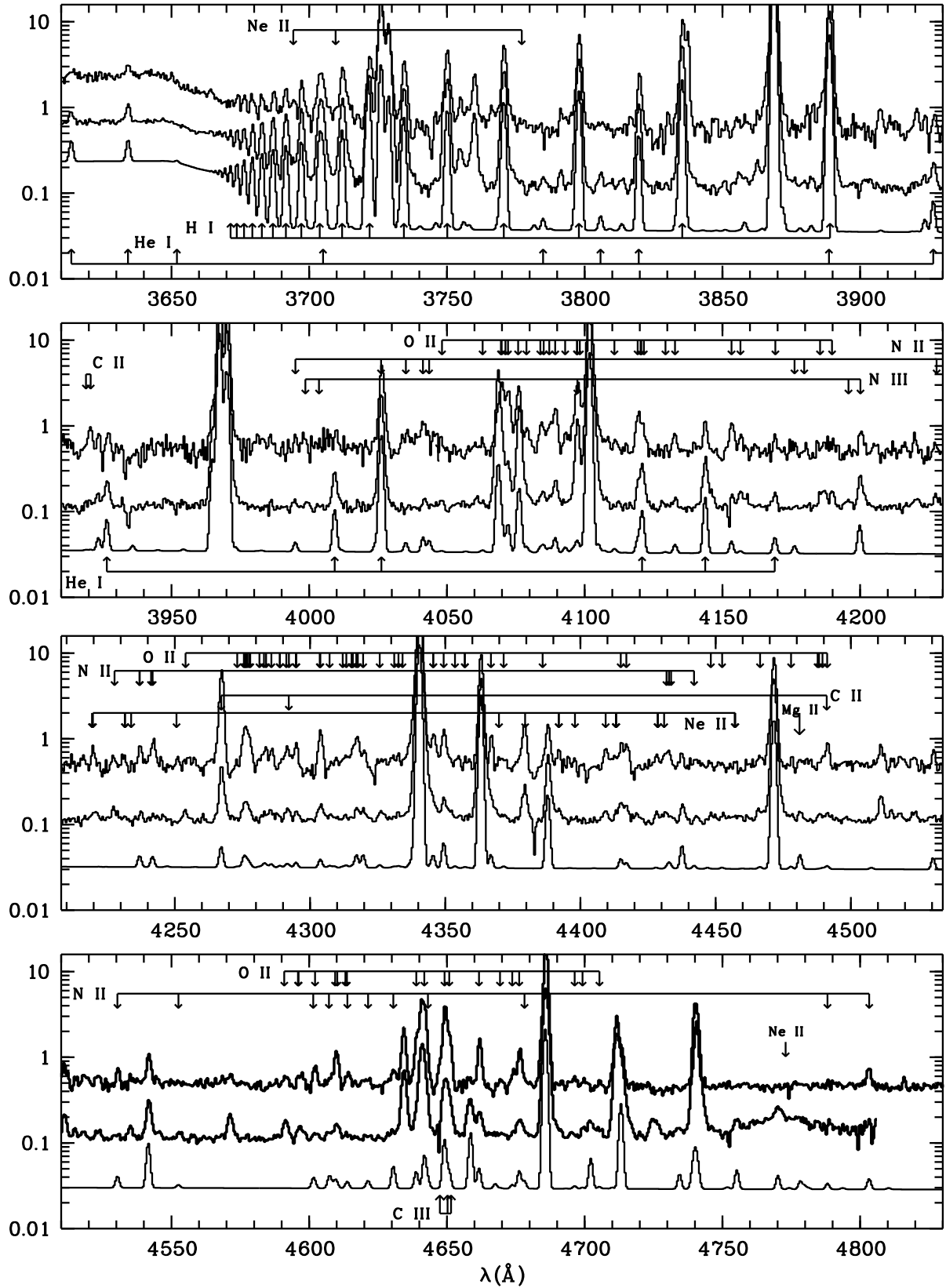, width=16.0cm, bbllx=56,
bblly=30, bburx=543, bbury=697,clip=,angle=0}
\caption{Optical spectra of M~3-32 (upper curve, scaled up by a factor of 2.5)
and M~3-21 (middle curve, scaled up by a factor of 1.5) from 3520 -- 4980~\AA, showing
the rich recombination lines from H~{\sc i}, He~{\sc i} and He~{\sc ii} as well
as C~{\sc ii}, N~{\sc ii}, O~{\sc ii}, Ne~{\sc ii}.  Also overplotted is a
synthetic spectrum for M~3-21 (lower curve), which includes contribution of
recombination continuum and line emission from hydrogen and helium as well as
CELs and ORLs from CNONe ions.  The observed spectra have been corrected to
laboratory wavelengths using H~{\sc i} Balmer lines and normalized such that
$I({\rm H}\beta) = 100$. The spectra have also been corrected for interstellar
extinction.}
\label{spectrum}
\end{figure*}

As an example illustrating the high quality of our data, we plot in
Fig.~\ref{spectrum} the spectra of M~3-32 and M~3-21 from 3520 to 4980~{\AA}
after integration along the slit.  Also overplotted in Fig.~\ref{spectrum} is a
synthetic spectrum for M~3-21, which includes contribution from recombination
lines and continua of H~{\sc i}, He~{\sc i} and He~{\sc ii} as well as from
CELs and ORLs emitted by ionized carbon, nitrogen, oxygen and neon ions,
assuming \Te\, and \Ne\, as well as ionic abundances deduced in Sections~3 and
4.

All fluxes were measured on sky-subtracted and extracted one-dimensional
spectra, using techniques of Gaussian line profile fitting in {\sc midas}. For
the strongest lines, however, fluxes obtained by simply integrating over the
observed line profile were adopted. 

\begin{figure}
\centering
\epsfig{file=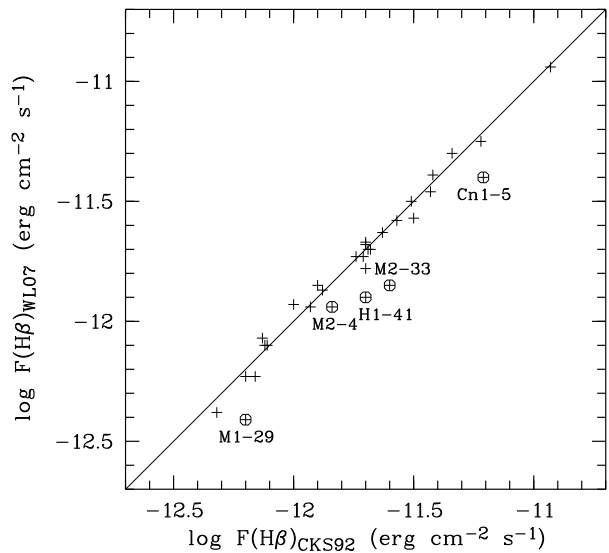, width=8cm, bbllx=62,
bblly=333,bburx=543, bbury=772,clip=,angle=0}
\caption{H$\beta$ fluxes derived from our own observations taken with an 
8\arcsec\ slit width (ordinate) are compared to values published in 
the literature as compiled by \citet{CKS92} (abscissa). The five nebulae
for which our measurements are more than 10\,per~cent lower than those given in 
\citet{CKS92} are marked with open circles and labelled by their names.
}
\label{hbeta}
\end{figure}

H$\beta$ fluxes derived from our wide slit observations are listed in
Table~\ref{obj} and compared to values published in the literature.  In
Fig.~\ref{hbeta}, we compare our own measurements with those compiled by
\citet{CKS92}. The agreement is quite good except for a few nebulae of angular
diameters larger than 8\arcsec. For those large PNe, our measurements should be
treated as lower limits.

Most ORLs from CNONe ions are quite weak and often blend together (c.f.
Fig.~\ref{spectrum}). To retrieve their intensities, multiple Gaussian fitting
was used.  Given that the typical expanding velocity of a PN is
$<25$~km~s$^{-1}$ (e.g.  \citealt{kwok1994}), much smaller than the width of
instrumental broadening ($\sim$ 55~km~s$^{-1}$ at H$\beta$ for a FWHM of
0.9~{\AA}, the best resolution of our optical spectra), all lines detected with
the same instrument setup were thus assumed to have the same line width, which
was dominated by instrumental broadening. This assumption, together with
accurately known laboratory wavelengths of identified lines, significantly
reduce the number of free (non-linear) parameters when fitting blended lines
using multiple Gaussian. Apart from this procedure, synthetic spectra (c.f.
Section 3.1), were also used to aid line identifications and flux measurements
in cases of serious line blending. The measured line fluxes, normalized such
that $F({\sc H}\beta) = 100$, are presented in Appendix~A, Table~\ref{allline}.
Fluxes of lines measured on high resolution blue spectra were normalized to
H$\beta$ via H$\gamma$, the flux of the latter relative to H$\beta$ was
obtained from low resolution spectra which covered nearly the whole optical
wavelength range.

Given the weakness of the nebular continuum and high-order Balmer lines, and
the fact that as $n$, the principal quantum number of the upper level, approaches
20, high-order Balmer lines start to blend together, good S/N's and spectral
resolution are essential for accurate measurements of the Balmer discontinuity
and decrement, and consequently for accurate determinations of the Balmer jump
temperature, \Te(BJ), and Balmer decrement density, \Ne(BD) (c.f. Section 3.4).
Among our sample of 31 PNe, deep high resolution spectra in the blue that cover
the wavelength region of the Balmer discontinuity and high-order Balmer lines
near 3646\AA\, are available for 8 GBPNe and 3 GDPNe. In subsequent analysis,
we shall call this sub-sample of PNe as `A' and the remaining nebulae (including
17 GBPNe and 3 GDPNe) as sub-sample~`B'. For PNe of sub-sample~B, values of
\Te(BJ) and \Ne(BD) determined are quite uncertain. Fortunately, given the fact
that emissivities of ORLs have only a weak power-law dependence on \Te
, quite similar to that of H~{\sc i} Balmer lines, and are nearly
independent on \Ne\, at low densities ($N_{\rm e} \la
10^6$\,cm$^{-3}$), the effects of uncertainties in \Te(BJ) and \Ne(BD) on ionic
abundances deduced from ORLs are minimal (c.f. Section 4.2).  

For a few PNe such as M~2-27 in our sample, only low resolution spectra were
available. Given the low spectral resolution and S/N ratio, the data were
insufficient to deblend and measure  weak lines.  In those cases, synthetic
spectra were used to fit the spectra in order to obtain crude estimates of 
\Te, \Ne\, and abundances from ORLs.  

\subsection{UV observations}
\label{obs:iue} 

\setcounter{table}{2}
\begin{table}
\centering
\caption{Journal of {\it IUE} observations.}
\label{iuej}
\begin{tabular}{l c r}
\hline
\hline
Name   &  Data set& Exp. Time \\
       &          & (sec)     \\
\hline
Cn~1-5 & SWP 39446  &  3000 \\   
       & SWP 39455  &  6000 \\   
       & LWP 18567  &  4000 \\   
       & LWP 18577  &  3000 \\   
\noalign{\smallskip}
H~1-42 & SWP 39447  & 12840 \\
\noalign{\smallskip}
H~1-50 & SWP 54686  & 23100 \\
\noalign{\smallskip}
IC~4846& SWP 33381  &  3600 \\
       & LWP 13130  &  4800 \\ 
\noalign{\smallskip}        
M~2-23 & SWP 44207  & 25380 \\
\noalign{\smallskip}
M~2-33 & SWP 55509  & 24000 \\
\noalign{\smallskip}
M~2-39 & SWP 33396  &  8580 \\
\noalign{\smallskip}
M~3-21 & SWP 39464  &  8460 \\
       & SWP 39465  & 10800 \\
\noalign{\smallskip}
M~3-29 & SWP 30480  &  1500 \\
       & SWP 54464  &  8400 \\
\noalign{\smallskip}
M~3-32 & SWP 54726  & 24000 \\
       & SWP 55541  & 24000 \\
\noalign{\smallskip}
M~3-33 & SWP 44199  &  7200 \\
\noalign{\smallskip}
NGC~6439&SWP 55484  & 19200 \\
\noalign{\smallskip}
NGC~6565&SWP 07984  &  2100 \\
       & SWP 35690  &  9000 \\
       & SWP 24266  &  3600 \\
       & LWP 04611  &  3600 \\
       & LWR 06955  &  2100 \\
       & LWP 15144  &  9000 \\
\noalign{\smallskip}
NGC~6567&SWP 17019  & 11160 \\
       & SWP 45362  &  7200 \\
       & SWP 45381  &  1320 \\
       & LWR 13307  &  3600 \\
       & LWP 23708  &  3600 \\
\noalign{\smallskip}
NGC~6620&SWP 45836  &  1800 \\
       & SWP 48320  &  8760 \\
       & SWP 54512  &  2400 \\
       & SWP 54700  &  3300 \\
\noalign{\smallskip}
Vy~2-1& SWP 44200  & 12240 \\
\noalign{\smallskip}
\hline
\end{tabular}
\end{table}

Sixteen among the 31 PNe have been observed with the {\it IUE} in the
ultraviolet.  The data were retrieved from the INES Archive Data Server in
Vilspa, Spain, processed with the final (NEWSIPS) extraction method.  All
spectra were obtained with the SWP and LWP/LWR cameras using the IUE large
aperture, a $10.3\arcsec\times 23\arcsec$ oval, large enough to encompass all
PNe of the current sample. The wavelength coverages of SWP and LWP/LWR spectra
are from 1150--1975 and from 1910--3300~{\AA}, respectively.  When several
spectra for a given nebula were available, they were co-added weighted by
integration time. The UV fluxes were normalized to $F({\rm H}\beta) = 100$
using the absolute H$\beta$ fluxes tabulated in Table~\ref{obj}. Absolute
H$\beta$ fluxes compiled by CKS92 were adopted, except for objects without
published reliable measurement, for which our own measurements were used (c.f.
Section 2.1 and 3.1).  A journal of {\it IUE} observation is given in
Table~\ref{iuej}. Corrections for interstellar extinction will be discussed in
Section~3.1.

Several emission lines from highly-ionized species, such as N~{\sc
v}~$\lambda$1240, have been detected in the {\it IUE} SWP spectrum of the
medium excitation class PN Cn~1-5. The nebula has a Wolf-Rayet central star
(\citealt{acker2003}). The lines show clearly P-Cygni profiles and therefore
most likely originate from wind emission of the central star.  Its optical
spectrum yields a He~{\sc ii}~$\lambda4686$ intensity of 0.9 on the scale where
$F({\rm H}\beta) = 100$.  The measurement was quite uncertain as the 4570 --
4720~{\AA} spectral region was severely contaminated by strong WR features from
C~{\sc iii} and C~{\sc iv}. At $T_{\rm e} = 10\,000$~K, the He~{\sc
ii}~$\lambda1640$/$\lambda$4686 intensity ratio has a predicted value of 6.6.
The measured optical flux of the He~{\sc ii}~$\lambda4686$ line then implies a
$\lambda1640$ line flux of 6, which is a factor of ten lower than measured from
the {\it IUE}\ spectrum (after reddening corrections).  This shows that He~{\sc
ii} emission detected in the {\it IUE}\ spectrum is indeed dominated by wind
emission, rather than from the gaseous envelope. In Section~4, we will use the
observed intensity of the He~{\sc ii}~$\lambda4686$ to obtain an estimate of
the He$^{++}$/H$^+$ abundance for the gaseous envelope of this PN.  We also
note that the blue absorption is quite weak for the C~{\sc iv} $\lambda$1550
feature and its observed flux will be used to obtain a rough estimate of the
C$^{3+}$/H$^+$ ionic abundance for this PN.

\subsection{Infrared observations}
\label{obs:iso} 

Three objects in our sample were observed with the Short Wavelength
Spectrometer (SWS) and Long Wavelength Spectrometer (LWS) on board {\it ISO}.
Cn~1-5 has one SWS01 and one LWS01 spectrum available. The SWS01 spectrum
covered the 2.4--45~$\mu$m wavelength range at a FWHM of approximately
0.3~$\mu$m, whereas the LWS01 covered from 43--197~$\mu$m at a FWHM of about
0.6~$\mu$m. For M\,2-23, several lines in the SWS wavelength range were scanned
using the SWS02 mode, which yielded a spectral resolving power between 1000 and
4000, depending on wavelength as well as on source angular extension. Finally
the SWS02 and LWS02 observing modes were used to scan a number of lines in
NGC\,6567 in both the SWS and LWS wavelength ranges. 

The angular sizes of the three nebulae are small enough such that both SWS and
LWS observations should capture the total flux from the entire nebula.  Line
fluxes measured from the infrared spectra were normalized to $F({\rm H}\beta) =
100$ via the absolute H$\beta$ flux listed in Table~\ref{obj} which were
compiled by CKS92. The normalized line fluxes are listed in
Table~\ref{isoline} and then dereddened as discussed in the next Section.

\section{Nebular analysis}
\label{anal}

Our analyses follow closely the procedures outlined in \cite{liu2001} who
presented detailed studies of two GBPNe, M~1-42 and M~2-36. Firstly, extinction
curves towards individual sightlines were derived from radio continuum flux
density and H~{\sc i}, He~{\sc i} and He~{\sc ii} recombination lines.  Plasma
diagnostic analyses were then carried out using both CELs and ORLs, followed by
ionic and total elemental abundance determinations. 

\subsection{Extinction correction}
\label{anal:ext}

Before any nebular analyses, measured line fluxes need to be corrected for
effects of extinction by intervening dust grains along the sightline.  Some
PNe, such as the nearby NGC~7027 and NGC~6302, are also known to harbour large
amounts of local dust (\citealt{seaton1979}; \citealt{middlemass1990};
\citealt{lester1984}). For GDPNe, it is generally sufficient to use the
standard Galactic extinction law for the general diffuse ISM,
which has a total-to-selective extinction ratio $R_{\rm v} = A({\rm
V})/E_{\rm B-V} = A({\rm V})/[A({\rm B})-A({\rm V})]$ = 3.1 (c.f.
\citealt{howarth83}; H83 hereafter). In that case, the amount of extinction
can be characterized by a single parameter, the logarithmic extinction at
H$\beta$, $c({\rm H}\beta)$. However, large variations in reddening curves
towards individual sightlines are also observed, particularly in the UV
(\citealt{massa1989}). It is found that the linear rise of extinction in the far-UV
decreases dramatically with $R_{\rm v}$ and suggests that in dense regions,
carriers of the far-UV extinction, including those of the 2175~\AA\ bump reside
in large grains (\citealt{cardelli1989}). Earlier work on extinction towards
the Galactic bulge in the optical and UV region indicates a steep reddening
law in the UV, with values of $R_{\rm v}$ ranging from 1.75 to 2.7
(\citealt{walton1993}; \citealt{ruffle2004}). 

\cite{liu2001} derived extinction curves towards M~1-42 and M~2-36 by comparing
the observed Balmer decrement, the He~{\sc ii} $\lambda1640/\lambda4686$ and
$\lambda3203/\lambda4686$ ratios, and the ratio of total H$\beta$ flux to radio
{\rm f-f} continuum flux density with the predictions of recombination theory
(\citealt{storey1995}). In the current work the same method was adopted.
Firstly, intrinsic intensities $I(\lambda)$ of H~{\sc i} Balmer lines were
calculated from the radio continuum flux density at 5\,GHz or 1.4\,GHz,
assuming that the nebula is optically thin at those frequencies.  Comparison
between the predicted and observed fluxes of individual lines yields the
extinction value $A(\lambda)$ at the wavelength of the line. Similarly, from
the observed He~{\sc ii}$\lambda$1640/4686 ratio, the differential extinction
between the two wavelengths, $A(\lambda1640) - A(\lambda4686)$, can be derived.
The value, when combined with $A(\lambda4686)$, obtained by interpolating
$A(\lambda4340)$ and $A(\lambda4861)$, then yields $A(\lambda1640)$. In this
way, extinction values at a set of wavelengths can be established, allowing
characterization of the extinction law towards individual objects.

Since the ratio of the radio continuum flux density to F(H$\beta$) depends
weakly on \Te\, and helium ionic abundances, a preliminary  plasma diagnostic
and abundance analysis were carried out assuming the standard ISM reddening law
(\citealt{howarth83}). Radio flux densities at 1.4\,GHz for our sample nebulae
were mostly taken from \cite{condon98}, whereas those of 5\,GHz were mainly
from \cite{ZPB89}.  For total H$\beta$ flux, $F({\rm H}\beta)$, we adopted
values compiled by \cite{CKS92}. For objects without published reliable
measurement of $F({\rm H}\beta)$, we have used our own measurements obtained
using an 8-arcsec wide slit. We check the accuracy of our measurements in
Fig.~\ref{hbeta} by comparing our wide-slit measurements and those from the
literature. The agreement is good. The four out of five objects for which our
measured values are over 10\,per~cent smaller than those given in the
literature have optical diameters larger than 7\arcsec~ and thus the
discrepancies are likely caused by our finite slit width.  For M~2-33, a
compact nebula with an optical diameter less than
6\arcsec~(\citealt{dopita1990}), we prefer to use our own measurement as the
value given by \cite{acker1992} was noted to be quite uncertain.
  
From the radio flux and total H$\beta$ flux and adopting preliminary values of
derived \Te\, and He ionic abundances, we obtain $A({\rm
H}\beta)$, the absolute extinction at H$\beta$, which was then used together
with the observed intensity ratios of H~{\sc i}, He~{\sc i} and He~{\sc ii}
lines to calculate extinction values at the corresponding wavelength. During
this process, some cautions have to be exercised. For example, in some PNe,
H$\epsilon$ was blueshifted to the position of the interstellar Ca~{\sc ii} H
line and absorbed by it. Similarly, in some PNe the He~{\sc ii}~$\lambda$1640
line is contaminated by wind emission from the central star.

\begin{figure}
\centering
\epsfig{file=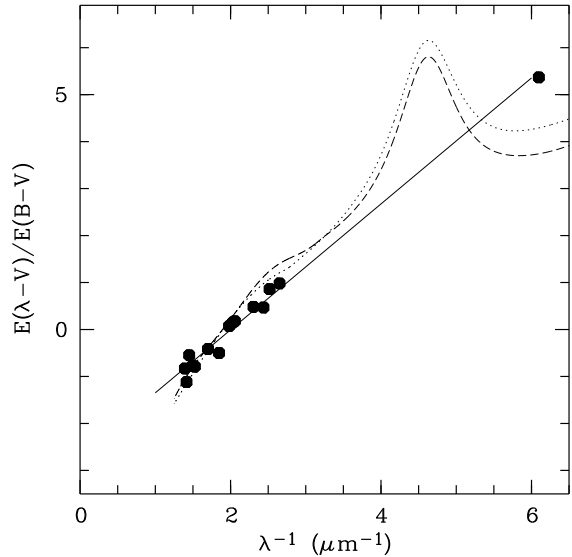, width=7.5cm, bbllx=100,
bblly=334, bburx=543, bbury=772,clip=,angle=0}
\caption{Normalized extinction $X(x)$ towards NGC~6565. The solid line 
is a linear fit to the data, $X(x) =  -2.69 + 1.34x$. The dotted line and 
long-dashed lines are standard Galactic reddening law for $R_{\rm V} = 3.1$ 
given by H83 and CCM89, respectively.}
\label{NGC6565extin}
\end{figure}

As an example, Fig.~\ref{NGC6565extin} plots $X$ towards NGC\,6565 as a
function of $x$, where $x \equiv 1/\lambda(\mu{\rm m})$ and $X(x) \equiv
E(\lambda - {\rm V})/E({\rm B} - {\rm V}) = [A(\lambda) - A({\rm V})]/A({\rm
B}) - A({\rm V})]$.  Also overplotted in the Figure are standard Galactic ISM
extinction laws given by \cite{howarth83} (H83; dotted line) and
\cite{cardelli1989} (CCM89; long-dashed line) for a total-to-selective ratio
$R_{\rm V} = 3.1$. The extinction curve deduced for the sightline towards
NGC\,6565 is quite similar to previously found for the bulge PN M\,1-42 by
\cite{liu2001} -- the extinction in the far ultraviolet as derived from the
He~{\sc ii}~$\lambda1640/\lambda4686$ ratio is much higher than predicted by
the standard Galactic ISM extinction law.  It is unfortunate that there are no
lines usable near the prominent 2175\,{\AA} bump present in the standard
extinction law. Following \cite{liu2001}, we fit $X$ as a function of $x$ with
a straight line, which seems to be appropriate for available data. For
NGC\,6565, the result is also shown in Fig.~\ref{NGC6565extin}.

\begin{table}
\caption{Extinction towards individual nebulae.}
\centering
\label{allextin}
\begin{tabular}{l c c c c c c r }
\hline
\noalign{\smallskip}
Name     &\multicolumn{4}{c}{$c({\rm H}\beta)$}& $R_{\rm V}$ & $f(\lambda)^d$ \\ 
\noalign{\smallskip}
\cline{2-5}
\noalign{\smallskip}
         &opt. & rad.  & UV & adopt &      &           \\ 
\noalign{\smallskip}
\hline
\noalign{\smallskip}
Cn~1-5       & 0.49 & 0.39 & 0.59$^b$ & 0.49 & 3.05  & $-$0.91,0.44\\
Cn~2-1       & 1.07 & 0.83 &          & 0.83 & 3.10  & CCM89 \\
H~1-35       & 1.51 & 0.96 &          & 1.31 & 3.27  & CCM89 \\
H~1-41       & 0.65 & 0.67 &          & 0.65 & 3.45  & $-$1.21,0.59\\ 
H~1-42       & 0.87 & 0.74 &          & 0.77 & 3.35  & $-$1.31,0.64\\ 
H~1-50       & 0.68 & 0.53 &          & 0.68 & 3.70  & $-$1.15,0.56\\
H~1-54       & 1.54 & 0.68 &          & 1.53 & 1.92  & $-$2.26,1.09\\
He~2-118$^a$ & 0.17 & 0.14 &          & 0.16 & 3.85  &           \\
IC~4699      & 0.23 & 0.45 & 0.16     & 0.22 & 3.21  & $-$1.42,0.69\\
IC~4846      & 0.69 & 0.34 &          & 0.65 & 2.79  &      CCM89  \\
M~1-20       & 1.40 & 1.02 &          & 1.13 & 3.40  &      CCM89  \\
M~1-29       & 2.03 & 1.63 &          & 1.83 & 3.10  &      CCM89  \\
M~1-61       & 0.92 & 0.92 &          & 0.92 & 1.81  & $-$2.31,1.12 \\
M~2-4        & 1.33 & 0.81 &          & 0.81 & 1.97  &      CCM89  \\
M~2-6        & 1.14 & 0.84 &          & 0.85 & 2.62  &      CCM89  \\
M~2-23       & 1.20 & 0.98 &          & 1.12 & 2.97  &      CCM89  \\
M~2-27       & 1.31 & 1.37 &          & 1.35 & 2.61  &      CCM89  \\
M~2-31       & 1.41 & 1.18 &          & 1.15 & 3.07  &      CCM89  \\
M~2-33       & 0.55 & 0.61 &          & 0.55 & 3.00  &      CCM89  \\
M~2-39       & 0.61 & 0.63 &          & 0.63 & 1.92  &      CCM89  \\
M~2-42       & 1.06 & 0.66 &          & 0.65 & 1.97  &      CCM89  \\
M~3-7        & 1.65 & 1.51 &          & 1.46 & 4.18  &      H83   \\
M~3-21       & 0.50 & 0.37 & 0.78     & 0.35 & 2.92  & $-$1.27,0.62\\
M~3-29       & 0.24 & 0.12 &          & 0.12 & 1.86  &      CCM89  \\
M~3-32       & 0.64 & 0.40 & 0.69     & 0.62 & 3.10  &      H83   \\
M~3-33       & 0.50 & 0.42 & 0.32     & 0.39 & 3.31  &      CCM89  \\
N~6439       & 1.10 & 0.93 & 1.15     & 0.93 & 2.81  &      CCM89  \\
N~6565       & 0.32 & 0.36 & 0.52     & 0.37 & 3.10  & $-$1.03,0.50\\
N~6567       & 0.90 & 0.66 & 0.73$^c$ & 0.73 & 3.36  &      H83   \\
N~6620       & 0.52 & 0.46 & 0.86     & 0.43 & 3.10  &      CCM89  \\
Vy~2-1       & 0.83 & 0.66 &          & 0.64 & 2.80  &      CCM89  \\
\hline
\end{tabular}
\begin{list}{}{}
\item[$^{a}$] Only a lower limit for the 5\,GHz flux is available from the 
literature. Thus $c({\rm H}\beta)$ derived from the radio continuum to 
H$\beta$ flux ratio is an upper limit, as is $R_{\rm V}$.
\item[$^{b}$] Derived from the He~{\sc ii} $\lambda$1640/$\lambda$3203 ratio.
\item[$^{c}$] Derived from the UV continuum.
\item[$^{d}$] Extinction law used to fit the extinction curve. 
H83 and CCM89 denote extinction laws given by \cite{howarth83} and 
\cite{cardelli1989}, respectively. When two numbers are listed, it denotes
a linear extinction law, $f(\lambda) = a + bx$, and the two numbers refer to
values of $a$ and $b$, respectively. $x \equiv 1/\lambda(\mu{\rm m})$.  
\end{list}
\end{table}

Excluding He~2-118, for which only a lower limit on its 5\,GHz flux density is
available from the literature, for the remaining 30 sample PNe, the normalized
extinction, $X(x)$, was derived for each object at wavelengths of all usable
lines. The logarithmic extinction, $f(\lambda)$, commonly used in nebular
analyses and defined such that  $f(\lambda) = 0$ at H$\beta$, was also
calculated.  $f(\lambda) = 0.4A(\lambda)/c({\rm H}\beta) - 1$, where $c({\rm
H}\beta) = 0.4A({\rm H}\beta)$ is the logarithmic extinction at H$\beta$. The
deduced normalized extinction curve was then compared with the H83 law and the
CCM89 law. For eight sample nebulae, reliable fluxes of the He~{\sc ii}
$\lambda$1640 line and/or the $\lambda$3203 line are available, providing an
estimate of the extinction in the UV.  For four of them, the extinction $X(x)$
shows a very good linear relation extending from optical to the UV. For the
four objects, a linear function was used to fit $X(x)$ as a function of $x$.
The resultant extinction curve was then used to deredden the observed line
flux.  For the other four PNe, either the H83 or the CCM89 law was used to fit
the data, depending on which one yielded a better fit. We note that the two
extinction laws have noticeable differences in the 3500--4500\,\AA\
wavelength range.

For the 22 remaining nebulae for which no estimate of UV extinction was
possible, the extinction curve was either fitted with a linear function, the
H83 law or the CCM89 law, whichever yielded the best fit.  The extinction law
used for each nebula is listed in the last column of Table~\ref{allextin}. If
an entry in this column consists of two numbers, it denotes a linear extinction
law, $f(x) = a + bx$, was used, and the two numbers refer to fitted values of
parameters $a$ and $b$, respectively. In Table~\ref{allextin}, we also list
values of $c({\rm H}\beta)$ derived from ratios of H$\alpha$/H$\beta$
(Column~2), $F_{\nu}$(5GHz)/H$\beta$ (Column~3) and of He~{\sc ii}
$\lambda$1640/$\lambda$4686 (Column~4), as well as our final adopted values
(Column~5) used to deredden the observed line fluxes. Finally, Column~6 of
Table~\ref{allextin} gives the total-to-selective extinction ratio $R_{\rm V}$.

\subsection{CEL plasma diagnostics}
\label{anal:celdiag}

\begin{table}
\caption{Plasma diagnostic line ratios and ionization potentials of the 
emitting ions.  Subscripts 'n', 'a' and 'f' represent nebular, auroral and 
fine-structure lines, respectively.}
\centering
\label{diagtab}
\begin{tabular}{l l c}
\hline
\noalign{\smallskip}
ID &  & IP (eV) \\
\hline
\noalign{\smallskip}
   & $T_{\rm e}$-diagnostics & \\          
\noalign{\smallskip}
 {[O\,{\sc ii}]}$_{\rm na}$  & $I(\lambda7320+\lambda7330)/I(\lambda3726+\lambda3729)$ & 13.6 \\
 {[N\,{\sc ii}]}$_{\rm na}$  & $I(\lambda6548+\lambda6584)/I(\lambda5754)$ & 14.5 \\
 {[O\,{\sc iii}]}$_{\rm na}$ & $I(\lambda4959+\lambda5007)/I(\lambda4363)$ & 35.1 \\
{[Ne\,{\sc iii}]}$_{\rm nf}$ & $I(15.5\mu{\rm m})/I(\lambda3868)$ & 41.0 \\
{[Ar\,{\sc iii}]}$_{\rm nf}$ & $I(9.0\mu{\rm m})/I(\lambda7135)$ & 27.6 \\
\\
\noalign{\smallskip}
    & $N_{\rm e}$-diagnostics & \\
\noalign{\smallskip}
{[S\,{\sc ii}]}$_{\rm nn}$   & $I(\lambda6731)/I(\lambda6716)$ & 10.4   \\
{[O\,{\sc ii}]}$_{\rm nn}$   & $I(\lambda3729)/I(\lambda3726)$ & 13.6   \\
{[Cl\,{\sc iii}]}$_{\rm nn}$ & $I(\lambda5537)/I(\lambda5517)$ & 23.8    \\
{[Ar\,{\sc iv}]}$_{\rm nn}$ & $I(\lambda4740)/I(\lambda4711)$  & 40.7    \\
{[Ne\,{\sc iii}]}$_{\rm ff}$  & $I(15.5\mu{\rm m})/I(36.0\mu{\rm m})$ & 41.0    \\
\hline
\end{tabular}
\end{table}

\begin{table*}
\caption{Electron temperatures and densities from CELs. `H' and `L' denote the
measured line ratio exceeds its high and low temperature/density limit,
respectively, whereas `E' indicates that recombination contribution as
estimated using Eqs.\,(1) and (2) exceeds the actual measured flux of the
[N~{\sc II}] $\lambda5755$ line or of the [O~{\sc II}] 
$\lambda\lambda7320,7330$ lines.}
\label{tenecel}
\begin{tabular} {l c c c c c | c c c c l}
\hline
\noalign{\smallskip}
Nebula    &\multicolumn{3}{c}{$T_{\rm e}$(CELs) (K)} & & \multicolumn{4}{c}{$N_{\rm e}$(CELs) (cm$^{-3}$)}\\
          &[O~{\sc iii}]$_{\rm na}$& [N~{\sc ii}]$_{\rm na}$&[O~{\sc ii}]$_{\rm na}$&&[O~{\sc ii}]$_{\rm nn}$ &[S~{\sc ii}]$_{\rm nn}$ &[Cl~{\sc iii}]$_{\rm nn}$ &[Ar~{\sc iv}]$_{\rm nn}$ \\  
\noalign{\smallskip}
\cline{2-4}
\cline{6-9}
\noalign{\smallskip}
\noalign{\smallskip}
  He 2-118 & 12630 &14960 &   H  &&4.03 &3.95 &4.06 &4.47 \\
  M 2-4    &  8570 & 9920 &14570 &&3.72 &3.82 &3.83 &3.98 \\
  M 2-6    & 10100 & 9730 &11650 &&     &3.90 &3.80 &3.92 \\
  M 3-7    &  7670 & 6900 & 5110 &&     &3.74 &3.43 &     \\
  M 1-20   &  9860 &11180 &16570 &&4.02 &4.00 &3.95 &4.05 \\
  NGC 6439 & 10360 & 9270 &11340 &&3.59 &3.70 &3.70 &3.83 \\
  H 1-35   &  9060 &12080 &13720 &&4.45 &4.69 &4.54 & L   \\
  M 1-29   & 10830 & 9020 & 9000 &&     &3.51 &3.66 &3.68 \\
  Cn 2-1   & 10250 &12030 &   H  &&3.82 &3.70 &3.89 &4.34 \\ 
  H 1-41   &  9800 & 9530 &12790 &&     &3.11 &3.07 &2.96 \\
  H 1-42   &  9690 & 9050 &13270 &&     &3.97 &3.72 &3.90 \\
  M 2-23   & 11980 &   H  &   H  &&     &4.23 &4.08 &4.76 \\
  M 3-21   &  9790 &12800 &19000 &&3.79 &4.08 &4.03 &4.41 \\
  H 1-50   & 10950 &12070 &15360 &&3.83 &3.87 &3.96 &4.15 \\
  M 2-27   & 11980 & 8650 & 8840 &&     &3.88 &4.11 &4.12 \\
  H 1-54   &  9540 &11600 &11360 &&     &4.11 &4.12 &     \\
  NGC 6565 & 10300 &10100 &10200 &&     &3.25 &3.19 &2.82 \\
  M 2-31   &  9840 &11370 &13470 &&     &3.82 &3.84 &3.69 \\
  NGC 6567 & 10580 &10016 &14360 &&3.93 &3.85 &3.89 &3.96 \\
  M 2-33   &  8040 & 9150 &11070 &&     &3.21 & L   &3.14 \\
  IC 4699  & 11720 &12490 & 9390 &&     &3.49 & L   &3.06 \\
  M 2-39   &  8050 & 8300 &11030 &&     &3.68 &3.17 &3.38 \\
  M 2-42   &  8470 & 9350 &11860 &&     &3.51 &3.46 &3.62 \\
  NGC 6620 &  9590 & 8630 & 9630 &&3.35 &3.39 &3.44 &3.43 \\
  Vy 2-1   &  7860 & 8580 & 8750 &&     &3.51 &3.69 &3.52 \\ 
  Cn 1-5   &  8770 & 8250 &10330 &&     &3.66 &3.52 &3.36 \\
  M 3-29   &  9190 & 8750 & 8930 &&     &2.91 &     &     \\
  M 3-32   &  8860 & 8230 &   E  &&3.55 &3.41 &2.94 &3.13 \\
  M 1-61   &  8900 &11510 & 8460 &&     &4.30 &4.24 &4.40 \\
  M 3-33   & 10380 &      & 7480 &&     &3.01 &3.19 &3.56 \\
  IC 4846  &  9930 &13280 &   H  &&     &3.82 &3.70 &3.92 \\
\noalign{\smallskip} 
\hline 
\end{tabular}
%\begin{list}{}{}
%\item $^a$ A superscript ``$^s$'' indicates that the measured line ratios were based on scanned spectra.
%\end{list}  
\end{table*}

Accurate determinations of abundances rely on the reliable measurements of
\Te\, and \Ne. In what follows, we present detailed plasma
diagnostics using both CELs and recombination lines/continua. Values of
\Ne\, and \Te, have been derived from CEL
diagnostic ratios by solving equations of statistical equilibrium of
multi-level ($\ge 5$) atomic models using {\sc equib}, a Fortran code
originally written by I. Howarth and S. Adams. CEL diagnostic lines used are
tabulated in Table~\ref{diagtab}, together with the ionization potentials
required to create the emitting ions. The atomic parameters used in the present
work are the same as those used by \cite{liu2000} in their case study of
NGC~6153.

\begin{figure}
\epsfig{file=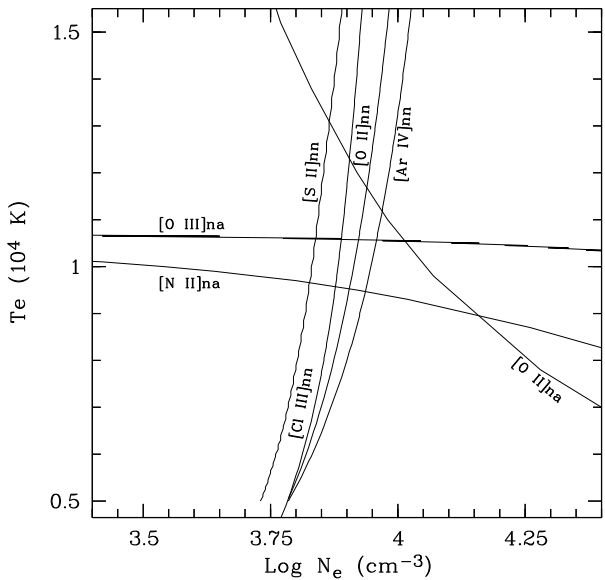, width=8cm, bbllx=56,
bblly=305, bburx=543, bbury=772,clip=,angle=0}
\caption{CEL plasma diagnostic diagram for NGC~6567. For diagnostic ID's, c.f.
Table~\ref{diagtab}.}
\label{diag_NGC6567}
\end{figure}

\begin{figure}
\epsfig{file=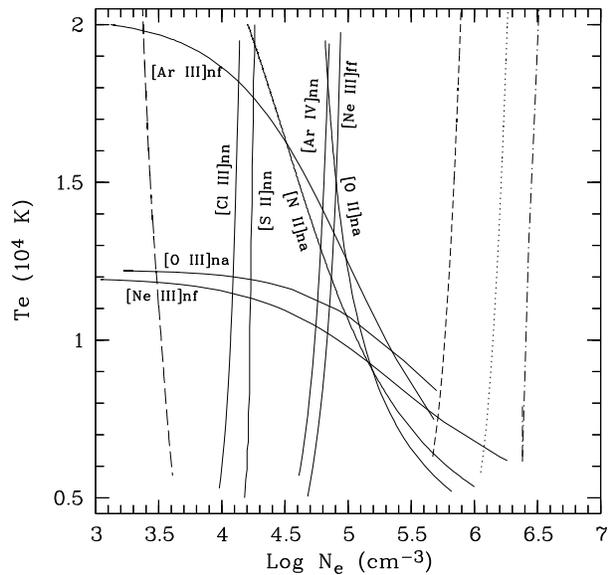, width=8cm, bbllx=56,
bblly=305, bburx=547, bbury=772,clip=,angle=0}
\caption{Plasma diagnostic diagram for M~2-23. For diagnostic ID's, c.f.
Table~\ref{diagtab}. The four broken lines represent density diagnostics from 
[Fe~{\sc iii}] lines (c.f. text for more details).} 
\label{diag_M2_23} 
\end{figure}

In order to obtain self-consistent results for \Ne\, and \Te\, probed by CEL
lines, we proceed as following. Firstly, \Ne([Cl~{\sc iii}]) and/or $N_{\rm
e}$([Ar~{\sc iv}]) were obtained assuming a \Te\, of 10\,000~K and then the
average value of the derived \Ne\, was adopted in calculating $T_{\rm
e}$([O~{\sc iii}]). The calculations were iterated to obtain final
self-consistent values of \Ne\, and \Te. We assume the results thus
obtained represent physical conditions of the high ionization regions.
Similarly, \Te([N~{\sc ii}]) was derived in combination with \Ne([O~{\sc ii}])
and \Ne([S~{\sc ii}]) to represent the physical conditions in the low
ionization zones. If all observed optical \Ne-indicators yielded similar
values of \Ne, the results were averaged and used for the whole nebulae, both
the high and low ionized regions. Recombination contributions to intensities of
the [O~{\sc ii}] and [N~{\sc ii}] auroral lines were corrected for using the
formulae derived by ~\cite{liu2000}, 
\begin{equation} \frac{I_{\rm R}({\lambda5755})}{I({\rm H}\beta)} = 
    3.19t^{0.30}\times \frac{{\rm N}^{2+}}{{\rm H}^+}, 
\end{equation} 
\begin{equation} 
\frac{I_{\rm R}({\lambda7320+\lambda7330})}{I({\rm H}\beta)} = 
   9.36t^{0.44}\times \frac{{\rm O}^{2+}}{{\rm H}^+}, 
\end{equation}
where  ${\rm O}^{2+}$/${\rm H}^+$
and ${\rm N}^{2+}$/${\rm H}^+$ are ionic abundances derived from ORLs, 
and $t\equiv T_{\rm e}/10^{4}$K. 

\cite{wang2004} have examined the four optical \Ne-diagnostics,
[O~{\sc ii}]~$\lambda3729/\lambda3726$, [S~{\sc ii}]~$\lambda6716/\lambda6731$,
[Cl~{\sc iii}]~$\lambda5517/\lambda5537$ and [Ar~{\sc
iv}]~$\lambda4711/\lambda4740$, for a sample of over 100 PNe, including 31 PNe
in the current sample. The atomic parameters recommended in their work were
adopted in the current study. For objects in common, \Ne\, presented here
differ by no more than 1\,per~cent from those published by \cite{wang2004}.

Fig.~\ref{diag_NGC6567} and \ref{diag_M2_23} present plasma diagnostic diagrams
for NGC~6567 and M~2-23, respectively, as two examples.  In the case of
NGC\,6567, various \Ne-diagnostic lines yield consistent values of \Ne, so a
constant \Ne\, has been adopted in calculating \Te\, from the [N~{\sc
ii}$]_{\rm na}$ and [O~{\sc iii}$]_{\rm na}$ nebular to auroral diagnostic line
ratios.  The two \Te's thus obtained were then adopted in calculating CEL
abundances for singly ionized species and for species of higher ionization
degrees, respectively.

Electron temperatures and densities derived from various CEL diagnostics are
tabulated in Table~\ref{tenecel}. In the Table, `H' and `L' denote the measured
line ratio exceeds its high and low temperature/density limit, respectively,
whereas `E' indicates that recombination contribution as estimated using
Eqs.(1) and (2) exceeds the actual measured flux of the [N~{\sc II}]
$\lambda5755$ line or of the [O~{\sc II}] $\lambda\lambda7320,7330$ lines.

\begin{figure}
\centering
\epsfig{file=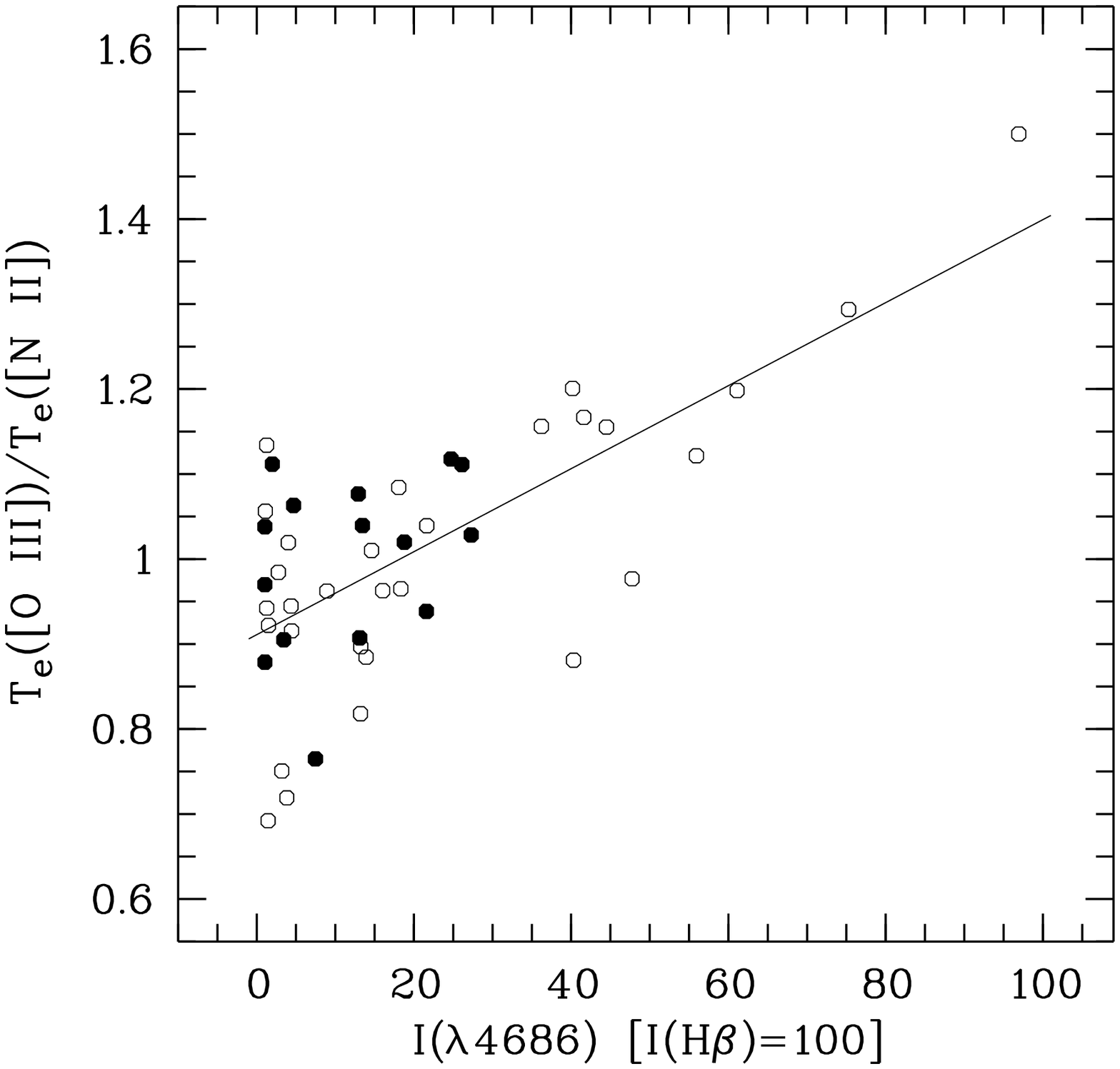, width=7.0cm, bbllx=32,
bblly=283, bburx=542, bbury=774,clip=,angle=0}
\caption{\Te([O~{\sc iii}])/\Te([N~{\sc ii}]) ratios against $I$($\lambda$4686) for 
GBPNe (filled circles) and GDPNe (open circles). The solid line is a linear fit to 
the 45 points in the Figure.}
\label{teoiiivsnii} 
\end{figure}

Kingsburgh \& Barlow (1994, KB94 hereafter) have examined the relation between
\Te([O~{\sc iii}]) and \Te([N~{\sc ii}]) for a sample of Galactic PNe. They
found that the ratio of the former to the later increases with increasing
$I$($\lambda$4686), the intensity of He~{\sc ii}~$\lambda$4686, 
\[ \frac{T_{\rm
e}(\mbox{[O\,{\sc iii}]})}{T_{\rm e}(\mbox{[N\,{\sc ii}]})} =
(1.15\pm0.06)+(0.0037\pm0.0009)I(\lambda4686){,} \] with a correlation
coefficient of 0.69. A similar analysis for GBPNe in the WL07 sample and GDPNe
in the TLW sample (c.f. \S~\ref{discussion}) exhibits similar positive
correlation, as shown in Fig.~\ref{teoiiivsnii}. A linear fit to the 45 points
in the Figure yields: \[ \frac{T_{\rm e}(\mbox{[O\,{\sc iii}]})}{T_{\rm
e}(\mbox{[N\,{\sc ii}]})} = (0.91\pm0.02)+(0.0049\pm0.0008)I(\lambda4686){,} \]
with a correlation coefficient of 0.70, which is roughly in agreement with that
obtained by \cite{KB94}. 

\subsection{Two interesting objects}
\label{anal:2obj}

\begin{table*}
\caption{Electron temperatures deduced from H~{\sc i} Balmer
discontinuity, from He~{\sc i} recombination lines and from O~{\sc ii} ORLs,
as well as electron density derived from the H~{\sc i} Balmer decrement.
An `L' denotes that the measured O~{\sc ii} $\lambda$4089/$\lambda$4649 exceeds
its low temperature limit of 0.41 at 288~K.}
\label{teneorl}
\begin{tabular} {l c c c c c c c c c}
\hline
\noalign{\smallskip}
Nebula    & \multicolumn{1}{c}{$T_{\rm e}$(H~{\sc i}) (K)} &  \multicolumn{4}{c}{$T_{\rm e}$(He~{\sc i}) (K)} & $T_{\rm e}$(O~{\sc ii}) (K) & & \multicolumn{1}{c}{$N_{\rm e}$(H~{\sc i}) (cm$^{-3}$)} \\
\noalign{\smallskip}
          &BJ &$\frac{\lambda6678}{\lambda4471}$&$\frac{\lambda5876}{\lambda4471}$ &$\frac{\lambda7281}{\lambda5876}$ &$\frac{\lambda7281}{\lambda6678}$&$\frac{\lambda4649}{\lambda4089}$ && BD  \\  
\noalign{\smallskip}
\noalign{\smallskip}
\hline 
\noalign{\smallskip}
  He 2-118 &14500 &12050 &12870  &5250 & 6800 &       &&5.0 \\
  M 2-4    & 7900 &      &14100  &6100 & 6100 &       &&4.6 \\
  M 2-6    &11700 & 4300 & 2860  &6200 & 6500 &   L   &&    \\
  M 3-7    & 6900 & 8100 & 2580  &1600 & 2500 &       &&    \\
  M 1-20   &12000 &      &       &5100 & 6000 &       &&4.2 \\
  NGC 6439 & 9900 & 2060 & 1810  &4810 & 4900 &  851  &&5.5 \\
  H 1-35   &12000 &      &       &5800 & 7300 &  334  &&4.8 \\
  M 1-29   &10000 & 6900 &       &4000 & 4100 &       &&    \\
  Cn 2-1   &10800 & 6750 &       &5050 & 5450 &  857  &&5.0 \\ 
  H 1-41   & 4500 & 7110 & 5770  &2690 & 2930 &   L   &&    \\
  H 1-42   &10000 & 3700 & 2250  &5890 & 6290 &   662 &&    \\
  M 2-23   & 5000 & 4900 & 1320  &6100 & 7330 &   410 &&    \\
  M 3-21   &10400 & 3480 & 2150  &5170 & 5520 &   602 &&5.0 \\
  H 1-50   &12000 & 9050 & 3100  &5420 & 6480 &   L   &&5.1 \\
  M 2-27   &14000 & 8410 &       &2280 & 2920 &       &&    \\
  H 1-54   &12500 &13600 &10400  &4270 & 5840 &       &&    \\
  NGC 6565 & 8500 & 3850 & 6400  &4620 & 4090 &  11722&&    \\
  M 2-31   &14000 & 2170 & 1220  &3820 & 4460 &       &&    \\
  NGC 6567 &14000 & 7850 & 2670  &6720 & 7790 &       &&4.0 \\
  M 2-33   & 7000 & 8510 & 5200  &4050 & 4650 &   L   &&    \\
  IC 4699  &12000 & 7300 & 5950  &3310 & 2460 &   L   &&    \\
  M 2-39   & 5500 & 6810 & 7070  &6610 & 6600 &   L   &&    \\
  M 2-42   &14000 & 8000 & 7700  &3600 & 3300 &       &&    \\
  NGC 6620 & 8200 & 2500 & 2400  &3800 & 3800 & 3162  &&4.0 \\
  Vy 2-1   & 8700 & 5500 & 4550  &4750 & 4970 &   L   &&    \\ 
  Cn 1-5   &10000 &11600 &       &2450 & 2940 & 18000 &&    \\
  M 3-29   &10700 & 6800 & 4750  &1600 & 1800 &       &&    \\
  M 3-32   & 4400 & 3860 & 3100  &1560 & 1710 &   L   &&3.6 \\
  M 1-61   & 9500 &18500 &       &3500 & 5490 &       &&    \\
  M 3-33   & 5900 & 6650 & 3940  &4380 & 5020 &  1465 &&    \\
  IC 4846  &18000 & 6530 & 2430  &6080 & 6970 &  9954 &&    \\
\noalign{\smallskip} 
\hline 
\end{tabular}
\end{table*}

For most nebulae in our sample, diagnostics that sample respectively the high
and the low ionized regions generally yield \Ne\, of relatively small
differences, as in the case of NGC~6567 (Fig.~\ref{diag_NGC6567}). There are
however a few exceptions, such as M~2-23. For this object, as shown in
Fig.~\ref{diag_M2_23}, lines emitted by ions of high ionization potentials
(i.e., [Ar~{\sc iv}]$_{\rm nn}$ and [Ne~{\sc iii}]$_{\rm ff}$) yield a \Ne\, of
$\sim 10^{4.7}$\,cm$^{-3}$, approximately 0.8\,dex higher than derived from
lines of low ionization potentials, such as [S~{\sc ii}]$_{\rm nn}$.  We note
that for this nebula, the [N~{\sc ii}]$_{\rm na}$ and [O~{\sc ii}]$_{\rm na}$
nebular to auroral line ratios become sensitive to \Ne\, as well as to \Te.
M~2-23 thus shares some similarities with Mz~3 and M~2-24, in that both have
been previously found to harbour a dense emission core surrounded by an outer
lobe of much lower densities (c.f.  \citealt{zhang02,zhang03}). In
\cite{zhang02}, high critical density [Fe~{\sc iii}] diagnostic lines were used
to probe the physical conditions in the dense central emission core of Mz~3.
The [Fe~{\sc iii}] line strengths observed in M~2-23 are on the average about a
third of those detected in Mz~3.  Unfortunately, lines sensitive to \Te, such
as the $\lambda$6096 or the $\lambda$7088 lines were all too faint to be
detectable in our current spectra of M~2-23. In Fig.~\ref{diag_M2_23}, we show,
from left to right, the loci of four [Fe~{\sc iii}] \Ne-sensitive diagnostic
line ratios, $\lambda4659/\lambda4770$, $\lambda4755/\lambda4881$, 
$\lambda4770/\lambda4778$ and $\lambda4702/\lambda4734$. Except for the first
line ratio, which is less reliable given the intrinsic weakness of the [Fe~{\sc
iii}] $\lambda4770$ line, all the other three consistently yield a
\Ne\,$\sim\,$10$^6$\,cm$^{-3}$, comparable to the value of $\sim
10^{6.5}$\,cm$^{-3}$ found for Mz~3 by \cite{zhang02} and considerably higher
than values given by other \Ne-diagnostics of lower critical densities.

In the {\it ISO} SWS spectrum of M~2-23, apart from the ionic fine-structure
lines from heavy elements, H~{\sc i} Br$\alpha$, Br$\beta$ and Pf$\alpha$ lines
were also well detected.  The relative strengths of those high excitation
H~{\sc i} recombination lines have a weak dependent on \Te.  Using the emission
coefficients given by \cite{storey1995}, we determined a \Te\, of 5\,000~K,
with an uncertainty about 1\,500~K. The value is very similar to that deduced
from the H~{\sc i} Balmer discontinuity.

Another object in our sample that exhibits some interesting characteristics is
M~2-39. For this nebula, all \Ne-diagnostic line ratios measurable in the
optical, [S~{\sc ii}]$_{\rm nn}$, [Cl~{\sc iii}]$_{\rm nn}$ and [Ar~{\sc
iv}]$_{\rm nn}$, consistently yield a low \Ne\, between 1\,500 and
5\,100\,cm$^{-3}$, whereas the [N~{\sc ii}]$_{\rm na}$ ratio gives a \Te\, of
8\,300~K (Table~\ref{tenecel}). Yet the observed [O~{\sc iii}]$_{\rm na}$ ratio
shows an abnormally low value of 25, indicating a \Te\, of about 28\,000~K for
a \Ne\, of below 10\,000\,cm$^{-3}$. The [O~{\sc iii}]$_{\rm na}$ temperature
could be lower if the density prevailing in the O$^{2+}$ zone is much higher
than 10\,000\,cm$^{-3}$. Some evidence in favour of a dense O$^{2+}$ zone is
provided by a well observed feature near 7171~{\AA}, which we tentatively
identified as the [Ar~{\sc iv}] $\lambda$7170.62 line. If the identification is
correct, then its intensity relative to the [Ar~{\sc iv}]
$\lambda\lambda$4711,4740 nebular lines would indicate a \Ne\, of
$10^{6.6}$\,cm$^{-3}$. For comparison, if we assume a \Te\, of 10\,000~K, as
indicated by [N~{\sc ii}]$_{na}$ and [O~{\sc ii}]$_{na}$, then a \Ne\, as high
as $10^{6.2}$\,cm$^{-3}$ would be required to reproduce the measured [O~{\sc
iii}]$_{\rm na}$ ratio. The identification of the feature at 7171~{\AA} as
[Ar~{\sc iv}] was however marred by the fact that another [Ar~{\sc iv}] line,
at 7263~{\AA}, expected to have a strength only 30\,per~cent lower than that of
the [Ar~{\sc iv}] $\lambda$7171, was either absent or had at most a strength
only one third of the measured strength of the 7171~{\AA} feature.  Clearly,
better data are needed to clarify whether M\,2-39 also contains a dense high
excitation emission core as in the case of Mz\,3, M\,2-24 and probably also of
M\,2-23. In our current analysis, we have adopted a \Ne\, of
$10^{6.6}$\,cm$^{-3}$ for the high ionization regions of M\,2-39, assuming that
the 7171~{\AA} is indeed entirely due to [Ar~{\sc iv}]. Then the observed
[O~{\sc iii}]$_{\rm na}$ ratio yields a \Te\, of 8\,050~K (c.f.
Table\,\ref{tenecel}). This \Ne\, and \Te\, have been used to calculate ionic
abundances of heavy element ions from the high ionization regions. We note that
the abundances thus obtained were abnormally high compared to other bugle PNe
in the sample, even higher than values deduced from ORLs for the same object in
the case of carbon and oxygen (c.f.  Table~\ref{totabun}). These abundances are
however quite uncertain and should be treated with caution.

Given their peculiarities, both M~2-23 and M~2-39 will be excluded in our
statistical analysis of the abundance patterns of GBPNe, leaving a sample size
of 23.

\subsection{Hydrogen recombination temperatures and densities}
\label{anal:diagorl}

Electron temperatures have been determined for all sample PNe from the Balmer
discontinuity of H~{\sc i} recombination spectra with various levels of
uncertainty depending on the resolution and S/N ratios of spectra available for
the wavelength range of the Balmer jump at 3650~\AA. For practical reasons of
measurement and following \cite{liu2001}, we define the Balmer jump, BJ, as the
difference of continuum flux densities measured at 3643 and 3681~\AA, i.e.
$I_{\rm c}(\lambda3643) - I_{\rm c}(\lambda3681)$.  Balmer jump
temperature, \Te(BJ), was then calculated from the ratio of Balmer jump to 
H\,11 using the formula provided by \cite{liu2001},
\begin{equation} 
T_{\rm e} = 368 \times(1 + 0.259y^++3.409y^{2+})
   (\frac{\rm BJ}{\rm H11})^{-\frac{3}{2}},
\end{equation} 
where BJ/H11 is in units of \AA$^{-1}$ and $y^+$ and $y^{2+}$ denote helium
ionic abundances, He$^+$/H$^+$ and He$^{2+}$/H$^+$, respectively.  Since $y^+$
and $y^{2+}$ derived from helium recombination lines also have a weak
dependence on adopted \Te, the process was iterated until self-consistent
values for $y^+$ and $y^{2+}$ and \Te(BJ) were achieved. The results are 
tabulated in the second column of Table~\ref{teneorl}.

Electron densities, \Ne(BD)'s, were also derived by comparing the observed
decrement of high-order Balmer lines with that predicted by recombination theory
and are presented in the last column of Table~\ref{teneorl}.

\cite{zhang04a} present determinations of \Te(BJ) and \Ne(BD) for a sample of
48 Galactic PNe, including some objects in the current sample, by fitting the
observed spectra with synthesized theoretical spectra with \Te\, and \Ne\, as
input parameters. The technique only works well for high resolution spectra.
For objects in common, our current results are in good agreement with those
presented in \cite{zhang04a}. For objects having high resolution spectra,
typical uncertainties in \Te(BJ) are about 1000~K. For those for which only low
resolution spectra are available, the uncertainties increase to about 2000~K.

\subsection{Helium temperatures}
\label{anal:tehe}

\begin{figure}
\centering
\epsfig{file=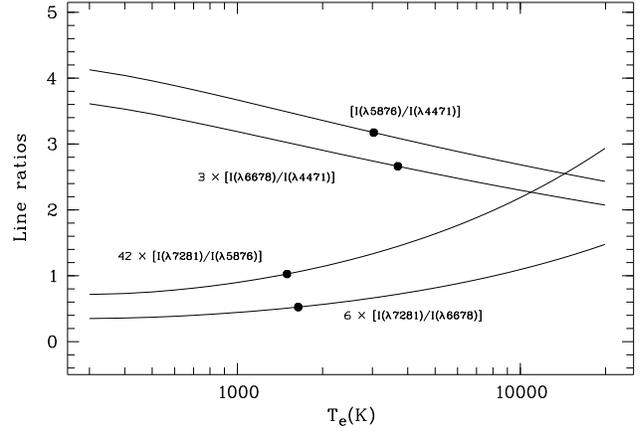, width=5.7cm, bbllx=52,
bblly=33, bburx=559, bbury=772,clip=,angle=-90}
\caption{He~{\sc i} line ratios as a function of \Te, assuming \Ne = 
4000\,cm$^{-3}$. The ratios observed in M~3-32 are marked.}
\label{tehei} 
\end{figure}

Electron temperatures have been derived from He~{\sc i} recombination line
ratios $\lambda6678/\lambda4471$, $\lambda5876/\lambda4471$,
$\lambda7281/\lambda5876$, $\lambda7281/\lambda6678$ and are presented in
Cols.\,3--6 of Table~\ref{teneorl}. Albeit the lines involved are weaker, \Te\,
determined from the fourth pair of diagnostic line ratio has a number of
advantages over those deduced from the other three pairs (c.f.
\citealt{zhang05a} for details) and will be adopted in the current
work\footnote {However, We note that the $\lambda$7281 line is quite sensitive
to the assumption of Case A or B recombination. Although Case B is generally
regarded as a good approximation, there is some evidence of departure from pure
Case B recombination (Liu et al. 2000, 2001). In that case, \Te(He{\sc i})
deduced from the $\lambda$7281/$\lambda$6678 and $\lambda$7281/$\lambda$5876
ratio would be underestimated}. In Fig.~\ref{tehei}, we show variations of the
four He~{\sc i} line ratios as a function of \Te\, for an assumed \Ne\,=
4000\,cm$^{-3}$.  Values measured in M\,3-32 are marked as an example.  The
parameters used to calculate emissivities of He~{\sc i} lines are taken from
Table~2 of \cite{zhang05a}. In their analysis, \cite{zhang05a} presented
He~{\sc i} temperatures deduced from the $\lambda7281/\lambda5876$ and
$\lambda7281/\lambda6678$ ratios for a large sample of Galactic PNe; 9 of them
are also in the current sample.  For objects in common, \Te\, obtained by them
are slightly higher than ours. The differences are mainly caused by the
differences in extinction corrections.  In their analysis, they had adopted the
standard H83 extinction law for the general ISM and determined extinction to
individual nebulae based on optical data alone. In the current work, a much
more detailed and thorough treatment of the extinction has been carried out
(c.f. Section~3.1).

\subsection{Temperatures from heavy element ORLs}
\label{anal:teorls}

\citet{liu2001} compared the observed relative intensities of O~{\sc ii} ORLs
with predictions of the recombination theory in four PNe exhibiting
particularly large adfs, M~1-42, M~2-36, NGC~7009 and NGC~6153. They found that
while the agreement was quite good in general, there were exceptions. For
example, the intensity of the strongest 3p-3s transition, 3p~$^4$D$^{\rm
o}_{7/2}$ -- 3s$^4$P$_{5/2}$ $\lambda$4649, appears to be too strong (by $\sim
40$\,per~cent) relative to the strongest 4f -- 3d transition, 4f\,G[5]$^{\rm
o}_{11/2}$ -- 3d$^4$F$_{9/2}$ $\lambda$4089, compared to the theoretical value
calculated at \Te\,(BJ). It was later realized that the discrepancies were
actually caused by the fact that those heavy element ORLs arise mainly from
another component of plasma of very low \Te, much lower than values deduced
from the Balmer discontinuity of H~{\sc i} recombination spectrum
(\citealt{liu2003}).  By comparing observed ratios of the two strongest \oii\,
ORLs, $\lambda$4089/$\lambda$4649, with the theoretical predictions calculated
down to a minimum \Te\, of 288~K, values of \Te(\oii), the average \Te\,
under which \oii~ORLs are emitted, have been determined for a large sample
of PNe (\citealt{tsamis2004}, \citealt{liuy2004b}, \citealt{wesson2005}). 

For the current sample of PNe, we have been able to determine \Te(\oii) for 11
objects from the $\lambda$4089/$\lambda$4649 ratio. The results are presented
in Col.\,7 of Table\,\ref{teneorl}. For another 8 nebulae in the sample
(flagged with an `L' in Table\,\ref{teneorl}) for which both lines have been
detected, the measured line ratios exceed 0.41, the maximum value at the low
\Te\, limit of 288~K. Part of the discrepancy was clearly caused by measurement
uncertainties, given the weakness of the lines. There is also an opportunity
that the O~{\sc ii} $\lambda$4089.3 may have been contaminated by the Si~{\sc
iv} $\lambda$4088.8 line, leading to artificially high
$\lambda$4089/$\lambda$4649 ratios in some PNe (\citealt{liu2006b}; c.f. also
Secction 6.1).  

\section{Ionic and total elemental abundances}
\label{abun}
\subsection{Ionic abundances from CELs}
\label{abun:cel}

\begin{table*}
\caption{Ionic and elemental abundances of helium from ORLs. Ionic and
elemental abundances of oxygen from CELs and from ORLs.}
\label{abunO}
\begin{tabular} {l c c c c c c c c c c }
\hline
\noalign{\smallskip}
\noalign{\smallskip}
        &\multicolumn{3}{c} {\bf He} & \multicolumn{4}{c} {\bf CELs}&  \multicolumn{3}{c}{\bf ORLs} \\
\noalign{\smallskip}
\noalign{\smallskip}
 Nebula &He$^+$/H$^{+}$ &He$^{2+}$/H$^{+}$ &He/H & O$^{+}$/H$^{+}$  & O$^{2+}$/H$^{+}$ & $f$(O)& O/H&O$^{2+}$/H$^{+}$ & $f$(O)& O/H\\
       *&average   & $\lambda$4686 & &  $\lambda\lambda$3726,29 & $\lambda\lambda4959,5007$ && &  average  &  &     \\
%        &          &               &&                           &                5007      &&&            &  & \\  
\noalign{\smallskip}
\noalign{\smallskip}
\hline
    He 2-118 & 8.00e$-$02 & 8.00e$-$03 & 8.80e$-$02 & 7.91e$-$06  & 1.99e$-$04 & 1.065 & 2.20e$-$04& 3.44e$-$04 & 1.108 & 3.81e$-$04 \\
    M 2-4    & 1.16e$-$01 &            & 1.16e$-$01 & 5.63e$-$05  & 4.76e$-$04 & 1.000 & 5.33e$-$04& 8.99e$-$04 & 1.118 & 1.00e$-$03 \\
    M 2-6    & 1.01e$-$01 &            & 1.01e$-$01 & 6.28e$-$05  & 2.37e$-$04 & 1.000 & 3.00e$-$04& 5.55e$-$04 & 1.264 & 7.02e$-$04 \\
    M 3-7    & 1.22e$-$01 & 1.50e$-$03 & 1.24e$-$01 & 2.54e$-$04  & 4.41e$-$04 & 1.008 & 7.00e$-$04& 1.94e$-$03 & 1.588 & 3.08e$-$03 \\
    M 1-20   & 9.50e$-$02 & 4.00e$-$05 & 9.50e$-$02 & 2.30e$-$05  & 3.57e$-$04 & 1.000 & 3.80e$-$04& 4.95e$-$04 & 1.064 & 5.27e$-$04 \\
    NGC 6439 & 1.12e$-$01 & 2.10e$-$02 & 1.33e$-$01 & 5.82e$-$05  & 3.96e$-$04 & 1.121 & 5.09e$-$04& 2.44e$-$03 & 1.286 & 3.14e$-$03 \\
    H 1-35   & 1.02e$-$01 & 2.00e$-$04 & 1.02e$-$01 & 5.21e$-$05  & 2.78e$-$04 & 1.001 & 3.31e$-$04& 5.71e$-$04 & 1.188 & 6.79e$-$04 \\
    M 1-29   & 1.12e$-$01 & 3.30e$-$02 & 1.45e$-$01 & 1.08e$-$04  & 3.83e$-$04 & 1.187 & 5.84e$-$04& 1.13e$-$03 & 1.523 & 1.72e$-$03 \\
    Cn 2-1   & 1.11e$-$01 & 8.00e$-$03 & 1.19e$-$01 & 6.93e$-$06  & 4.79e$-$04 & 1.047 & 5.09e$-$04& 1.40e$-$03 & 1.062 & 1.49e$-$03 \\
    H 1-41   & 8.10e$-$02 & 2.10e$-$02 & 1.02e$-$01 & 1.99e$-$05  & 3.48e$-$04 & 1.166 & 4.29e$-$04& 1.78e$-$03 & 1.232 & 2.20e$-$03 \\
    H 1-42   & 1.09e$-$01 & 7.00e$-$04 & 1.10e$-$01 & 3.21e$-$05  & 4.19e$-$04 & 1.004 & 4.53e$-$04& 9.64e$-$04 & 1.081 & 1.04e$-$03 \\
    M 2-23   & 1.12e$-$01 &            & 1.12e$-$01 & 5.67e$-$06  & 2.58e$-$04 & 1.000 & 2.64e$-$04& 3.73e$-$04 & 1.021 & 3.81e$-$04 \\
    M 3-21   & 1.14e$-$01 & 6.40e$-$03 & 1.20e$-$01 & 1.37e$-$05  & 6.30e$-$04 & 1.037 & 6.68e$-$04& 1.66e$-$03 & 1.059 & 1.76e$-$03 \\
    H 1-50   & 9.50e$-$02 & 1.10e$-$02 & 1.06e$-$01 & 1.86e$-$05  & 4.28e$-$04 & 1.075 & 4.80e$-$04& 1.22e$-$03 & 1.122 & 1.37e$-$03 \\
    M 2-27   & 1.27e$-$01 & 8.00e$-$04 & 1.28e$-$01 & 5.33e$-$05  & 6.86e$-$04 & 1.004 & 7.42e$-$04& 1.50e$-$03 & 1.082 & 1.62e$-$03 \\
    H 1-54   & 8.70e$-$02 & 3.00e$-$05 & 8.70e$-$02 & 5.43e$-$05  & 1.87e$-$04 & 1.000 & 2.42e$-$04& 4.77e$-$04 & 1.290 & 6.15e$-$04 \\
    NGC 6565 & 9.90e$-$02 & 1.50e$-$02 & 1.14e$-$01 & 1.50e$-$04  & 3.83e$-$04 & 1.098 & 5.86e$-$04& 6.47e$-$04 & 1.530 & 9.90e$-$04 \\
    M 2-31   & 1.14e$-$01 &            & 1.14e$-$01 & 1.94e$-$05  & 4.37e$-$04 & 1.000 & 4.57e$-$04&          &       &          \\
    NGC 6567 & 1.02e$-$01 & 9.00e$-$04 & 1.03e$-$01 & 1.24e$-$05  & 2.76e$-$04 & 1.005 & 2.90e$-$04& 6.04e$-$04 & 1.051 & 6.35e$-$04 \\
    M 2-33   & 1.05e$-$01 &            & 1.05e$-$01 & 2.03e$-$05  & 4.93e$-$04 & 1.000 & 5.13e$-$04& 1.06e$-$03 & 1.041 & 1.10e$-$03 \\
    IC 4699  & 8.00e$-$02 & 1.80e$-$02 & 9.80e$-$02 & 5.10e$-$06  & 2.67e$-$04 & 1.144 & 3.11e$-$04& 1.66e$-$03 & 1.166 & 1.94e$-$03 \\
    M 2-39   & 1.12e$-$01 &            & 1.12e$-$01 & 6.55e$-$05  & 2.58e$-$03 & 1.000 & 2.64e$-$03& 9.14e$-$04 & 1.025 & 9.37e$-$04 \\
    M 2-42   & 1.07e$-$01 & 3.00e$-$04 & 1.07e$-$01 & 3.21e$-$05  & 5.27e$-$04 & 1.001 & 5.60e$-$04& 1.10e$-$03 & 1.062 & 1.17e$-$03 \\
    NGC 6620 & 1.11e$-$01 & 2.10e$-$02 & 1.32e$-$01 & 1.82e$-$04  & 5.11e$-$04 & 1.122 & 7.78e$-$04& 1.63e$-$03 & 1.521 & 2.48e$-$03 \\
    VY 2-1   & 1.29e$-$01 & 5.00e$-$04 & 1.30e$-$01 & 1.09e$-$04  & 5.50e$-$04 & 1.002 & 6.60e$-$04& 1.11e$-$03 & 1.200 & 1.33e$-$03 \\
    Cn 1-5   & 1.25e$-$01 & 8.00e$-$04 & 1.26e$-$01 & 2.02e$-$04  & 4.89e$-$04 & 1.004 & 6.93e$-$04& 7.94e$-$04 & 1.418 & 1.13e$-$03 \\
    M 3-29   & 1.00e$-$01 &            & 1.00e$-$01 & 5.88e$-$05  & 2.63e$-$04 & 1.000 & 3.22e$-$04& 5.76e$-$04 & 1.223 & 7.05e$-$04 \\
    M 3-32   & 1.14e$-$01 & 1.05e$-$02 & 1.24e$-$01 & 2.58e$-$05  & 3.83e$-$04 & 1.060 & 4.34e$-$04& 6.80e$-$03 & 1.131 & 7.70e$-$03 \\
    M 1-61   & 1.04e$-$01 &            & 1.04e$-$01 & 3.66e$-$05  & 4.85e$-$04 & 1.000 & 5.22e$-$04& 9.49e$-$04 & 1.075 & 1.02e$-$03 \\
    M 3-33   & 8.80e$-$02 & 1.70e$-$02 & 1.05e$-$01 & 6.87e$-$06  & 3.43e$-$04 & 1.125 & 3.94e$-$04& 2.25e$-$03 & 1.147 & 2.58e$-$03 \\
    IC 4846  & 7.90e$-$02 & 2.00e$-$04 & 7.92e$-$02 & 5.01e$-$06  & 3.83e$-$04 & 1.001 & 3.88e$-$04& 5.91e$-$04 & 1.014 & 6.00e$-$04 \\
\noalign{\smallskip}
\hline
\end{tabular}
\end{table*}

\begin{table*}
\caption{Ionic and elemental abundances of carbon from CELs and from ORLs.}
\label{abunC}
\begin{tabular} {l c c c c c c c c c  }
\hline
         & \multicolumn{4}{c} {\bf CELs}& \hspace{1.8cm}& \multicolumn{3}{c}{\bf ORLs} \\
\noalign{\smallskip}
\noalign{\smallskip}
 Nebula  &C$^{2+}$/H$^{+}$  & C$^{3+}$/H$^{+}$ & $f$(C)& C/H& & C$^{2+}$/H$^{+}$&C$^{3+}$/H$^{+}$ & $f$(C)& C/H\\
  & $\lambda$1908 & $\lambda 1550$ & & &&   $\lambda4267$ & $\lambda\lambda$4187,4650 &     \\
\noalign{\smallskip}
\noalign{\smallskip}
\hline
    He 2-118 &           &           &      &           &&6.03e$-$05 &           &1.108 &6.68e$-$05\\
    M 2-4    &           &           &      &           &&3.10e$-$04 &6.98e$-$05 &1.118 &4.25e$-$04\\
    M 2-6    &           &           &      &           &&8.22e$-$05 &           &1.264 &1.04e$-$04\\
    M 3-7    &           &           &      &           &&           &5.81e$-$04 &      &        \\
    M 1-20   &           &           &      &           &&3.45e$-$04 &1.71e$-$04 &1.064 &5.49e$-$04\\
    NGC 6439 &2.12e$-$04 &           &1.286 &2.73e$-$04 &&7.89e$-$04 &1.98e$-$04 &1.147 &1.13e$-$03\\
    H 1-35   &           &           &      &           &&1.32e$-$04 &3.16e$-$04 &1.187 &5.32e$-$04\\
    M 1-29   &           &           &      &           &&5.15e$-$04 &           &1.523 &7.84e$-$04\\
    Cn 2-1   &           &           &      &           &&3.82e$-$04 &4.92e$-$04 &1.014 &8.87e$-$04\\
    H 1-41   &           &           &      &           &&3.11e$-$04 &           &1.232 &3.83e$-$04\\
    H 1-42   &5.76e$-$05 &           &1.081 &6.23e$-$05 &&1.03e$-$04 &1.68e$-$04 &1.076 &2.92e$-$04\\
    M 2-23   &4.30e$-$05 &           &1.021 &4.40e$-$05 &&6.01e$-$05 &1.30e$-$04 &1.021 &1.94e$-$04\\
    M 3-21   &1.22e$-$04 &           &1.059 &1.29e$-$04 &&3.40e$-$04 &1.27e$-$04 &1.021 &4.77e$-$04\\
    H 1-50   &6.41e$-$05 &           &1.122 &7.20e$-$05 &&4.14e$-$04 &6.49e$-$04 &1.043 &1.11e$-$03\\
    M 2-27   &           &           &      &           &&8.19e$-$04 &           &1.082 &8.86e$-$04\\
    H 1-54   &           &           &      &           &&8.70e$-$05 &4.93e$-$04 &1.289 &7.48e$-$04\\
    NGC 6565 &1.98e$-$04 &           &1.530 &3.03e$-$04 &&3.27e$-$04 &5.47e$-$05 &1.392 &5.32e$-$04\\
    M 2-31   &           &           &      &           &&2.96e$-$04 &3.38e$-$03 &1.044 &3.84e$-$03\\
    NGC 6567 &7.80e$-$04 &           &1.051 &8.20e$-$04 &&1.61e$-$03 &7.00e$-$03 &1.044 &9.00e$-$03\\
    M 2-33   &3.94e$-$04 &           &1.041 &4.10e$-$04 &&1.91e$-$04 &1.10e$-$04 &1.041 &3.13e$-$04\\
    IC 4699  &1.02e$-$04 &           &1.166 &1.19e$-$04 &&5.18e$-$04 &6.82e$-$05 &1.019 &5.97e$-$04\\
    M 2-39   &6.09e$-$03 &           &1.025 &6.25e$-$03 &&4.07e$-$04 &5.91e$-$04 &1.025 &1.02e$-$03\\
    M 2-42   &           &           &      &           &&1.58e$-$04 &5.94e$-$04 &1.060 &7.98e$-$04\\
    NGC 6620 &1.45e$-$04 &           &1.521 &2.21e$-$04 &&8.25e$-$04 &8.44e$-$05 &1.355 &1.23e$-$03\\
    VY 2-1   &3.96e$-$04 &           &1.200 &4.75e$-$04 &&3.06e$-$04 &1.08e$-$04 &1.197 &4.96e$-$04\\
    Cn 1-5   &1.05e$-$03 &1.70e$-$04 &1.412 &1.72e$-$03 &&1.49e$-$03 &           &1.418 &2.10e$-$03\\
    M 3-29   &           &           &      &           &&3.33e$-$04 &           &1.223 &4.08e$-$04\\
    M 3-32   &2.61e$-$04 &           &1.131 &2.95e$-$04 &&2.93e$-$03 &2.30e$-$04 &1.067 &3.37e$-$03\\
    M 1-61   &           &           &      &           &&3.91e$-$04 &1.79e$-$04 &1.075 &6.13e$-$04\\
    M 3-33   &9.96e$-$05 &           &1.147 &1.14e$-$04 &&3.72e$-$04 &2.62e$-$04 &1.020 &6.47e$-$04\\
    IC 4846  &1.43e$-$04 &           &1.014 &1.45e$-$04 &&1.11e$-$04 &1.52e$-$04 &1.013 &2.66e$-$04\\
\noalign{\smallskip}
\hline
\end{tabular}
\end{table*}

\begin{table*}
\caption{Ionic and elemental abundances of nitrogen from CELs and from ORLs.}
\label{abunN}
\begin{tabular} {l c c c c c c  c c c c }
\hline
         &\multicolumn{4}{c}{\bf CELs} & \hspace{1.5cm}& \multicolumn{4}{c}{\bf ORLs}\\
\noalign{\smallskip}
\noalign{\smallskip}
 Nebula  & N$^{+}$/H$^{+}$ &N$^{2+}$/H$^{+}$& $f$(N)&N/H& & N$^{2+}$/H$^{+}$& N$^{3+}$/H$^{+}$ & $f$(N) & N/H \\
         & $\lambda\lambda6548,6584$ & $\lambda$1750 &  & & & average & $\lambda$4379 &  &  \\
\noalign{\smallskip}
\noalign{\smallskip}
\hline
    He 2-118 & 1.54e$-$06 &                 &27.85 & 4.29e$-$05 & &1.56e$-$04 &           &1.108 &1.73e$-$04\\
    M 2-4    & 1.82e$-$05 &                 &9.464 & 1.72e$-$04 & &3.84e$-$04 &           &1.118 &4.29e$-$04\\
    M 2-6    & 1.12e$-$05 &                 &4.782 & 5.36e$-$05 & &           &           &      &        \\
    M 3-7    & 3.29e$-$05 &                 &2.759 & 9.08e$-$05 & &1.07e$-$03 &           &1.588 &1.70e$-$03\\
    M 1-20   & 5.29e$-$06 &                 &16.51 & 8.74e$-$05 & &4.31e$-$04 &8.58e$-$05 &1.064 &5.50e$-$04\\
    NGC 6439 & 3.14e$-$05 &3.38e$-$04       &0.304 & 4.72e$-$04 & &6.92e$-$04 &2.11e$-$04 &0.092 &9.67e$-$04\\
    H 1-35   & 8.16e$-$06 &                 &6.346 & 5.18e$-$05 & &2.46e$-$04 &1.06e$-$04 &1.187 &4.18e$-$04\\
    M 1-29   & 7.34e$-$05 &                 &5.393 & 3.96e$-$04 & &6.82e$-$04 &2.92e$-$04 &1.227 &1.20e$-$03\\
    Cn 2-1   & 4.28e$-$06 &                 &73.43 & 3.14e$-$04 & &2.98e$-$04 &4.92e$-$04 &1.013 &8.01e$-$04\\
    H 1-41   & 3.50e$-$06 &                 &21.53 & 7.55e$-$05 & &6.62e$-$04 &1.06e$-$04 &1.048 &8.05e$-$04\\
    H 1-42   & 4.94e$-$06 &                 &14.12 & 6.98e$-$05 & &6.88e$-$04 &           &1.081 &7.44e$-$04\\
    M 2-23   & 2.18e$-$06 &                 &46.57 & 1.02e$-$04 & &7.81e$-$04 &6.05e$-$05 &1.021 &8.60e$-$04\\
    M 3-21   & 8.25e$-$06 &8.38e$-$05       &0.420 & 1.27e$-$04 & &2.14e$-$04 &8.99e$-$05 &0.098 &3.25e$-$04\\
    H 1-50   & 7.62e$-$06 &                 &25.84 & 1.97e$-$04 & &           &8.92e$-$05 &      &        \\
    M 2-27   & 4.56e$-$05 &                 &13.91 & 6.34e$-$04 & &1.50e$-$03 &           &1.082 &1.62e$-$03\\
    H 1-54   & 8.39e$-$06 &                 &4.450 & 3.73e$-$05 & &4.15e$-$04 &1.22e$-$05 &1.289 &5.51e$-$04\\
    NGC 6565 & 6.98e$-$05 &                 &3.895 & 2.72e$-$04 & &9.03e$-$04 &3.87e$-$05 &1.345 &1.27e$-$03\\
    M 2-31   & 9.61e$-$06 &                 &23.49 & 2.26e$-$04 & &           &           &      &        \\
    NGC 6567 & 2.20e$-$06 &3.51e$-$04       &0.024 & 3.62e$-$04 & &3.59e$-$04 &8.94e$-$06 &0.006 &3.70e$-$04\\
    M 2-33   & 2.23e$-$06 &                 &25.25 & 5.63e$-$05 & &4.96e$-$04 &           &1.041 &5.16e$-$04\\
    IC 4699  & 3.57e$-$07 &                 &60.98 & 2.18e$-$05 & &2.32e$-$04 &1.05e$-$04 &1.016 &3.43e$-$04\\
    M 2-39   & 1.18e$-$05 &                 &40.28 & 4.77e$-$04 & &7.84e$-$04 &           &1.025 &8.04e$-$04\\
    M 2-42   & 1.03e$-$05 &                 &17.45 & 1.80e$-$04 & &4.39e$-$04 &           &1.062 &4.67e$-$04\\
    NGC 6620 & 8.39e$-$05 &                 &4.278 & 3.59e$-$04 & &6.41e$-$04 &1.47e$-$04 &1.305 &1.03e$-$03\\
    Vy 2-1   & 2.68e$-$05 &2.08e$-$04       &1.002 & 2.36e$-$04 & &5.51e$-$04 &           &1.200 &6.61e$-$04\\
    Cn 1-5   & 1.32e$-$04 &7.29e$-$04$^{a}$ &1.004 & 8.65e$-$04 & &4.93e$-$03 &           &1.418 &6.99e$-$03\\
    M 3-29   & 1.74e$-$05 &                 &5.468 & 9.50e$-$05 & &           &           &      &        \\
    M 3-32   & 5.27e$-$06 &                 &16.83 & 8.87e$-$05 & &2.24e$-$03 &3.04e$-$04 &1.063 &2.70e$-$03\\
    M 1-61   & 8.97e$-$06 &                 &14.23 & 1.28e$-$04 & &4.70e$-$04 &8.51e$-$05 &1.075 &5.97e$-$04\\
    M 3-33   & 4.19e$-$07 &                 &57.27 & 2.40e$-$05 & &           &9.80e$-$05 &      &        \\
    IC 4846  & 1.58e$-$06 &                 &77.45 & 1.22e$-$04 & &1.25e$-$04 &           &1.014 &1.27e$-$04\\
\noalign{\smallskip}
\hline
\end{tabular}
\begin{list}{}{}
\item $^{a}$ Derived from the [N~{\sc iii}] 57$\mu$m infrared fine-structure line.
\end{list}
\end{table*}

\begin{table*}
\caption{Ionic and elemental abundances of neon from CELs and from ORLs. Ionic and elemental abundances of argon from CELs.}
\label{abunNeAr}
\begin{tabular}{lcccccccccc}
\hline
        & \multicolumn{3}{c} {\bf CELs}& \multicolumn{2}{c}{\bf ORLs} & \multicolumn{5}{c}{\bf CELs}\\
\noalign{\smallskip}
\noalign{\smallskip}
Nebula  &Ne$^{2+}$/H$^{+}$ & $f$(Ne)& Ne/H&  Ne$^{2+}$/H$^{+}$ & Ne/H & Ar$^{2+}$/H$^{+}$ & Ar$^{3+}$/H$^{+}$ & Ar$^{4+}$/H$^{+}$ & $f$(Ar) & Ar/H\\
        &$\lambda$3868     &        &     &  average   &      & $\lambda$7136     & $\lambda\lambda$4711,4740 & $\lambda$7006  &         &     \\
\noalign{\smallskip}
\noalign{\smallskip}
\hline
    He 2-118 &4.46e$-$05  &1.108  &4.94e$-$05  &1.55e$-$03&1.72e$-$03& 4.29e$-$07   &7.64e$-$08  &           &1.037  &5.24e$-$07 \\
    M 2-4    &1.22e$-$04  &1.118  &1.37e$-$04  &          &          & 2.56e$-$06   &1.07e$-$07  &           &1.118  &2.98e$-$06 \\
    M 2-6    &4.30e$-$05  &1.264  &5.44e$-$05  &          &          & 6.42e$-$07   &3.10e$-$08  &           &1.264  &8.50e$-$07 \\
    M 3-7    &6.77e$-$05  &1.588  &1.08e$-$04  &          &          & 6.33e$-$07   &            &           &1.568  &9.93e$-$07 \\
    M 1-20   &6.04e$-$05  &1.064  &6.43e$-$05  &          &          & 5.87e$-$07   &7.33e$-$08  &           &1.064  &7.03e$-$07 \\
    NGC 6439 &1.26e$-$04  &1.286  &1.61e$-$04  &1.35e$-$03&1.74e$-$03& 2.27e$-$06   &8.06e$-$07  &1.01e$-$07 &1.071  &3.40e$-$06 \\
    H 1-35   &3.50e$-$05  &1.188  &4.16e$-$05  &8.57e$-$05&1.02e$-$04& 1.35e$-$06   &1.88e$-$08  &           &1.187  &1.63e$-$06 \\
    M 1-29   &9.39e$-$05  &1.523  &1.43e$-$04  &          &          & 1.95e$-$06   &8.96e$-$07  &           &1.227  &3.49e$-$06 \\
    Cn 2-1   &9.46e$-$05  &1.062  &1.01e$-$04  &4.99e$-$04&5.30e$-$04& 1.24e$-$06   &4.22e$-$07  &           &1.013  &1.68e$-$06 \\
    H 1-41   &8.30e$-$05  &1.232  &1.02e$-$04  &          &          & 7.26e$-$07   &4.27e$-$07  &           &1.048  &1.21e$-$06 \\
    H 1-42   &8.56e$-$05  &1.081  &9.25e$-$05  &1.24e$-$03&1.34e$-$03& 9.03e$-$07   &2.61e$-$07  &           &1.076  &1.25e$-$06 \\
    M 2-23   &3.45e$-$05  &1.021  &3.52e$-$05  &1.45e$-$04&1.48e$-$04& 6.44e$-$07   &7.62e$-$08  &           &1.021  &7.36e$-$07 \\
    M 3-21   &1.72e$-$04  &1.059  &1.83e$-$04  &8.65e$-$04&9.17e$-$04& 1.81e$-$06   &8.56e$-$07  &4.80e$-$08 &1.069  &2.90e$-$06 \\
    H 1-50   &1.10e$-$04  &1.122  &1.23e$-$04  &          &          & 9.87e$-$07   &7.10e$-$07  &5.69e$-$08 &1.040  &1.82e$-$06 \\
    M 2-27   &1.90e$-$04  &1.082  &2.06e$-$04  &          &          & 3.08e$-$06   &4.61e$-$07  &           &1.077  &3.81e$-$06 \\
    H 1-54   &2.59e$-$05  &1.290  &3.34e$-$05  &1.86e$-$04&2.40e$-$04& 6.45e$-$07   &1.17e$-$08  &           &1.289  &8.47e$-$07 \\
    NGC 6565 &1.09e$-$04  &1.530  &1.67e$-$04  &8.20e$-$04&1.26e$-$03& 1.74e$-$06   &4.04e$-$07  &7.03e$-$08 &1.345  &2.99e$-$06 \\
    M 2-31   &9.37e$-$05  &1.044  &9.79e$-$05  &          &          & 1.42e$-$06   &3.68e$-$07  &           &1.044  &1.86e$-$06 \\
    NGC 6567 &4.71e$-$05  &1.051  &4.94e$-$05  &2.56e$-$04&2.69e$-$04& 3.74e$-$07   &1.20e$-$07  &           &1.006  &4.96e$-$07 \\
    M 2-33   &9.88e$-$05  &1.041  &1.03e$-$04  &7.04e$-$04&7.33e$-$04& 1.20e$-$06   &6.91e$-$08  &           &1.041  &1.32e$-$06 \\
    IC 4699  &5.32e$-$05  &1.166  &6.20e$-$05  &5.48e$-$04&6.39e$-$04& 1.46e$-$06   &4.10e$-$07  &           &1.016  &1.90e$-$06 \\
    M 2-39   &2.10e$-$04  &1.025  &2.15e$-$04  &          &          & 3.01e$-$06   &1.46e$-$06  &           &1.025  &4.58e$-$06 \\
    M 2-42   &1.19e$-$04  &1.062  &1.27e$-$04  &          &          & 1.44e$-$06   &1.59e$-$07  &           &1.060  &1.70e$-$06 \\
    NGC 6620 &1.44e$-$04  &1.521  &2.19e$-$04  &1.26e$-$03&1.92e$-$03& 2.97e$-$06   &6.23e$-$07  &1.16e$-$07 &1.305  &4.84e$-$06 \\
    VY 2-1   &1.41e$-$04  &1.200  &1.70e$-$04  &4.04e$-$04&4.85e$-$04& 2.41e$-$06   &1.08e$-$07  &           &1.128  &2.84e$-$06 \\
    Cn 1-5   &1.59e$-$04  &1.418  &2.26e$-$04  &5.06e$-$04&7.18e$-$04& 2.34e$-$06   &7.19e$-$08  &           &1.181  &2.85e$-$06 \\
    M 3-29   &5.81e$-$05  &1.223  &7.10e$-$05  &          &          & 6.33e$-$07   &            &           &1.223  &7.74e$-$07 \\
    M 3-32   &1.13e$-$04  &1.131  &1.28e$-$04  &2.00e$-$03&2.26e$-$03& 1.31e$-$06   &4.12e$-$07  &           &1.063  &1.83e$-$06 \\
    M 1-61   &1.21e$-$04  &1.075  &1.31e$-$04  &1.49e$-$03&1.60e$-$03& 1.43e$-$06   &1.06e$-$07  &           &1.075  &1.65e$-$06 \\
    M 3-33   &8.87e$-$05  &1.147  &1.02e$-$04  &          &          & 6.17e$-$07   &5.49e$-$07  &           &1.017  &1.19e$-$06 \\
    IC 4846  &5.74e$-$05  &1.014  &5.83e$-$05  &          &          & 8.00e$-$07   &2.29e$-$07  &           &1.013  &1.04e$-$06 \\
\noalign{\smallskip}
\hline
\end{tabular}
\end{table*}

\begin{table*}
\caption{Ionic and elemental abundances of sulphur and chlorine from CELs.}
\label{abunSCl}
\begin{tabular} {l c c c c c  c c c}
\hline
         & \multicolumn{4}{c} {\bf CELs}&\hspace{1mm} &\multicolumn{3}{c}{\bf CELs} \\
\noalign{\smallskip}
\noalign{\smallskip}
 Nebula  &S$^{+}$/H$^{+}$  & S$^{2+}$/H$^{+}$ & $f$(S)& S/H&& Cl$^{2+}$/H$^{+}$ & $f$(Cl)& Cl/H\\
 & $\lambda\lambda$6716,6731 & $\lambda6312$ & & &&  $\lambda\lambda5517,5537$ &  &     \\
\noalign{\smallskip}
\noalign{\smallskip}
\hline
   He 2-118 &  1.11e$-$07 & 1.77e$-$07 & 2.127 & 6.13e$-$07 & & 2.71e$-$08 & 3.461 & 9.37e$-$08\\
   M 2-4    &  5.87e$-$07 & 1.92e$-$05 & 1.520 & 3.00e$-$05 & & 1.69e$-$07 & 1.566 & 2.65e$-$07\\
   M 2-6    &  2.76e$-$07 & 3.29e$-$06 & 1.255 & 4.48e$-$06 & & 6.35e$-$08 & 1.360 & 8.64e$-$08\\
   M 3-7    &  7.24e$-$07 & 8.09e$-$06 & 1.105 & 9.74e$-$06 & & 1.64e$-$07 & 1.204 & 1.98e$-$07\\
   M 1-20   &  1.68e$-$07 & 2.48e$-$06 & 1.802 & 4.77e$-$06 & & 4.46e$-$08 & 1.924 & 8.58e$-$08\\
   NGC 6439 &  1.12e$-$06 & 6.14e$-$06 & 1.485 & 1.08e$-$05 & & 1.28e$-$07 & 1.755 & 2.25e$-$07\\
   H 1-35   &  2.46e$-$07 & 6.94e$-$06 & 1.354 & 9.74e$-$06 & & 9.25e$-$08 & 1.402 & 1.30e$-$07\\
   M 1-29   &  2.39e$-$06 & 6.64e$-$06 & 1.295 & 1.17e$-$05 & & 1.15e$-$07 & 1.762 & 2.03e$-$07\\
   Cn 2-1   &  1.80e$-$07 & 4.18e$-$06 & 2.916 & 1.27e$-$05 & & 6.95e$-$08 & 3.042 & 2.11e$-$07\\
   H 1-41   &  1.43e$-$07 & 2.67e$-$06 & 1.959 & 5.52e$-$06 & & 6.76e$-$08 & 2.064 & 1.40e$-$07\\
   H 1-42   &  2.90e$-$07 & 3.53e$-$06 & 1.716 & 6.57e$-$06 & & 8.18e$-$08 & 1.857 & 1.52e$-$07\\
   M 2-23   &  1.16e$-$07 & 2.40e$-$06 & 2.512 & 6.33e$-$06 & & 7.82e$-$08 & 2.633 & 2.06e$-$07\\
   M 3-21   &  3.29e$-$07 & 6.00e$-$06 & 2.551 & 1.62e$-$05 & & 1.22e$-$07 & 2.691 & 3.27e$-$07\\
   H 1-50   &  3.57e$-$07 & 3.51e$-$06 & 2.076 & 8.02e$-$06 & & 7.26e$-$08 & 2.288 & 1.66e$-$07\\
   M 2-27   &  1.04e$-$06 & 1.04e$-$05 & 1.708 & 1.95e$-$05 & & 2.01e$-$07 & 1.879 & 3.78e$-$07\\
   H 1-54   &  1.93e$-$07 & 4.00e$-$06 & 1.232 & 5.17e$-$06 & & 6.62e$-$08 & 1.292 & 8.55e$-$08\\
   NGC 6565 &  2.52e$-$06 & 7.01e$-$06 & 1.192 & 1.14e$-$05 & & 1.12e$-$07 & 1.622 & 1.81e$-$07\\
   M 2-31   &  5.02e$-$07 & 6.05e$-$06 & 2.014 & 1.32e$-$05 & & 1.01e$-$07 & 2.181 & 2.21e$-$07\\
   NGC 6567 &  7.89e$-$08 & 1.36e$-$06 & 2.012 & 2.90e$-$06 & & 3.48e$-$08 & 2.129 & 7.40e$-$08\\
   M 2-33   &  5.95e$-$08 & 4.70e$-$06 & 2.061 & 9.80e$-$06 & & 1.03e$-$07 & 2.087 & 2.15e$-$07\\
   IC 4699  &  3.02e$-$08 & 7.59e$-$07 & 2.744 & 2.16e$-$06 & & 3.96e$-$08 & 2.853 & 1.13e$-$07\\
   M 2-39   &  3.81e$-$07 & 4.69e$-$06 & 2.396 & 1.21e$-$05 & & 4.43e$-$06 & 2.591 & 1.15e$-$05\\
   M 2-42   &  4.96e$-$07 & 6.46e$-$06 & 1.833 & 1.28e$-$05 & & 1.35e$-$07 & 1.974 & 2.70e$-$07\\
   NGC 6620 &  3.25e$-$06 & 9.40e$-$06 & 1.220 & 1.54e$-$05 & & 1.82e$-$07 & 1.641 & 2.98e$-$07\\
   VY 2-1   &  8.88e$-$07 & 1.00e$-$05 & 1.339 & 1.46e$-$05 & & 1.53e$-$07 & 1.457 & 2.23e$-$07\\
   Cn 1-5   &  2.55e$-$06 & 8.35e$-$06 & 1.158 & 1.26e$-$05 & & 1.59e$-$07 & 1.511 & 2.40e$-$07\\
   M 3-29   &  3.30e$-$07 & 3.48e$-$06 & 1.300 & 4.96e$-$06 & & 8.38e$-$08 & 1.423 & 1.19e$-$07\\
   M 3-32   &  2.46e$-$07 & 3.23e$-$06 & 1.813 & 6.29e$-$06 & & 8.09e$-$08 & 1.951 & 1.58e$-$07\\
   M 1-61   &  2.82e$-$07 & 2.65e$-$06 & 1.720 & 5.04e$-$06 & & 9.30e$-$08 & 1.904 & 1.77e$-$07\\
   M 3-33   &  1.93e$-$08 & 1.21e$-$06 & 2.688 & 3.30e$-$06 & & 4.30e$-$08 & 2.731 & 1.17e$-$07\\
   IC 4846  &  1.09e$-$07 & 3.36e$-$06 & 2.968 & 1.03e$-$05 & & 7.09e$-$08 & 3.064 & 2.17e$-$07\\
\noalign{\smallskip}
\hline
\end{tabular}
\end{table*}

\begin{table*}
\large
\caption{Ionization correction factors (ICFs).}
\label{ICFs}
\begin{tabular}{l c c c c l l}
\hline
\noalign{\smallskip}
%Element & CELsor ORLs&X$^{+}$/H$^+$ & X$^{2+}$/H$^+$ & X$^{3+}$/H$^+$ & & Equations& ICF \\
 X & Lines &\multicolumn{3}{c}{Detected ions} & X/H & Correction factors \\
\noalign{\smallskip}
\hline
\noalign{\smallskip}
 C & ORLs \& CELs &  & C$^{2+}$ & C$^{3+}$ & ICF$\times$($\frac{{\rm C}^{2+}}{{\rm H}^+}$ + $\frac{{\rm C}^{3+}}{{\rm H}^+}$) &  $(\frac{{\rm O}^{+}+{\rm O}^{2+}}{{\rm O}^{2+}})_{\rm CELs}$ \\
\noalign{\smallskip}
   &              &  & C$^{2+}$ &          & ICF$\times(\frac{{\rm C}^{2+}}{{\rm H}^+})$ & $(\frac{{\rm O}}{{\rm O}^{2+}})_{\rm CELs}$ \\
\noalign{\smallskip} 
\noalign{\smallskip} 
 N   & CELs    &  N$^{+}$& N$^{2+}$ &          & $(1+f_{1})\times\frac{{\rm N}^{2+}}{{\rm H}^+} + \frac{{\rm N}^{+}}{{\rm H}^+}$  & $f_1 = (\frac{{\rm N}^{3+}}{{\rm N}^{2+}})_{\rm ORLs}$ \\
\noalign{\smallskip}     
     &         &  N$^{+}$& N$^{2+}$ &          & ICF$\times(\frac{{\rm N}^{2+}}{{\rm H}^+} + \frac{{\rm N}^{+}}{{\rm H}^+})$  &   $(\frac{{\rm He}^{+}+{\rm He}^{2+}}{{\rm He}^{+}})^{2/3}$ \\
\noalign{\smallskip}     
     &         &  N$^{+}$&          &          & ICF$\times(\frac{{\rm N}^{+}}{{\rm H}^+})$ & $(\frac{{\rm O}}{{\rm O}^{+}})_{\rm CELs}$ \\
\noalign{\smallskip}
     & ORLs    &          & N$^{2+}$ & N$^{3+}$ & $(1 + f_2)\times\frac{{\rm N}^{2+}}{{\rm H}^+} + \frac{{\rm N}^{3+}}{{\rm H}^+}$ & $f_2 = (\frac{{\rm N}^{+}}{{\rm N}^{2+}})_{\rm CELs}$\\
\noalign{\smallskip}     
     &         &          & N$^{2+}$ & N$^{3+}$ & ICF$\times(\frac{{\rm N}^{2+}}{{\rm H}^+} + \frac{{\rm N}^{3+}}{{\rm H}^+})$ &  $[1- (\frac{{\rm O}^{+}}{{\rm O}})_{\rm CELs}]^{-1}$ \\           
\noalign{\smallskip}
     &         &          & N$^{2+}$ &          & ICF$\times(\frac{{\rm N}^{2+}}{{\rm H}^+})$ & $(\frac{{\rm O}}{{\rm O}^{2+}})_{\rm CELs}$ \\
\noalign{\smallskip}
\noalign{\smallskip}
 O   &  CELs   & O$^{+}$ &  O$^{2+}$ & &  ICF$\times(\frac{{\rm O}^{+}}{{\rm H}^+} + \frac{{\rm O}^{2+}}{{\rm H}^+})$ & $(\frac{{\rm He}^{+}+{\rm He}^{2+}}{{\rm He}^{+}})^{2/3}$ \\
\noalign{\smallskip}
     &  ORLs   &         &  O$^{2+}$ & &  ICF$\times\frac{{\rm O}^{2+}}{{\rm H}^+}$   & $(\frac{{\rm He}^{+}+{\rm He}^{2+}}{{\rm He}^{+}})^{2/3}\times[1+(\frac{{\rm O}^{+}}{{\rm O}^{2+}})_{\rm CELs}]$ \\
\noalign{\smallskip}
\noalign{\smallskip}
 Ne  & ORLs \& CELs     &         & Ne$^{2+}$  & &  ICF$\times\frac{{\rm Ne}^{2+}}{{\rm H}^+}$ & $(\frac{{\rm O}}{{\rm O}^{2+}})_{\rm CELs}$ \\ 
\noalign{\smallskip}
\noalign{\smallskip}
 S   & CELs     & S$^{+}$ & S$^{2+}$   & & ICF$\times(\frac{{\rm S}^{+}}{{\rm H}^+} + \frac{{\rm S}^{2+}}{{\rm H}^+})$ & $[1 - (1 - \frac{{\rm O}^{+}}{{\rm O}})^3]_{\rm CELs}^{-1/3}$ \\      
\noalign{\smallskip}
\noalign{\smallskip}
 Ar & CELs     & Ar$^{2+}$ & Ar$^{3+}$ & Ar$^{4+}$ & ICF$\times(\frac{{\rm Ar}^{2+}}{{\rm H}^+} + \frac{{\rm Ar}^{3+}}{{\rm H}^+} + \frac{{\rm Ar}^{4+}}{{\rm H}^+})$ & $[1 - (\frac{{\rm N}^{+}}{{\rm N}})_{\rm CELs}]^{-1}$ \\
\noalign{\smallskip}
   &          &  Ar$^{2+}$ & Ar$^{3+}$ &           & ICF$\times(\frac{{\rm Ar}^{2+}}{{\rm H}^+} + \frac{{\rm Ar}^{3+}}{{\rm H}^+})$ &   $[1 - (\frac{{\rm N}^{+}}{{\rm N}})_{\rm CELs}]^{-1}$ \\
\noalign{\smallskip}
\noalign{\smallskip}
 Cl & CELs    &            & Cl$^{2+}$ &           & ICF$\times\frac{{\rm Cl}^{2+}}{{\rm H}^+}$  & $(\frac{{\rm S}}{{\rm S}^{2+}})_{\rm CELs}$\\
\noalign{\smallskip}
\hline
\end{tabular}
\end{table*}

Abundances for ionic species of C, N, O, Ne, S, Ar, Cl and Fe have been
determined from CELs for 31 sample nebulae. In the calculations, we have
adopted \Te\ derived from [N~{\sc ii}]$_{\rm na}$ nebular to auroral line ratio
and \Ne\ derived from the [S~{\sc ii}]$_{\rm nn}$ nebular line ratio (or the
average of values derived from the [S~{\sc ii}]$_{\rm nn}$ and [O~{\sc
ii}]$_{\rm nn}$ ratios if the latter is available).  For ionic species of
higher ionization stages, \Te\, deduced from the [O~{\sc iii}]$_{\rm na}$ ratio
and the average \Ne\, derived from the [Cl~{\sc iii}]$_{\rm nn}$ and [Ar~{\sc
iv}]$_{\rm nn}$ ratios were used.  Note that for most objects, we have
corrected for a recombination contribution to the [N~{\sc ii}] $\lambda$5755
auroral line, except for a few cases where the contamination as estimated using
the aforementioned formula (and adopting an ionic abundance N$^{2+}$/H$^+$ as
yielded by N~{\sc ii} ORLs) was found to be unphysically large, becoming
comparable to or even larger than the actually measured total flux of the
$\lambda$5755 feature. In the case of M~3-33, the [N~{\sc ii}] $\lambda$5755
line was too faint to have a reliable measurement. For this nebula \Te[O~{\sc
iii}] has been used to calculate abundances of all ionic species.

Ionic abundance ratios, ${\rm X}^{\rm i+}/{\rm H}^+$, derived from
collisionally excited UV, optical and IR lines are tabulated in
Tables~\ref{abunO} (oxygen), \ref{abunC} (carbon), \ref{abunN} (nitrogen),
\ref{abunNeAr} (neon and argon) and \ref{abunSCl} (sulfur and chlorine). 

\subsection{Ionic abundances from ORLs}
\label{abun:orl}

\subsubsection{He}
\label{abun:orl:he}

Singly and doubly ionized helium abundances were derived from the He\,{\sc i}
$\lambda$4471, $\lambda$5876 and $\lambda$6678 recombination lines and from the
He\,{\sc ii} $\lambda$4686 recombination line, respectively.  The results from
the three  He\,{\sc i} lines were averaged, weighted by 1:3:1.  The total
helium abundances relative to hydrogen were obtained by summing abundances of
singly and doubly ionized species. The results are tabulated in Cols.\,2--4 of
Table~\ref{abunO}.

\subsubsection{C, N, O, Ne and Mg}
\label{abun:orl:cnone}

Ionic and elemental abundances have been derived from ORLs for some or all of
the elements C, N, O, Ne and Mg for our sample of 31 PNe.  Electron
temperatures derived from the H~{\sc i} Balmer discontinuity, \Te(BJ), were
adopted in the calculations. For the 11 PNe of sub-sample\,A, \Ne(BD) were
derivable and were adopted. For the remaining 20 PNe, the average \Ne\, derived
from various optical CEL diagnostics were used.  For individual ionic species,
we adopt the same effective recombination coefficients and assume the same case
[Case A and B] as \cite{liu2000}, except for low $l$ transitions of N~{\sc ii},
for which the more recent calculations of \citet{kisielius2002} are used.  For
ionic species for which a host of lines have been detected, such as N~{\sc ii},
O~{\sc ii} and Ne~{\sc ii}, the results derived from individual transitions
were averaged, according to procedures outlined in \citet{liu1995} and
\citet{liu2000}.  The results are tabulated in Tables~\ref{abunO}, \ref{abunC},
\ref{abunN} and \ref{abunNeAr}.  

\subsection{Elemental abundances}
\label{abun:tot}

\begin{table*}
\caption{Total elemental abundances on a logarithmic scale where H = 12.}
\label{totabun}
\begin{tabular}{lccccccccccccc}
\hline
Nebula & {\bf He}   & \multicolumn{2}{c} {\bf C}& \multicolumn{2}{c}{\bf N} & \multicolumn{2}{c}{\bf O} & \multicolumn{2}{c}{\bf Ne} & {\bf S} & {\bf Cl} & {\bf Ar} &{\bf Mg}\\
\noalign{\smallskip}
\noalign{\smallskip}
             & ORLs   & CELs & ORLs  & CELs & ORLs  & CELs & ORLs  & CELs   &  ORLs  & CELs  & CEl & CELs & ORLs \\
\noalign{\smallskip}
\noalign{\smallskip}
\hline
\multicolumn{14}{l}{Bulge PNe}\\
   M 2-4    &  11.06 &      & 8.63 & 8.24 & 8.63 & 8.73 & 9.00 & 8.14 &      & 7.48 & 5.42 & 6.47 & 7.70 \\
   M 2-6    &  11.00 &      & 8.02 & 7.73 &      & 8.48 & 8.85 & 7.74 &      & 6.65 & 4.94 & 5.93 &      \\
   M 3-7    &  11.09 &      & 8.96 & 7.96 & 9.23 & 8.85 & 9.49 & 8.03 &      & 6.99 & 5.30 & 6.00 &      \\
   M 1-20   &  10.98 &      & 8.74 & 7.94 & 8.74 & 8.58 & 8.72 & 7.81 &      & 6.68 & 4.93 & 5.85 &      \\
   NGC 6439 &  11.12 & 8.44 & 9.05 & 8.67 & 8.99 & 8.71 & 9.50 & 8.21 & 9.24 & 7.03 & 5.35 & 6.53 &      \\
   Cn 2-1   &  11.08 &      & 8.95 & 8.50 & 8.90 & 8.71 & 9.17 & 8.00 & 8.72 & 7.10 & 5.33 & 6.23 &      \\
   H 1-41   &  11.01 &      & 8.58 & 7.88 & 8.91 & 8.63 & 9.34 & 8.01 &      & 6.74 & 5.14 & 6.08 &      \\
   H 1-42   &  11.04 & 7.79 & 8.46 & 7.84 & 8.87 & 8.66 & 9.02 & 7.97 & 9.13 & 6.82 & 5.18 & 6.10 &      \\
M 2-23$^a$  &  11.05 & 7.64 & 8.29 & 8.01 & 8.93 & 8.42 & 8.58 & 7.55 & 8.17 & 6.80 & 5.31 & 5.87 &      \\
   M 3-21   &  11.08 & 8.11 & 8.68 & 8.10 & 8.51 & 8.82 & 9.25 & 8.26 & 8.96 & 7.21 & 5.51 & 6.45 & 7.53 \\
   H 1-50   &  11.03 &      & 9.05 & 8.29 &      & 8.68 & 9.14 & 8.09 &      & 6.90 & 5.22 & 6.26 &      \\
   M 2-27   &  11.11 &      & 8.95 & 8.80 & 9.21 & 8.87 & 9.21 & 8.31 &      & 7.29 & 5.58 & 6.58 &      \\
   H 1-54   &  10.94 &      & 8.87 & 7.57 & 8.74 & 8.38 & 8.79 & 7.52 & 8.38 & 6.71 & 4.93 & 5.93 &      \\
NGC 6565    &  11.06 & 8.48 & 8.73 & 8.43 & 9.10 & 8.77 & 9.00 & 8.22 & 9.10 & 7.06 & 5.26 & 6.48 & 7.68 \\
   M 2-31   &  11.06 &      & 9.58 & 8.35 &      & 8.66 &      & 7.99 &      & 7.12 & 5.34 & 6.27 &      \\
   M 2-33   &  11.02 & 8.61 & 8.50 & 7.75 & 8.71 & 8.71 & 9.04 & 8.01 & 8.87 & 6.99 & 5.33 & 6.12 &      \\
   IC 4699  &  10.99 & 8.08 & 8.78 & 7.34 & 8.53 & 8.49 & 9.29 & 7.79 & 8.81 & 6.34 & 5.05 & 6.28 & 7.86 \\
  M 2-39$^a$&  11.05 & 9.80 & 9.01 & 8.68 & 8.91 & 9.42 & 8.97 & 8.33 &      & 7.08 & 7.06$^{e}$ & 6.66 &      \\
   M 2-42   &  11.03 &      & 8.90 & 8.26 & 8.67 & 8.75 & 9.07 & 8.10 &      & 7.11 & 5.43 & 6.23 &      \\
   NGC 6620 &  11.12 & 8.34 & 9.09 & 8.56 & 9.01 & 8.89 & 9.39 & 8.34 & 9.28 & 7.19 & 5.47 & 6.69 &      \\
   Vy 2-1   &  11.11 & 8.68 & 8.70 & 8.37 & 8.82 & 8.82 & 9.12 & 8.23 & 8.69 & 7.17 & 5.35 & 6.45 &      \\
   Cn 1-5   &  11.10 & 9.24 & 9.32 & 8.94 & 9.84 & 8.84 & 9.05 & 8.35 & 8.86 & 7.10 & 5.38 & 6.45 & 7.37 \\
   M 3-29   &  11.00 &      & 8.61 & 7.98 &      & 8.51 & 8.85 & 7.85 &      & 6.70 & 5.08 & 5.89 &      \\
   M 3-32   &  11.10 & 8.47 & 9.53 & 7.95 & 9.43 & 8.64 & 9.89 & 8.11 & 9.35 & 6.80 & 5.20 & 6.26 & 7.93 \\
   M 3-33   &  11.02 & 8.06 & 8.81 & 7.38 &      & 8.59 & 9.41 & 8.01 &      & 6.52 & 5.07 & 6.07 &      \\
\noalign{\smallskip}
\multicolumn{14}{l}{Disk PNe}\\
He 2-118 &  10.94 &      & 7.80 & 7.63 & 8.24 & 8.34 & 8.58 & 7.69 & 9.23 & 5.79 & 4.97 & 5.72 &      \\
H 1-35   &  11.01 &      & 8.73 & 7.71 & 8.62 & 8.52 & 8.83 & 7.62 & 8.01 & 6.99 & 5.11 & 6.21 &      \\
M 1-29   &  11.16 &      & 8.89 & 8.60 & 9.08 & 8.77 & 9.24 & 8.16 &      & 7.07 & 5.31 & 6.54 &      \\
NGC 6567 &  11.01 & 8.91 & 9.95 & 8.56 & 8.57 & 8.46 & 8.80 & 7.69 & 8.43 & 6.46 & 4.87 & 5.70 & 7.80 \\
M 1-61   &  11.02 &      & 8.79 & 8.11 & 8.78 & 8.72 & 9.01 & 8.12 & 9.20 & 6.70 & 5.25 & 6.22 &      \\
IC 4846  &  10.90 & 8.16 & 8.43 & 8.09 & 8.10 & 8.59 & 8.78 & 7.77 &      & 7.01 & 5.34 & 6.02 &      \\
\noalign{\smallskip}
Average$^{b}$  & 11.05& 8.57 & 8.99 & 8.32 & 9.09 & 8.71 & 9.26 & 8.09 & 9.03 & 7.02 & 5.29& 6.31 & 7.72  \\
% s.d.          &  0.05& 0.18 & 0.86 & 0.15 & 0.17 & 0.13 & 0.54 & 0.20 & 0.30 & 0.30 & 0.19& 0.27 & 0.21  \\
~~~~ $\sigma$      &  0.05& 0.18 & 0.86 & 0.15 & 0.17 & 0.13 & 0.54 & 0.20 & 0.30 & 0.30 & 0.19& 0.27 & 0.21  \\
%$\sigma$      &  0.05& 0.18 & 0.86 & 0.15 & 0.17 & 0.13 & 0.54 & 0.20 & 0.30 & 0.30 & 0.19& 0.27 & 0.21  \\
\noalign{\smallskip}
Average$^{c}$  & 11.06& 8.56 & 9.03 & 8.34 & 9.14 & 8.70 & 9.32 & 8.13 & 9.07 & 7.05 & 5.29& 6.34 & 7.71  \\
% s.d.          &  0.06& 0.17 & 0.65 & 0.79 & 0.16 & 0.13 & 0.55 & 0.26 & 0.33 & 0.33 & 0.18& 0.27 & 0.18  \\
~~~~ $\sigma$      &  0.05& 0.18 & 0.86 & 0.15 & 0.17 & 0.13 & 0.54 & 0.20 & 0.30 & 0.30 & 0.19& 0.27 & 0.21  \\
%$\sigma$      &  0.05& 0.18 & 0.86 & 0.15 & 0.17 & 0.13 & 0.54 & 0.20 & 0.30 & 0.30 & 0.19& 0.27 & 0.21  \\
\noalign{\smallskip}
\noalign{\smallskip}
\multicolumn{14}{l}{Selected bulge samples from the recent literature}\\
RPDM97     &  11.08&      &      & 8.50 &      & 8.72 &       & 7.99 &       & 7.00      &5.70       &6.58&     \\
CMKAS00    & 10.98&        &       & 8.29 &       & 8.74 &       &       &       & 6.92      &5.10       &6.35&     \\ 
EC01       &  11.09&        &       & 7.84 &       & 8.47 &       &       &       & 6.98      &            &6.32&     \\
ECM04      &  11.04&       &       & 8.40 &       & 8.65 &       & 8.13 &       & 6.89      &            &6.73&     \\
EBW04      &  11.05&       &       & 8.43 &       & 8.66 &       & 8.03 &       & 7.05      &            &6.60&     \\       
\noalign{\smallskip}
\multicolumn{14}{l}{Selected disk samples from the recent literature}\\
KB94      &  11.06& 8.74 &       & 8.35 &       & 8.68 &       & 8.09 &       & 6.92 &      & 6.39 &       \\
%TLW$^{d}$ &  11.02& 8.52 & 9.09 & 8.20 & 8.94 & 8.60 & 9.21 & 7.99 & 9.06 & 6.86 & 5.34 & 6.20 & 7.56 \\
TLW$^{d}$ &  11.02& 8.52 & 9.09 & 8.17 & 8.94 & 8.60 & 9.22 & 7.99 & 9.06 & 6.84 & 5.35 & 6.20 & 7.56 \\
\noalign{\smallskip}
Solar$^{e}$ &  10.90& \multicolumn{2}{c}{8.39} & \multicolumn{2}{c}{7.83} & \multicolumn{2}{c}{8.69} & \multicolumn{2}{c}{7.87} & 7.19 & 5.26 &6.55  & 7.55 \\
\noalign{\smallskip}
\hline
\end{tabular}
\begin{list}{}{}
\item $^{a}$ Excluded from statistical analyses (c.f. Section~\ref{anal:2obj}).
\item $^{b}$ Average abundances for 23 GBPNe of the current sample.
\item $^{c}$ Average abundances for 23 GBPNe of the current sample plus
M~1-42 and M~2-36 previously analyzed by \citet{liu2001}.
\item $^{d}$ Combined sample of 58 GDPNe (c.f. Section~\ref{discussion}).         
\item $^{e}$ \citet{lodders2003}.
\end{list}
\end{table*}

Total elemental abundances were derived from ionic abundances (both from CELs
and ORLs) using the same procedures and ionization correction factors (ICFs)
adopted in \cite{wesson2005}. They are tabulated in Table~\ref{ICFs}. Ionic and
elemental abundances as well as ICFs for individual elements, $f$(X), of all
PNe are presented in Tables~\ref{abunO}, \ref{abunC}, \ref{abunN},
\ref{abunNeAr} and \ref{abunSCl}. 

No [Fe~{\sc iv}] lines have been detected in any of the nebulae and no
ICF formula is available to obtain Fe/H elemental abundance from the
Fe$^{2+}$/H$^+$ ionic abundance alone. As such, we have excluded Fe in our
analysis. The Mg~{\sc ii} $\lambda4481$ ORL has been detected in seven 
nebulae.  Following \cite{BPL2003}, we assumed that the Mg~{\sc ii}
$\lambda4481$ has an effective recombination coefficient equal to that of the C~{\sc ii}
$\lambda4267$ line in calculating Mg$^{2+}$/H$^{+}$ from the observed strength
of the $\lambda4481$ line. In addition, considering the unusually wide
ionization potential interval occupied by Mg$^{2+}$, we assumed that Mg/H =
Mg$^{2+}$/H$^{+}$, i.e. no need of applying any ICF corrections. As a
consequence, no Mg$^{2+}$/H$^{+}$ ionic abundances are tabulated separately.
Instead, only the total elemental abundances of magnesium are given in
Table~\ref{totabun} (see below).

Total elemental abundances deduced for each nebula are presented in
Table~\ref{totabun}.  For ease of comparison, we list abundances derived from
both CELs and ORLs in the same Table.  Also tabulated in the
Table are average abundances of individual elements derived from the current
sample of GBPNe, from other samples of GBPNe published in the literature, from
the TLW (c.f. Section~\ref{discussion}) and KB94 sample of GDPNe, as
well as the solar photospheric abundances. Note that average abundances listed
in the Table are logarithmic abundance averages (in units where H = 12), not
averages of logarithmic abundances. 

\subsection{Abundance uncertainties}
\label{abun:err}

There are two major sources of uncertainty that affecting the derived element
abundances: those propagating from line flux measurements and those arising from
the analyzing methodology (e.g. those introduced via the application of empirical
ICFs).  Another potential source of error affecting in particular the current
sample of objects is the uncertainty in the extinction law towards the Galactic
center.  The latter is difficult to characterize due to the lack of
other independent means of measuring the extinction curve towards the Galactic
center.

Uncertainties in line fluxes will also affect abundance determinations via
derivations of \Te\, and \Ne.  For most of our 
objects, we expect that flux measurements of strong lines (strengths
larger than 0.008 relative to H$\beta$) should be accurate to better than 5
per~cent. This encompasses essentially all diagnostic CELs, including the
relatively weak [O~{\sc iii}]~$\lambda4363$ and [N~{\sc ii}]~$\lambda5755$
auroral lines, important for \Te\, determinations. We estimated that
uncertainties in plasma diagnostics of CELs should typically be no more than
4\,per~cent in \Te\, and 30\,per~cent in \Ne, and their total effect on
deduced ionic abundances should be less than 6\,per~cent. There are a few
exceptions where the abundances were based on low critical density far-IR
fine-structure lines, which are more sensitive to \Ne\, variations.      

Emissivities of ORLs have only a weak dependence on \Te\, and \Ne.  However,
those lines are generally much weaker than CELs and are often affected by
serious line blending.  For a few objects lacking high S/N, high
resolution spectra, such as M~1-29, uncertainties in the derived ORL abundances
of N, O and Ne could be as large as 30\,per~cent.  The situation is much better
for carbon, which has a few relatively strong well observed ORLs in the
optical. Thus in most cases, we expect carbon abundances derived from ORLs
to be accurate to better than 5\,per~cent. Comparable or even better
accuracies are expected for helium abundances. 

\citet{alexander1997} have investigated potential errors in abundance
determinations that may result from applying the KB94's ICF scheme when only
optical lines are observed, by constructing a series of constant-density PN
models.  They found that the errors are smaller for small unresolved nebulae,
and typically amount to $\la 10$\,per~cent for helium and oxygen, $\la
25$\,per~cent for nitrogen, sulfur and neon, but the abundances of argon are
probably systematically overestimated. Given the large distance to the Galactic
bulge, most PNe in our sample are quite compact, less than 10\arcsec\, in
diameter. A few objects in our sample have  UV and/or IR spectra available.
For those objects, we expect a better accuracy for nitrogen abundances.

In summary, we estimate that the total error budgets of our current abundance
determinations are: 0.12\,dex for He/H from ORLs and O/H from CELs, 0.26 to
0.30\,dex for N/H, S/H, Ne/H, Ar/H and Cl/H from CELs, 0.12\,dex for C/H from
ORLs, 0.25 to 0.29\,dex for O/H, N/H and Ne/H from ORLs.

\section{Comparisons with previous studies}
\label{abundis:b2b}

Hitherto several spectroscopic studies presenting approximately 350 abundance
determinations for more than 250 GBPNe have been published. In this section, we
will compare our results with those from some of the most recent ones,
including W88~\citep{webster1988}, RPDM97~\citep{ratag1997}, SRM98~
\citep{stasinska1998}, CMKAS00~\citep{cuisinier2000}, EC01~ \citep{escudero01},
ECM04~\citep{escudero04} and  EBW04~ \citep{exter2004}. We will compare the
general patterns of abundance obtained in Section~\ref{abundis:b2b:aver} as
well as individual measurements for common objects in
Section~\ref{abundis:b2b:indiv}.  Unless otherwise specified, the current
sample of 25 GBPNe, WL07, refers to the 23 GBPNe presented in the current paper
(c.f. the first part of Table~\ref{totabun}, with two peculiar objects,
M\,2-23 and M\,2-39 excluded) plus M~1-42 and M~2-36 previously analyzed by
\cite{liu2001}. 

Except for \cite{liu2001} presenting heavy element abundances deduced both from
CELs and from ORLs for two GBPNe, essentially all abundance determinations of
heavy elements for GBPNe published in the literature have been based on strong
CELs only. Thus, comparisons will only be made for abundances derived from CELs.
Also, carbon abundances have previously been measured only for a very small
number of GBPNe, so the comparisons will concentrate on helium, oxygen, 
nitrogen and neon. 

Note that much of the earlier work on GBPNe used the standard extinction law of
the general diffuse ISM, which is known to be inapplicable for the Galactic
centre (e.g. \citealt{udalski2003}; \citealt{ruffle2004}), in particular in the
UV.  However, for analyses based on optical or infrared observations only, the
uncertainties thus introduced were likely to be small.  The recombination
contribution to the [N~{\sc ii}] $\lambda$5755 auroral line was also
uncorrected, leading to overestimated \Te([N~{\sc ii}]) and consequently
underestimated N$^{+}$/H$^{+}$, O$^{+}$/H$^{+}$ and S$^{+}$/H$^{+}$ ionic
abundance ratios in some nebulae. 

\subsection{Comparisons of general patterns of abundances}
\label{abundis:b2b:aver}

W88 presented elemental abundances for a sample of 49 GBPNe, amongst which five
are in common with our sample: M~2-27, M~2-33, M~3-7, M~3-33 and Cn~2-1. Her
sample yields an average oxygen abundance of 8.73, very similar to the value of
8.70 deduced from the current sample (c.f.  Table~\ref{totabun}). For helium,
her sample yields an average abundance of 11.06 (11.05 for non-Type\,I PNe),
identical to the value deduced  from the current sample.  Note however,
collisional excitation of He~{\sc i} was not corrected for in W88 -- the
effects typically amount to a few per cent.  She did not present nitrogen
abundances separately. From the N/O abundance ratios listed in her paper, we
derive an average N/O ratio of $0.24$, much less than the average value of 0.40
deduced from the current sample, suggesting that nitrogen abundances obtained
by W88 are systematically lower than ours by almost a factor of two.

RPDM97 presented abundances for a sample of about 110 GBPNe, deduced from their
own observations or from line fluxes published in the literature. They used a
grid of photoionization models to estimate ICFs, which were then used to
calculate elemental abundances from abundances of observed ionic species. The
average abundances calculated from values of individual nebulae presented in
Table~5 of RPDM97 are listed in Table~\ref{totabun}. For He, O and S, their
results are in excellent agreement with ours (within 0.05~dex).  For N, Cl and
Ar, the differences are somewhat larger. For Cl and Ar, measurement
uncertainties may play a role -- weak lines of [Cl~{\sc iii}] and [Ar~{\sc iv}]
might not be well measured in their low-resolution spectra. For nitrogen, parts
of the discrepancy may arise from uncertainties in ICFs. For CEL abundance
analyses, only the trace ionic species N$^+$ is seen in the optical wavelength
range. More abundant N$^{++}$ is only observable in the UV or the far infrared.
Thus ICF corrections for nitrogen are often quite large and uncertain when only
optical data are available.

SRM98 combined spectroscopic data published by \citet{aller1987}, W88 and
RPDM97 and selected 90 objects with a 6~cm radio flux density less than
100\,mJy, considered to belong to the Bulge population at the 90--95 per~cent
confidence level. They used an empirical abundance determination scheme similar
to our CEL analyses and adopted the same set of ICFs given by KB94. However,
they applied \Te\ derived from either the [O~{\sc iii}]$_{\rm na}$ or the
[N~{\sc ii}]$_{\rm na}$ ratios for all ionic species in abundance calculations.
For their GBPNe sample containing 85 objects with oxygen abundance measured,
they derived a mean oxygen abundance 8.48 with a standard deviation ($\sigma$)
of 0.434. For 32 brighter objects in their sample, the average abundance is
8.67 ($\sigma=0.213$). The latter is in good agreement with our result (c.f.
Table~\ref{totabun}). 

CMKAS00 presented spectrophotometric observations and determined element
abundances for a sample of 30 GBPNe.  \Te's derived from the [N~{\sc ii}]$_{\rm
na}$ and [O~{\sc iii}]$_{\rm na}$ ratios were adopted for low- and
high-ionization zones, respectively, and \Ne's were derived from the [S~{\sc
ii}]$_{\rm nn}$ ratio. They used ICFs given by \citet{koppen1991}, which are
the same as ours for He, N and Ne.  Mean abundances of He, O, N, S, Cl and Ar,
calculated from their Table~3, are presented in Table~\ref{totabun}.  For N, O
and Ar, their results are almost identical to ours, with differences less than
0.05~dex. For S and Cl, the differences are larger, but still within the
uncertainties ($\sim 0.2$~dex).  For helium their sample yields an average
abundance of 10.98, 0.08\,dex lower than our value, and is the lowest amongst
all GBPNe samples listed in Table~\ref{totabun}. CMKAS00 used the
criteria identified by \citet{acker1991} to select GBPNe, being similar to
ours. There are however only two objects in common between the two samples.

EC01 and ECM04 present element abundances for samples of 45 and 57 GBPNe,
respectively. Unlike other studies, EC01 and ECM04 determined ionic abundances
from measured line strengths using analytic functions deduced by
\citet{alexander1997}, rather than by solving level populations of the emitting
ions. The only exception was Cl$^{++}$, not dealt with by
\citet{alexander1997}; for this they solved statistical balance equations to
obtain emissivities of the [Cl ~{\sc iii}] $\lambda\lambda$5517,5537 lines as a
function of \Te\, and \Ne.  Average abundances from these two samples of GBPNe
were calculated from Table~6 of EC01 and Table~4 of ECM04 and tabulated in
Table~\ref{totabun}.

The sample of EC01 seems to yield much lower average abundances for nitrogen
and oxygen, by respectively as much as 0.60 and 0.23~dex compared to our sample
and that of ECM04, too large to accounted for by observational uncertainties
alone.  The large discrepancies are difficult to understand given that nebulae
in the EC01 sample were observed and analyzed in the same way as those in the
ECM04 sample, and the latter yields average abundances which are in excellent
agreement with ours for all elements, with the possible exception of Ar for
which our value is 0.39\,dex lower\footnote{If we exclude the 4 PNe in the
ECM04 sample that show an Ar abundance either lower than 5 or higher than 7.6,
which seems unrealistic, then the difference in the average Ar abundances
deduced from the two samples drops to 0.11\,dex}. It is not clear to us whether
there were any systematic errors in the analysis of EC01. There are no objects
in common between the EC01 and other samples, including ours, and EC01 did not
publish line fluxes. Hence it is difficult to check whether the differences are
real, caused by e.g. a different population being sampled by EC01 or are
possibly artifacts.

\begin{figure}
\centering
\epsfig{file=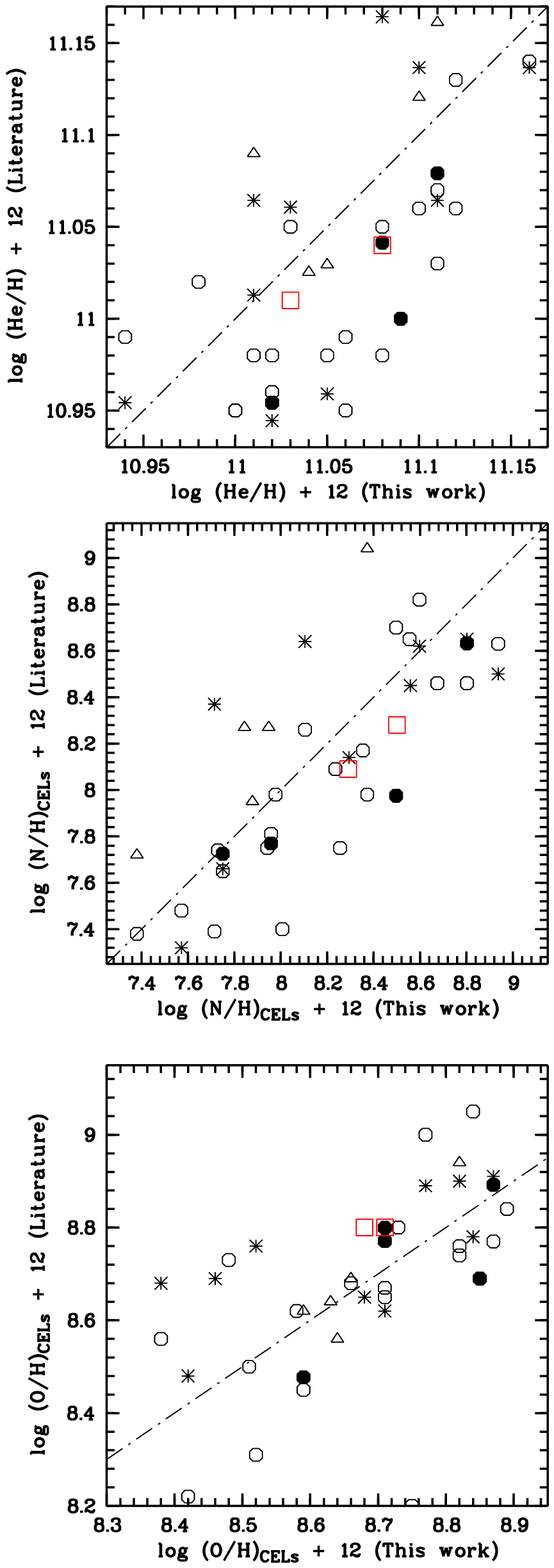, width=7.5cm, bbllx=37,
bblly=35, bburx=286, bbury=740,clip=,angle=0}
\caption{He, N and O abundances derived in the current work are compared to
those in the literature for common objects. {\it Filled circles:}\ W88;
{\it Open circles:}\ RPDM97; {\it Open squares:}\ CMKAS00; {\it Open 
triangles:}\ ECM04; {\it Asterisks:} EBW04.}
\label{comm_lit_heno}
\end{figure}

Finally in a recent study, EBW04 determined abundances for a sample
of 45 GBPNe, using a reduction procedure, set of atomic parameters and an ICF
scheme that are almost identical to ours.  Average abundances derived from
their sample are in excellent agreement with ours, with differences less than
0.04~dex for He, O, S, about 0.1~dex for N and Ne and about 0.26~dex for Ar
(c.f. Table~\ref{totabun}).  

\subsection{Comparisons of abundances for common objects}
\label{abundis:b2b:indiv}

Fig.~\ref{comm_lit_heno} compares abundances of He, N and O determined in the
current work with those published in the literature for common objects that
were also observed and analyzed in the five other samples of GBPNe discussed
above, but excluding the sample of EC01. CMKAS00 have only two objects in
common with us, and the He and N abundances they obtained are about 0.03 and
0.21\,dex lower than ours, whereas those of oxygen are 0.10\,dex higher. The
level of agreement is quite good, considering the measurement uncertainties
involved. The numbers of common objects are 6, 20, 5 and 10 for the W88,
RPDM97, ECM04 and EBW04 samples, respectively.

In term of systematic errors, abundances obtained by EBW04 are in best
agreement with out results, although the scatter is also relatively large.
The average abundance differences between the two samples are $0.00\pm 0.07$,
$-0.09\pm 0.13$ and $0.10\pm 0.46$\,dex for He, O and N, respectively,
where the errors are standard deviations. For ECM04, the abundances they
deduced for He and O for the five objects in common are in excellent agreement
with our results, with average differences of $0.00\pm 0.10$ and $-0.02\pm
0.07$\,dex, respectively. However, our nitrogen abundances are significantly
lower than their results, by $-0.37\pm 0.21$\,dex on average.

For objects in common with the W88 and with the RPDM97 samples, abundances of
oxygen and nitrogen obtained in those studies are in good agreement with our
results. For the W88 sample, the average differences are $0.03\pm 0.10$ and
$0.26\pm 0.19$\,dex for oxygen and nitrogen, respectively. For the RPDM97
samples, the corresponding values are $0.01\pm 0.22$ and $0.14\pm 0.23$\,dex.
However, both W88 and RPDM97 yield helium abundances that are systematically
lower than our values, by $0.05\pm 0.03$ and $0.04\pm 0.04$\,dex on average,
respectively. The differences are found to mainly arise from measured line
fluxes -- for the same object, helium line fluxes measured by W88 and by RPDM97
are systematically lower than our values. For objects in common with RPDM97, if
we recalculate helium abundances with our own method but using their line
fluxes, the results are almost identical to their values. For the W88 sample,
if we apply our analysis to her line data, the resultant helium abundances
decrease by approximately 0.03\,dex. This is caused by the fact W88 did not
make any correction for the effects of collisional excitation of He~{\sc i}
lines. Both W88 and RPDM97 applied non-linearity corrections to their line
fluxes, measured, respectively, with the 3.9m Anglo-Australian Telescope using
an Image Dissector Scanner as detector and with the ESO 3.6m or 1.52m
telescopes also using an IDS. It is quite possible that both W88 and RPDM97
have over-corrected the effects of non-linearity, leading to underestimated
fluxes for lines of low or intermediate strengths. If this was indeed the case,
then intensities of the [O~{\sc iii}] $\lambda$4363 and [N~{\sc ii}]
$\lambda$5754 auroral lines were also likely to have been underestimated,
leading to underestimated \Te, with apparent consequences to ionic abundances
deduced for heavy elements.

\section{CELs versus ORLs}
\label{orl2cel}

As described in Section~\ref{intr}, plasma diagnostic analyses and elemental
abundance determinations of PNe and H~{\sc ii} regions using CELs and
ORLs/continua have ubiquitously showed dichotomy between results deduced from
those two types of emission excited by different mechanisms. The discrepancies
are generally characterized by $\Delta$\Te, defined as \Te([\oiii])$-$\Te(BJ),
the difference between \Te\, deduced from the [O~{\sc iii}] collisionally
excited lines and from the Balmer discontinuity of H~{\sc i} recombination
spectrum, and by abundance discrepancy factor, adf, defined as the ratio of
ionic abundances X$^{\rm i+}$/H$^+$ derived from ORLs and from CELs.  Deep
optical spectroscopic surveys for dozens of GDPNe and a number of Galactic and
extragalactic H~{\sc ii} regions have shown that values of \Te(BJ) are
systematically lower than those of \Te([\oiii]) deduced for the same objects
and that adf's are always larger than unity, falling typically in the range
1.6--3, but with a significant tail extending to higher values, reaching $\sim
70$ in the most extreme case. Except for the two GBPNe analyzed by
\cite{liu2001}, ORL studies of PNe have so far concentrated on relatively
bright GDPNe.  The sample of GBPNe presented in the current work thus provides
an interesting environment to probe the behaviours of the aforementioned
fundamental discrepancies for a different stellar population.  In particular,
previous studies of GDPNe show that relative abundances of heavy elements, such
as the C/O ratios deduced from ORLs are compatible to those derived from CELs.
It is important to test whether the same result also holds for GBPNe.

\subsection{Electron temperatures} 
\label{orl2cel:tene}

\begin{figure*}
\centering
\epsfig{file=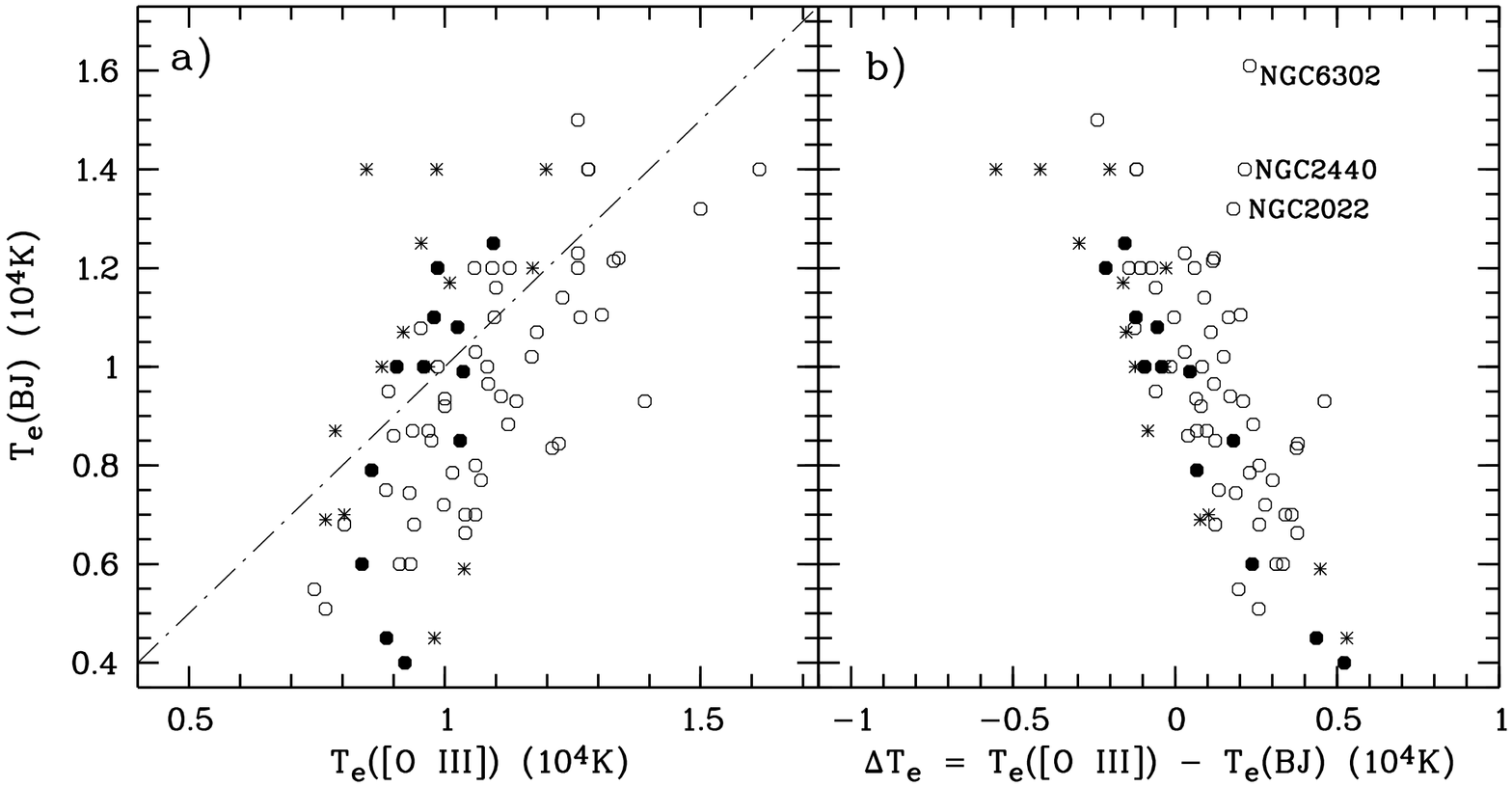, width=7.0cm, 
bbllx=181, bblly=53, bburx=525, bbury=714, clip=, angle=-90}
\caption{\Te(BJ) plotted against a) \Te([O~{\sc iii}]) and b) $\Delta T_{\rm e}
\equiv$ \Te([O~{\sc iii}])$-$\Te(BJ) for GBPNe (filled circles and asterisks) and
GDPNe (open circles). Filled circles and asterisks represent respectively GBPNe of
sub-samples~A and B, defined in Section~\ref{obs:opt}.  In a), the
dotted-dashed diagonal line represents identical \Te(BJ) and \Te([O~{\sc
iii}]).  }
\label{BJvsOIII} 
\end{figure*}

\begin{figure} 
\centering
\epsfig{file=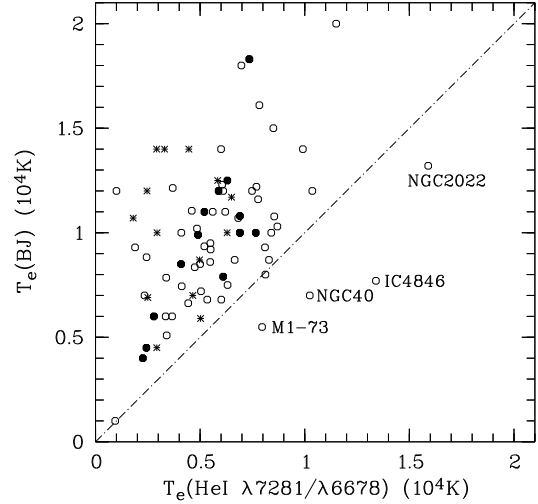, width=7.0cm, 
bbllx=32, bblly=283, bburx=543, bbury=772, clip=, angle=0}
\caption{\Te(BJ) plotted against \Te(He~{\sc i}~$\lambda7281/\lambda6678$),
using the same symbols as in Fig.~\ref{BJvsOIII}. The solid diagonal
line represents $y = x$.}
\label{TeBJHeI} 
\end{figure}

\begin{figure} 
\centering
\epsfig{file=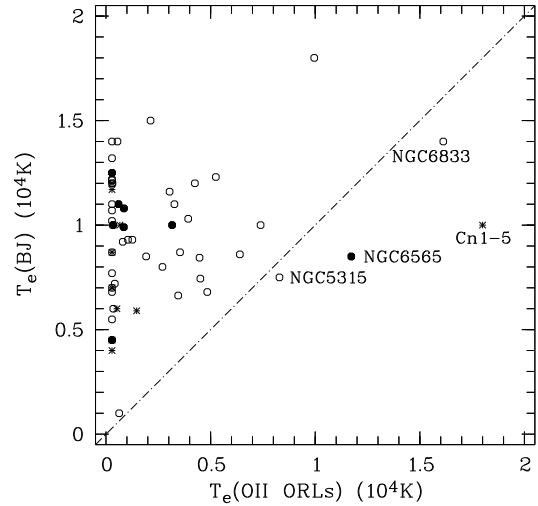, width=7.0cm, 
bbllx=32, bblly=283, bburx=543, bbury=772, clip=, angle=0}
\caption{\Te(BJ) plotted against \Te(O~{\sc ii} ORLs) using the same symbols 
as in Fig.~\ref{BJvsOIII}. 
}
\label{TeBJOIIr} 
\end{figure}

In Figs.~\ref{BJvsOIII}, \ref{TeBJHeI} and \ref{TeBJOIIr}, $T_{\rm e}$(BJ) is
plotted against \Te([O~{\sc iii}]),  $\Delta$\Te, \Te(He~{\sc
i}~$\lambda7281/\lambda6678$) and against \Te(O~{\sc ii}~ORLs). Nebulae of
sub-sample~B, for which only low resolution spectra are available near the
Balmer discontinuity region, do not behave obviously differently compared to
objects of sub-sample~A, although measurements for sub-sample~B clearly suffer
from larger uncertainties compared to the latter sub-sample.  As a consequence,
in analyses below we shall concentrate on patterns and relations revealed by
sub-sample~A.

A majority of our sample GBPNe (15 out of 25) have \Te(BJ) higher than
\Te[O~{\sc iii}] (c.f. Fig.\,\ref{BJvsOIII}). For the sub-sample of high data
quality, exactly half of the objects have \Te(BJ) higher than \Te([O~{\sc
iii}]) while the other half have lower \Te(BJ). The result appears to
contradict to what previously found for GDPNe where for majority of them,
\Te(BJ) is systematically lower than \Te([O~{\sc iii}]). Note that the
magnitude of Balmer jump is inversely proportional to \Te, thus as \Te\,
increases, the Balmer jump becomes increasing difficult to measure. And because
of the nonlinear dependence of \Te(BJ) on the magnitude of Balmer jump, the
resulting error of \Te(BJ) for a given uncertainty of BJ is not symmetric -- it
biases towards higher values. It seems to us, at least parts, if not all, of
the discrepant behaviour of GBPNe compared to GDPNe as seen in
Fig.\,\ref{BJvsOIII}, are due to lower quality of the data available for GBPNe,
which are generally fainter than nearby GDPNe. 

While the left panel of Fig.\,\ref{BJvsOIII} shows no obvious correlation
between \Te(BJ) and \Te([O~{\sc iii}]), the right panel shows that there is a
tight anti-correlation between \Te(BJ) and $\Delta$\Te\, and different samples
seem to yield similar results. Both panels however reflect the fact that while
\Te(BJ) varies over a wide range of values, \Te([O~{\sc iii}]) falls within a
much narrow range, as previously observed in GDPNe (c.f.
\citealt{liu2006a,liu2006b} and references therein).

Fig.~\ref{TeBJHeI} compares \Te(BJ) with \Te(He~{\sc i}
$\lambda7281/\lambda6678$), \Te\, derived from the ratio of He~{\sc
i} singlet recombination lines $\lambda7281$ and $\lambda6678$. For all PNe
plotted in the Figure, both disk and bulge ones, \Te(He~{\sc i}
$\lambda7281/\lambda6678$) is invariably lower than \Te(BJ), in consistent with
the expectations of the two-abundance model proposed by \cite{liu2000}, which
predicts that \Te(O~{\sc ii} ORLs) $\la $ \Te(He~{\sc i} ORLs) $\la$ \Te(H~{\sc
i} BJ) $\la$ \Te([O~{\sc iii}])
(\citealt{pequignot2002,pequignot2003,liu2003}), exactly opposite to the
expectations of the paradigm of temperature fluctuations which predict
\Te(He~{\sc i} $\lambda7281/\lambda6678$) $\ga$ \Te(BJ) (\citealt{zhang05a}).

Fig.~\ref{TeBJOIIr} plots \Te(O~{\sc ii} ORLs) against \Te(BJ), and again, in
consistent with what previously found for GDPNe, values of \Te(O~{\sc ii} ORLs)
are considerably lower. 

About half PNe in our GBPNe sample have an [O~{\sc ii}]
$\lambda$4089/$\lambda$4649 line ratio higher than the low \Te\, limit of
288~K, the lowest \Te\, at which the effective recombination coefficients for
O~{\sc ii} are available. \cite{tsamis2004} detected a faint emission line at
4116\,\AA\, in three bright PNe, NGC\,3242, NGC\,5882 and NGC\,6818, and
attributed it to Si~{\sc iv}~4s$^2$S$_{1/2}-$4p$^2$P$_{1/2}$ $\lambda$4116.1.
They corrected for the contribution of the other component of this multiplet,
the 4s$^2$S$_{1/2}-$4p$^2$P$_{3/2}$ $\lambda$4088.8 line, to the observed
intensity of the O~{\sc ii} $\lambda$4089.3 line. After the corrections they
obtained an \Te(\oii\, ORLs) of 2600, 8700 and 2900~K for NGC\,3242, NGC\,5882
and NGC\,6818, respectively. For comparison, the H~{\sc i} Balmer jump
temperatures they obtained for the three PNe are 10,200, 7800 and 12,100~K,
respectively, and the corresponding values for \Te([\oiii])  are 11,700, 9400
and 13,300~K. The 4116\,\AA\, feature has not been detected in NGC\,6153
(\citealt{liu2000}; adf = 9.7), in bulge PNe M\,1-42 and M\,2-36
(\citealt{liu2001}; adf = 22.0 and 6.9, respectively) and in Hf\,2-2
(\citealt{liu2006}; adf = 71.2). For NGC\,7009 (\citealt{liu1995}; adf = 4.7),
our unpublished deep spectra reveals a weak feature at 4116\,\AA.  And if we
apply similar corrections, then \Te(\oii\ ORLs) yielded by the
$\lambda4089/\lambda4649$ ratio increases from 420\,K before the correction to
1310~K.  For the sample of PNe observed and analyzed by
\cite{liuy2004a,liuy2004b}, a line 4116.24\,\AA\, rest wavelength was detected
with a dereddened intensity of 0.029 (in units where H$\beta = 100$) in the
spectrum of NGC\,6884 and was identified as [Ni~{\sc v}] $\lambda$4117.20.  If
we instead attribute it to Si~{\sc iv} $\lambda$4116, then \Te(\oii\, ORLs)
yielded by the $\lambda4089/\lambda4649$ ratio will increase from the value of
3040~K before the correction to 22,100~K after the correction.  Finally,
amongst the large number of sample PNe studied by \cite{wesson2005}, the only
PN that has a feature detected at 4116\,\AA\, is Vy\,2-2, with an observed
intensity of $0.178$ in units of H$\beta = 100$ after reddening corrections.
However, the feature is unlikely to be Si~{\sc iv} $\lambda$4116 for two
reasons: 1) If it is Si~{\sc iv} $\lambda$4116, then the other component of the
multiplet at $\lambda$4089 should have an intensity of $2\times 0.178 = 0.356$,
which is over a factor of three higher than the measured total intensity of the
$\lambda$4089 feature (0.098); 2) The O$^{++}$/H$^+$ ionic abundance ratio
derived from the O~{\sc ii} $\lambda$4089 line, assuming no contamination from
Si~{\sc iv}, is in good agreement with values deduced from other O~{\sc ii}
ORLs. The case of Vy\,2-2 suggests that the identification of the 4116\,\AA\,
features detected in some PNe as the Si~{\sc iv} $\lambda$4116 is not
completely secure and further observations of better S/N's and spectral
resolutions are needed to clarify the situation.

\subsection{Abundance discrepancy factors}
\label{orl2cel:adf}

The abundance discrepancy factor (adf) for a heavy element ionic species
X$^{\rm i+}$ is defined as ratio of ionic abundance ratios X$^{\rm i+}$/H$^+$
derived from ORLs and from CELs, adf(X$^{\rm i+}$) $\equiv$ (X$^{\rm
i+}$/H$^{+}$)$_{\rm ORLs}$/(X$^{\rm i+}$/H$^{+}$)$_{\rm CELs}$. Previous
studies of several samples of GDPNe show that adf varies over a wide range of
values, from near unity (i.e. the ionic abundance ratios derived from the two
types of emission lines agree) to over a factor of ten, peaking around a factor
of two but with a tail extending to much higher values, reaching a factor of
$\sim 70$ in the most extreme case (c.f. \citealt{liu2006a,liu2006b}).

\begin{figure} \centering
\epsfig{file=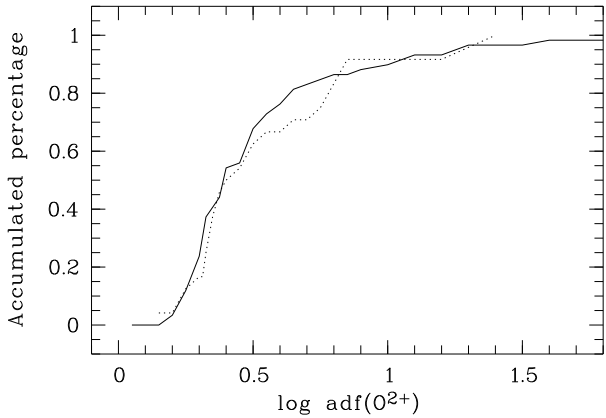, width=5.5cm,
bbllx=52, bblly=37, bburx=559, bbury=772, clip=, angle=-90}
\caption{Comparison of the accumulative distributions of 
$\log\,{\rm adf}({\rm O}^{2+})$ of GDPNe (solid line) and of
GBPNe (dotted line).} 
\label{adfb2d}
\end{figure}

In Fig.~\ref{adfb2d} we plot and compare the accumulative distributions of adf
for disk PNe from the TLW (c.f. Section~\ref{discussion}) sample and for bulge
PNe from the current sample.  The two distributions show some differences for
adf(O$^{2+}$) between 0.5 and 0.8\,dex. A Kolmogorov-Smirnov Two-sample test,
with the Null Hypothesis that the two samples have the same distributions of
adfs, yields a Kolmogorov-Smirnov statistic $D$ of 0.1808 with a probability of
the two distributions having a difference exceeding $|D|$ under the Null
Hypothesis, $P(>|D|) = 0.59$, which is equal to 1 minus the confidence level.
In other words, the risk is high, nearly 60~per~cent, to reject the Null
Hypothesis.  The analysis thus shows that to a fairly good confidence level,
both GDPNe and GBPNe seem to yield similar distributions of adf(O$^{2+}$). 

\begin{figure} \centering 
\epsfig{file=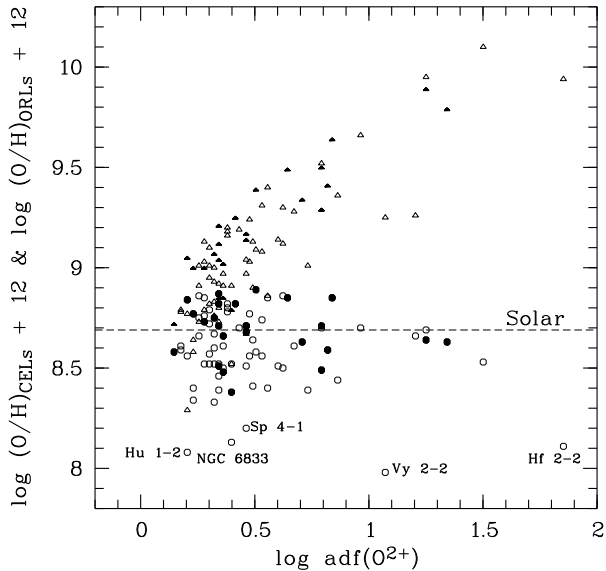, width=8.0cm, bbllx=37, bblly=288, bburx=548, 
bbury=772, clip=, angle=0} 
\caption{Elemental abundances of oxygen derived from ORLs (triangles) and from
CELs (circles) plotted against adf(O$^{2+}$) for GBPNe (filled symbols) and
GDPNe (open symbols). The solar photospheric oxygen abundance
(\citealt{lodders2003}) is marked.} 
\label{adf2O} 
\end{figure}

\begin{figure} 
\epsfig{file=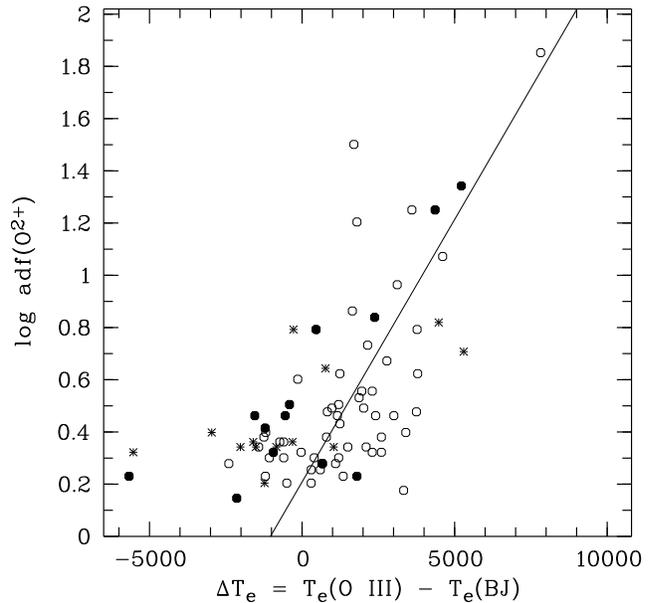, width=8.cm, bbllx=51,
bblly=36, bburx=537, bbury=550, clip=, angle=-90}
\caption{adf(O$^{2+}$) plotted against $\Delta$\Te. For the meaning of
different symbols, c.f. Fig.~\ref{BJvsOIII}. The solid line shows the linear
fit obtained by \citet{liu2001}.} 
\label{adfDT} \end{figure}

In Fig.~\ref{adf2O}, we plot total elemental abundances of oxygen derived from
CELs and from ORLs against adf(O$^{2+}$). The Figure shows that except for a
few PNe with O/H less than 8.2, oxygen abundances derived from CELs fall into a
relatively narrow range, consistent with the general consensus that oxygen is
neither created nor destroyed by any significant amount in LIMS. By contrast,
oxygen abundances yielded by ORLs, apart from being always higher than the
corresponding values derived from CELs for the same object, cover a wide range
of values, reaching about a factor of 25 higher than the solar value in the
most extreme case. Such high abundances are extremely difficult to explain in
terms of the current theory of evolution of low- and intermediate-mass (single)
stars, and argue strongly against the paradigm of temperature fluctuations as
the underlying physical cause of the large adf values, as the scenario
implicitly assumes that the higher abundance values yielded by ORLs, being
insensitive to the presence of temperature fluctuations, are the correct values
(c.f. \citealt{liu2006b}).  Together with Fig.~\ref{BJvsOIII}, Fig.~\ref{adf2O}
also implies that whatever effects are causing the large values of adf observed
in PNe, it seems they manifest themselves mainly by affecting emission of
ORLs/continua, leading to large variations in \Te\, and ionic abundances
that are inferred from emission of ORLs/continua.

Fig.~\ref{adfDT} plots $\log$~adf(O$^{2+}$) against $\Delta$\Te.  Also
overplotted is the linear fit obtained for a much smaller sample of PNe by
\cite{liu2001} who first noticed that the two quantities are positively correlated.
The result suggests that both phenomena, i.e. that \Te\,
derived from the [O~{\sc iii}] forbidden lines are systematically higher that
those derived from the Balmer discontinuity of the H~{\sc i} recombination
spectrum, and that ionic abundance ratios of heavy elements derived from ORLs
are systematically higher than deduced from CELs, are related and probably
caused by the same underlying physical mechanism. Furthermore, from the
discussion above, we may conclude that the mechanism, whatever it is, affects
mainly the emission of ORL/continua, rather than that of CELs. The results fit
nicely with the expectations of the two-abundance model proposed by
\citet{liu2000}, but contradict those of the paradigm of temperature
fluctuations which affect mainly the production of emission of CELs.

\begin{figure} \epsfig{file=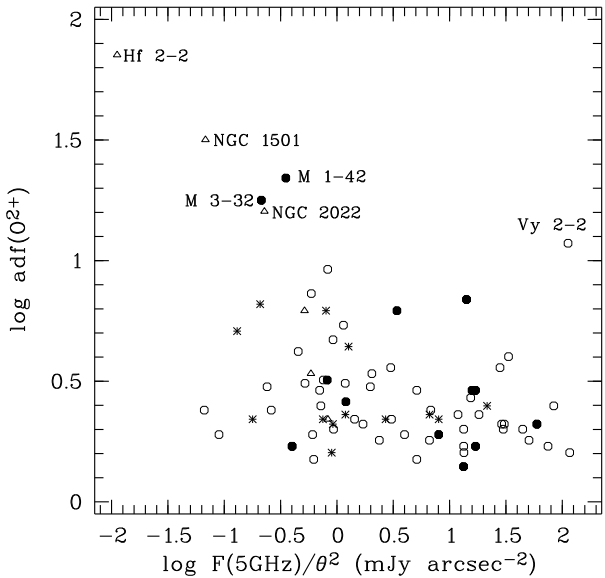, width=8.0cm, bbllx=37,
bblly=288, bburx=543, bbury=772, clip=, angle=0}
\caption{adf(O$^{2+}$) plotted against radio band surface brightness at 6~cm.
For the meaning of different symbols, c.f. Fig.~\ref{BJvsOIII}. For a few PNe
that have no measurements of 5\,GHz radio flux density, we 
have adopted 1.4\,GHz fluxes and converted the values to 5\,GHz fluxes 
assuming optically thin emission. Those objects are marked with 
triangles.}
\label{adfF5G} \end{figure}

The mean surface brightness reflects the evolutionary stage of a PN. As the PN
evolves and expands, its density drops and thus the surface brightness.
\citet{garnett01} found that adf(O$^{2+}$) tends to be larger for bigger, lower
surface brightness PNe. A similar trend was found by \citet{liuy2004b}.  In
Fig.~\ref{adfF5G}, we plot adf(O$^{2+}$) against the nebular 6\,cm radio
surface brightness, defined as $F(5\,{\rm GHz})/\theta^2$, where $F(5\,{\rm
GHz})$ is nebular 6~cm radio flux density in mJy and $\theta$ is angular
diameter in arcsec (c.f. Table~\ref{obj}). Using radio flux density instead of
H$\beta$ flux has the advantage that the former is not affected by interstellar
extinction. Although with large scatter, the diagram shows some evidence
that adf increases with decreasing surface brightness, consistent with
previous results.

\begin{figure} \epsfig{file=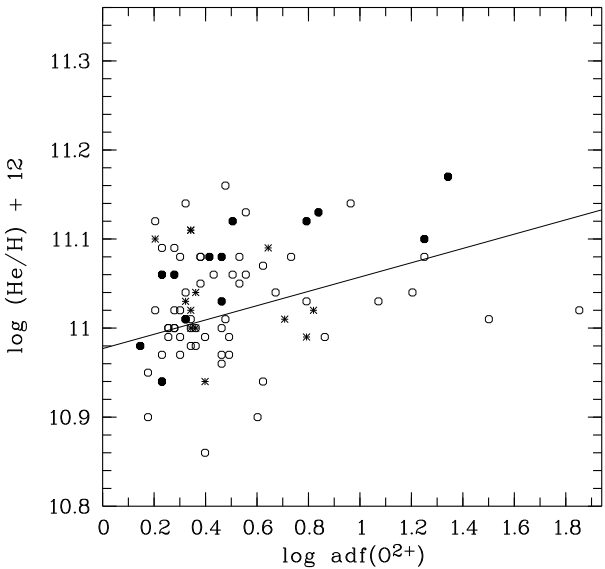, height=8.0cm,
bbllx=52, bblly=29, bburx=536, bbury=543, clip=, angle=-90}
\caption{Helium abundance plotted against adf(O$^{2+}$). For the meaning of 
different symbols, c.f. Fig.~\ref{BJvsOIII}. The solid line represents the 
linear fit obtained by \citet{liuy2004b}.}
\label{ADFOvsHe} \end{figure}

\citet{zhang04a} found a positive correlation between $\Delta T_{\rm e}$ and
helium abundance, in accordance with the expectations of the two-abundance
model. Our current data for GBPNe exhibit a similar trend, albeit with
larger scatter due to the larger uncertainties associated with the current
determinations of \Te(BJ) (c.f. \S\ref{obs:opt}).  Fig.~\ref{ADFOvsHe} plots He
abundance against adf(O$^{2+}$). The solid line in the Figure shows the linear
regression obtained by \citet{liuy2004b} for disk PNe. Again, while
the scatter is large, there is a discernible positive correlation between the
two quantities.

\begin{figure} \epsfig{file=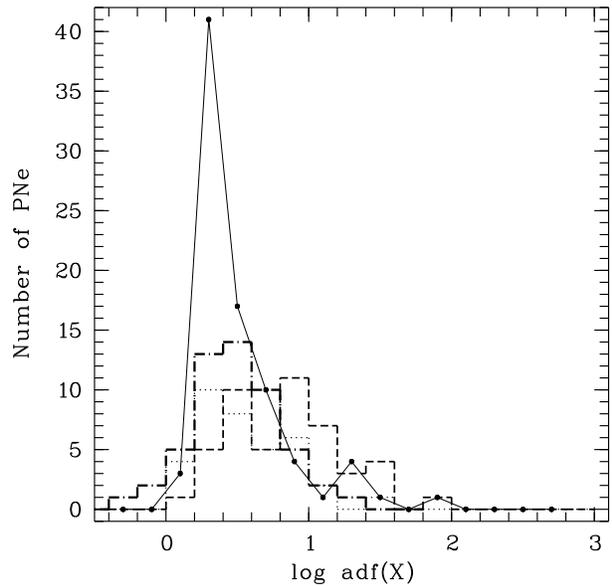, width=8.0cm,
bbllx=44, bblly=288, bburx=543, bbury=772, clip=, angle=0}
\caption{Comparison of the distributions of $\log$~adf for O$^{2+}$/H$^+$ 
(solid curve), C$^{2+}$/H$^+$ (dashed-dotted histogram), N$^{2+}$/H$^+$ 
(dotted histogram) and Ne$^{2+}$/H$^+$ (dashed histogram) ionic abundance 
ratios.} 
\label{DF_ADF_NONeC} \end{figure}

\begin{figure*}
\centering
\epsfig{file=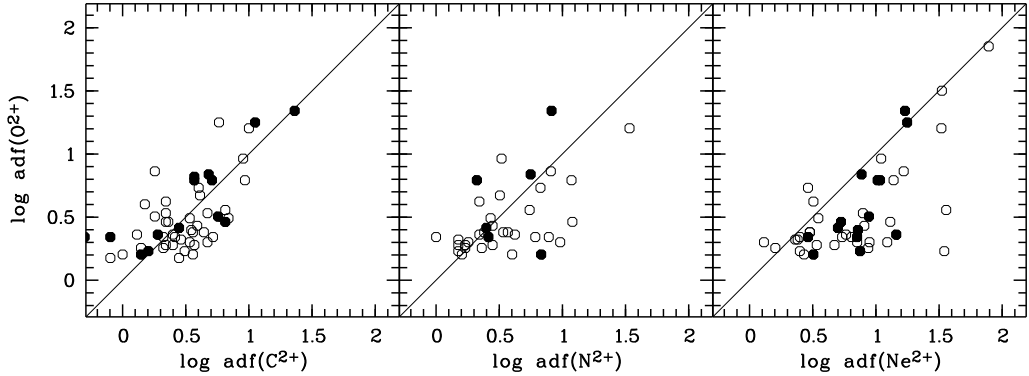, width=5.0cm, bbllx=295,
bblly=47, bburx=533, bbury=697, clip=, angle=270}
\caption{$\log$~adf of O$^{2+}$ plotted against those of C$^{2+}$, N$^{2+}$ and Ne$^{2+}$
for GBPNe (filled circles) and GDPNe (open circles). The solid 
lines represent $y=x$.} 
\label{adfNNeCvsO}
\end{figure*}

\begin{figure*}
\centering
\epsfig{file=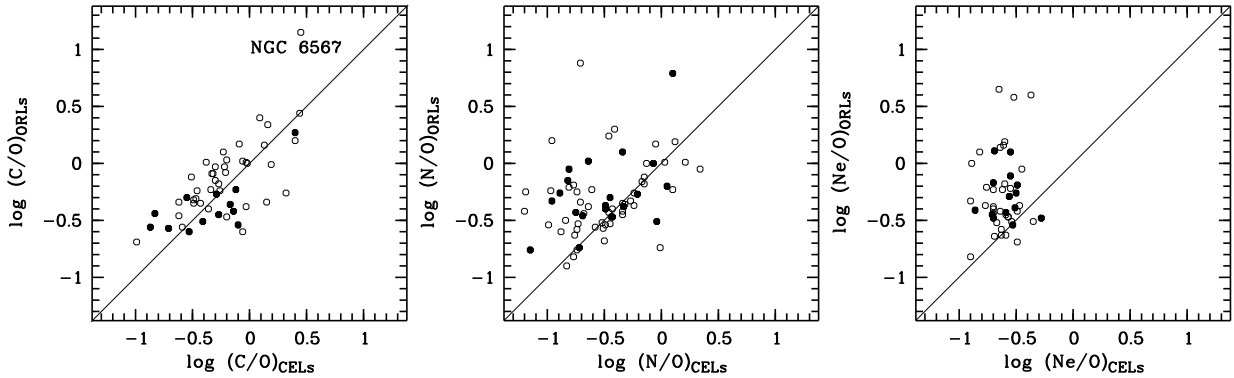, width=5.0cm, bbllx=308,
bblly=47, bburx=528, bbury=768, clip=, angle=270}
\caption{Comparison of heavy element abundance ratios C/O, N/O and Ne/O
derived from ORLs and from CELs, for GBPNe (filled circles) and GDPNe (open 
circles). The solid lines represent $y=x$.}
\label{C2O_N2O_Ne2O}
\end{figure*}

It has been shown that apart from oxygen, the large discrepancies between ionic
abundances relative to hydrogen derived from ORLs and from CELs exist for the
other three abundant second-row elements, carbon, nitrogen and neon
(\citealt{liu1995,luo2001}). In addition, for a given PN, values of the adf
deduced for the four most abundant second-row heavy elements are comparable
(\citealt{liu2000}). In other words, both CEL and ORL analyses yield similar
{\em relative}\ abundance ratios of heavy elements, such as C/O, N/O and Ne/O.
Fig.~\ref{DF_ADF_NONeC} shows the distributions of $\log$\,adf(X), where X =
C$^{2+}$/H$^+$, N$^{2+}$/H$^+$, O$^{2+}$/H$^+$ and Ne$^{2+}$/H$^+$, for all
Galactic PNe for which a measurement of the quantity is available. The
distribution of adf(O$^{2+}$) peaks strongly near 0.3\,dex, with a 
FWHM of 0.26\,dex and an extended tail towards higher values.
For C$^{2+}$/H$^+$ and N$^{2+}$/H$^+$, the distributions peak around 0.5\,dex
with a FWHM of about 0.7 and 1.0, respectively. For Ne$^{2+}$/H$^+$, it peaks
around 0.8\,dex with a FWHM of 1.2\,dex.  In Fig.~\ref{adfNNeCvsO}, we plot the
adf deduced for O$^{2+}$/H$^+$ against those of C$^{2+}$/H$^+$, N$^{2+}$/H$^+$
and Ne$^{2+}$/H$^+$. The Figure shows that adf(C$^{2+}$) follows closely
adf(O$^{2+}$), whereas adf(N$^{2+}$) and adf(Ne$^{2+}$) are systematically
higher than adf(O$^{2+}$). Heavy element abundance ratios, C/O, N/O and Ne/O,
derived from CELs and from ORLs, are compared in Fig.~\ref{C2O_N2O_Ne2O}.  In
line with the patterns revealed in Fig.~\ref{adfNNeCvsO}, both CELs and ORLs
yield compatible C/O abundance ratio, whereas for N/O and Ne/O, the ratios
deduced from ORLs are systematically higher, a direct consequence of higher
values of adf found for N$^{2+}$/H$^+$ and Ne$^{2+}$/H$^+$. The average C/O
ratios found from ORLs are $-0.11\pm 0.05$ for GDPNe (39 objects) and $-0.38\pm
0.06$ for GBPNe (13 objects). The corresponding results from CELs are $-0.19\pm
0.05$ and $-0.35\pm 0.10$ for GDPNe and GBPNe, respectively.  In the scenario
of two-abundance scenario proposed by \cite{liu2000}, ORLs and CELs trace
respectively the cold, H-deficient inclusions and the hot, diffuse medium of
normal composition. The fact that both ORLs and CELs yield compatible C/O
ratios suggests that the postulated H-deficient inclusions have the same C/O
ratio as the diffuse medium and therefore unlikely to be nucleo-processed
material as proposed by \cite{iben1983} for H-deficient knots observed in
Abell\,30 and Abell\,78. Deep spectroscopy and ORL abundance analysis of
H-deficient knots in Abell\,30 (\citealt{wesson2003}) also show that the knots
are O-rich rather than C-rich, in contradiction to predictions of the
born-again scenario proposed by \cite{iben1983}.

Fig.~\ref{C2O_N2O_Ne2O} also shows the C/O ratio varies over a factor of ten,
from $-0.5$ to $+0.5$\,dex, and that C/O ratios of GBPNe are systematically
lower than those of GDPNe. This important result will be discussed further in
the next Section.

For the four abundant second-row elements, adf(O$^{2+}$) is probably the best
determined. First of all, high quality calculations of effective recombination
coefficients for the whole range of O~{\sc ii} ORLs are available
(\citealt{liu1995}). Those calculations have been extended down to a
\Te\, as low as 1000~K and even to 288~K for some selective transitions
(\citealt{bastin2006}). Secondly, most prominent O~{\sc ii} ORLs fall within
the narrow 4000--5000\,\AA\, optical wavelength range where one also finds the
three [O~{\sc iii}] CELs, the $\lambda\lambda$4959,5007 nebular lines and the
$\lambda$4363 auroral line, upon which CEL analysis relies. All those lines can
thus be measured simultaneously using a single instrument and sampling the same
area of a given nebula. This avoids or minimizes errors caused by effects such
as uncertainties in flux calibration and reddening corrections. For carbon,
good atomic data also exist for C~{\sc ii} ORLs (\citealt{davey2000}) and the
lines are generally well measured, in particular the relatively strong 3d--4f
$\lambda$4267 transition, yielding fairly accurate C$^{2+}$/H$^+$ ratios. On
the other hand, for CEL analysis, the derivation of carbon ionic abundance
ratios has to rely on UV CELs, most importantly, the C~{\sc iii}]
$\lambda\lambda$1907,1909 for C$^{2+}$/H$^+$ and C~{\sc iv}
$\lambda\lambda$1548,1550 for C$^{3+}$/H$^+$.  Thus except for a number of
extended PNe for which the optical spectra have been obtained by scanning the
whole nebula (e.g. \citealt{liu2000,tsamis2004}), and for those nebulae of
small angular sizes ($\la $ a few arcsec, such as most GBPNe) and nearly
symmetric, the adf(C$^{2+}$) deduced may suffer from errors. These
uncertainties result from the fact that the UV and optical observations may not
be sampling the same portions of the nebula as well as from potential
uncertainties in flux normalization and reddening corrections.

\begin{figure} \centering \epsfig{file=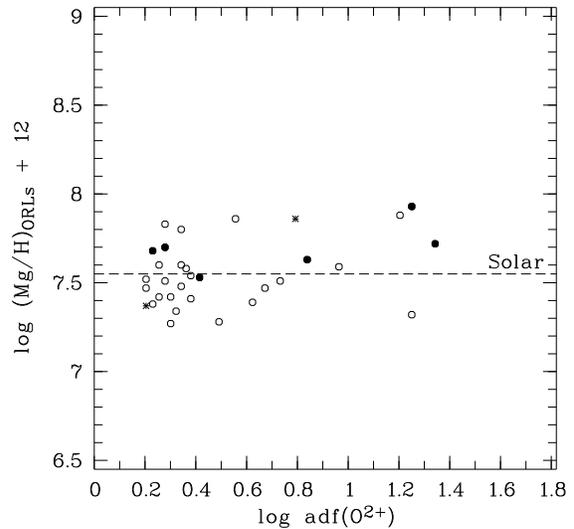, width=7.0cm,
bbllx=52, bblly=35, bburx=554, bbury=565, clip=, angle=-90}
\caption{Magnesium abundance, Mg/H $\approx$ Mg$^{2+}$/H$^+$, with the later
derived from the Mg~{\sc ii} $\lambda$4481 ORL, plotted against 
adf(O$^{2+}$). The solar photospheric magnesium abundance 
(\citealt{lodders2003}) is marked. For an explanation of different symbols,
c.f. Fig.\,\ref{BJvsOIII}.} 
\label{ADFOvsMg} 
\end{figure}

Good atomic data now exist for $l\leq 2$ transitions of N~{\sc ii}
(\citealt{kisielius2002}) and Ne~{\sc ii} (\citealt{kisielius1998}) ORLs, but
not those arising from levels of higher orbital angular momenta, such as 3d--4f
lines.  N~{\sc ii} and Ne~{\sc ii} ORLs are generally much more difficult to
observe than O~{\sc ii} ORLs due to lower abundances of nitrogen and neon
compared to oxygen. Most of the strongest Ne~{\sc ii} 3--3 transitions fall at
shorter wavelengths below 4000\,\AA. Whereas Ne$^{2+}$/H$^+$ can be deduced
from strong optical [Ne~{\sc iii}] $\lambda\lambda$3868,3967 nebular lines,
N$^{2+}$/H$^+$ can only be obtained from either the relatively weak N~{\sc
iii}] $\lambda$1751 intercombination line or from the [N~{\sc iii}] 57$\mu$m
far-infrared lines. All those factors make the current estimates of
adf(N$^{2+}$) and adf(Ne$^{2+}$) subject to large uncertainties. We believe
that a major fraction, if not all, of the larger values of adf found for
N$^{2+}$/H$^+$ and Ne$^{2+}$/H$^+$ compared to adf(O$^{2+}$)
(Fig.\,\ref{adfNNeCvsO}) and consequently, higher N/O and Ne/O ratios yielded
by ORLs compared to those obtained from CELs (Fig.\,\ref{C2O_N2O_Ne2O}), are
caused by those uncertainties and are therefore artificial. Better observations
and atomic data for N~{\sc ii} and Ne~{\sc ii} ORLs are clearly needed. More
accurate determinations of N/O and Ne/O ratios using ORLs are of great interest
to unveil the nature and origins of the postulated H-deficient inclusions
embedded in the diffuse nebula that are responsible for large values of adf
observed in PNe.

Magnesium is the only third-row element for which an ORL from one of its
dominant ionic species has been detected, i.e. the Mg~{\sc ii} 3d--4f
$\lambda$4481 line yielding Mg$^{2+}$/H$^+$. To a good approximation, Mg/H
$\sim$ Mg$^{2+}$/H$^+$ (\citealt{BPL2003}). Mg/H abundances thus deduced are
now available for a total of 32 PNe, including 24 GDPNe and 8 GBPNe (excluding
the peculiar M\,2-24; \citealt{zhang03}). The results are plotted in
Fig.~\ref{ADFOvsMg} against adf(O$^{2+}$). Mg/H abundance is particularly
difficult to obtain from CELs, as it has to rely on weak lines in the UV or in
the infrared. Thus adf(Mg$^{2+}$) cannot be deduced.  However, as previously
noticed by \cite{BPL2003}, in stark contrast to C, N, O and Ne, for a wide
range of adf(O$^{2+}$), Mg/H deduced from the Mg~{\sc ii} $\lambda$4481 ORL
fall in a narrow range compatible with the solar photospheric value. The 24
GDPNe plotted in Fig.\,\ref{ADFOvsMg} yield an average Mg/H ratio of 7.56\,dex
with a standard deviation of 0.22\,dex, almost identical to the solar value of
7.55 (\citealt{lodders2003}). The 8 GBPNe yield a slightly higher average
value, of 7.71\,dex.  We will discuss magnesium abundance in the context of the
chemical evolution of the Galaxy in the next Section.

\subsection{Summary}

In summary, the current detailed plasma diagnostics and abundance analyses
contrasting results obtained from CELs and those deduced from ORLs and continua
for a large sample of GBPNe have yielded the same phenomena and strengthen the
conclusions previously found for GDPNe:

1) Nebular electron temperatures deduced from recombination lines or continua
are systematically lower than those deduced from the classic [O~{\sc iii}]
nebular to auroral forbidden line ratio. In addition, there is strong evidence
that \Te\, derived from He~{\sc i} ORL ratios, $T_{\rm e}$(He~{\sc i}), are
systematically lower than $T_{\rm e}$(BJ), \Te\, deduced from the H~{\sc i}
recombination continuum Balmer discontinuity, and that \Te\, given by
recombination line ratios of heavy element ions, such as from O~{\sc ii} ORL
ratios, are even lower, yielding values that are just a few hundred K in some
cases.

2) Heavy element abundances relative to hydrogen derived from ORLs are {\em
always}\ higher than values derived from CELs. This is true for all abundant
second-row elements, C, N, O and Ne. For a given nebula, values of the adf
found for those individual elements are comparable, leading to similar values
of abundance ratios, C/O, N/O and Ne/O deduced from ORLs and from CELs. By
contrast, no enhancement of ORL abundances relative to CEL values is apparent
for magnesium, the only third-row element that has been studied using ORLs.
There is also evidence that objects exhibiting particularly large values of adf
also have higher helium abundances. 
 
3) The observed discrepancies between \Te\, and heavy element
abundances derived from ORLs and from CELs are correlated. Objects showing
large values of adf tend to have large \Te\, discrepancies, suggesting
that the two phenomena are related, and are likely caused by the same
underlying physical process.

4) The large discrepancies between \Te\, derived from CELs and from ORLs appear
to be mainly caused by abnormally low values yielded by recombination lines or
continua. For the large number of nebulae analyzed, \Te\,       derived from
CELs fall in a narrow range around 10,000~K, consistent with what is predicted
by photoionization models for nebulae of ``normal'' chemical composition, i.e.
that of Sun, a typical G-dwarf in the Galaxy.  Similarly, the large
discrepancies between heavy element abundances deduced from ORLs and from CELs
are largely caused by the abnormally high values obtained from ORLs. Excluding
a few nebulae that are known to belong to the halo population and thus have
lower heavy element abundances, abundances of most PNe deduced from CELs fall
in a narrow range of slightly sub-solar. By contrast, ORLs yield much higher
heavy element abundances, reaching more than a factor of ten higher than the
solar values in some extreme cases. It appears that whatever is causing the
large differences between results obtained from ORLs and from CELs, the main
effect of the process is to enhance emissivities of recombination lines and
continua, but hardly affect those of CELs. The results also suggest that heavy
element abundances deduced from ORLs are unlikely to be real, or representative
of the bulk composition of the whole nebula. This is in stark contradiction to
the prediction of the paradigm of temperature fluctuations, which implicitly
assumes that the higher abundances yielded by ORLs are the correct ones.

5) All the above results point to the existence of a cool component of
metal-rich plasma embedded in the diffuse nebula, as first proposed by
\citet{liu2000}. The cool inclusions are responsible for most of the observed
fluxes of ORLs emitted by heavy element ions, yet contribute little to the
fluxes of CELs. As the nebula expands and ages, the diffuse emission fades and
emission from the embedded cool inclusions becomes more apparent.

\section{Discussion}
\label{discussion}

In this Section, we compare elemental abundances deduced for the current
sample of GBPNe with those derived previously for GDPNe. We investigate
the abundance patterns and discuss the results
in the context of the nucleosynthesis and dredge-up processes in LIMS and their
roles in the chemical evolution of the Galaxy.  Of the GBPNe observed in the
current work (Table~\ref{obj}), we have excluded in our analysis the peculiar
nebulae M\,2-23 and M\,2-39 (c.f.  \S\,\ref{anal:2obj}), but included M\,2-36
and M\,1-42 previously studied by \cite{liu2001}. The sample, WL07, consists
of 25 bulge PNe in total.
 
As described in \S\,\ref{intr}, deep spectroscopy, contrasting plasma
diagnostic and abundance analysis results obtained from the traditional method
based on strong CELs with those based on much fainter ORLs, has been carried
out for three samples of GDPNe (\citealt{tsamis2003,tsamis2004}, 12 PNe, TBLS04
hereafter; \citealt{liuy2004a,liuy2004b}, 12 PNe, LLBL04 hereafter;
\citealt{wesson2005}, 23 PNe, WLB05 hereafter). In addition, case studies for a
number of individual PNe have been published, some of them showing particularly
prominent and rich ORL spectra from heavy element ions.  They include NGC\,7009
(\citealt{liu1995,luo2001}), NGC\,6153 (\citealt{liu2000}), NGC\,6543
(\citealt{wesson2004}), Me\,1-1 (\citealt{shen2004}), NGC\,1501
(\citealt{ercolano2004}), NGC\,7027 (\citealt{zhang05b}) and Hf\,2-2
(\citealt{liu2006}). In the current work, six PNe observed belong to the disk
population (c.f. Table~\ref{obj}).  In all the studies, the nebulae were
observed using low- to intermediate-resolution long-slit CCD spectrophotometry
and then analyzed in a similar manner (plasma diagnostics, atomic data and ICF
scheme). All those disk PNe, including six from the current work, are grouped
to form a homogeneous sample of GDPNe to compare with the current sample of
GBPNe. A few nebulae observed and analyzed using similar techniques were
excluded, either because of their peculiarities, e.g. Abell\,30
(\citealt{wesson2003}), Mz\,3 (\citealt{zhang02}), M\,2-24 (\citealt{zhang03}),
or because the objects actually belong to a different population, e.g.
NGC\,4361 (\citealt{liu1998}) and DdDm\,1 (WLB05); both are believed to be halo
PNe. We note that fluxes for a number of lines published by WLB05 for PN
M\,3-27 seem to be incorrect. We have therefore remeasured the lines and
repeated the analysis.  In total, our final sample of GDPNe consists of 58 PNe
and will be referred to hereafter as TLW (`T' for TBLS04, `L' for LLBL04 and
`W' for WLB05 -- the three samples constituting the majority of the TLW
sample).  

We note that most of the objects observed by TBLS04 are within the solar
circle, whereas the northern PNe in the WLB05 sample are mostly outside the
solar circle, and the LLBL04 sample is a mixture. The mean O/H abundances
deduced from the three samples are 8.70, 8.58 and 8.60, respectively. The
corresponding values after excluding Type-I PNe (and the very low excitation PN
in the WLB05 sample) are 8.74, 8.57 and 8.59, respectively. The small
variations of mean oxygen abundances between the samples thus reflect an
abundance gradient across the Galactic disk.  We also note that the average
metal abundances yielded by the KB94 sample of disk PNe are systematically
higher than those from the TLW sample, varying from about 0.1\,dex for oxygen,
neon and sulfur to about 0.2\,dex for carbon, nitrogen and argon. Part of the
differences may be caused by abundance gradients across the disk. Most PNe in
the KB94 sample are southern hemisphere nebulae and therefore within the solar
circle. For the 38 and 16 PNe within and outside the solar circle in the KB94
sample, the mean oxygen abundances are 8.71 and 8.60, respectively.

\subsection{Comparison between bulge and disk samples}
\label{abundis:b2d}

\begin{figure*}
\centering
\epsfig{file=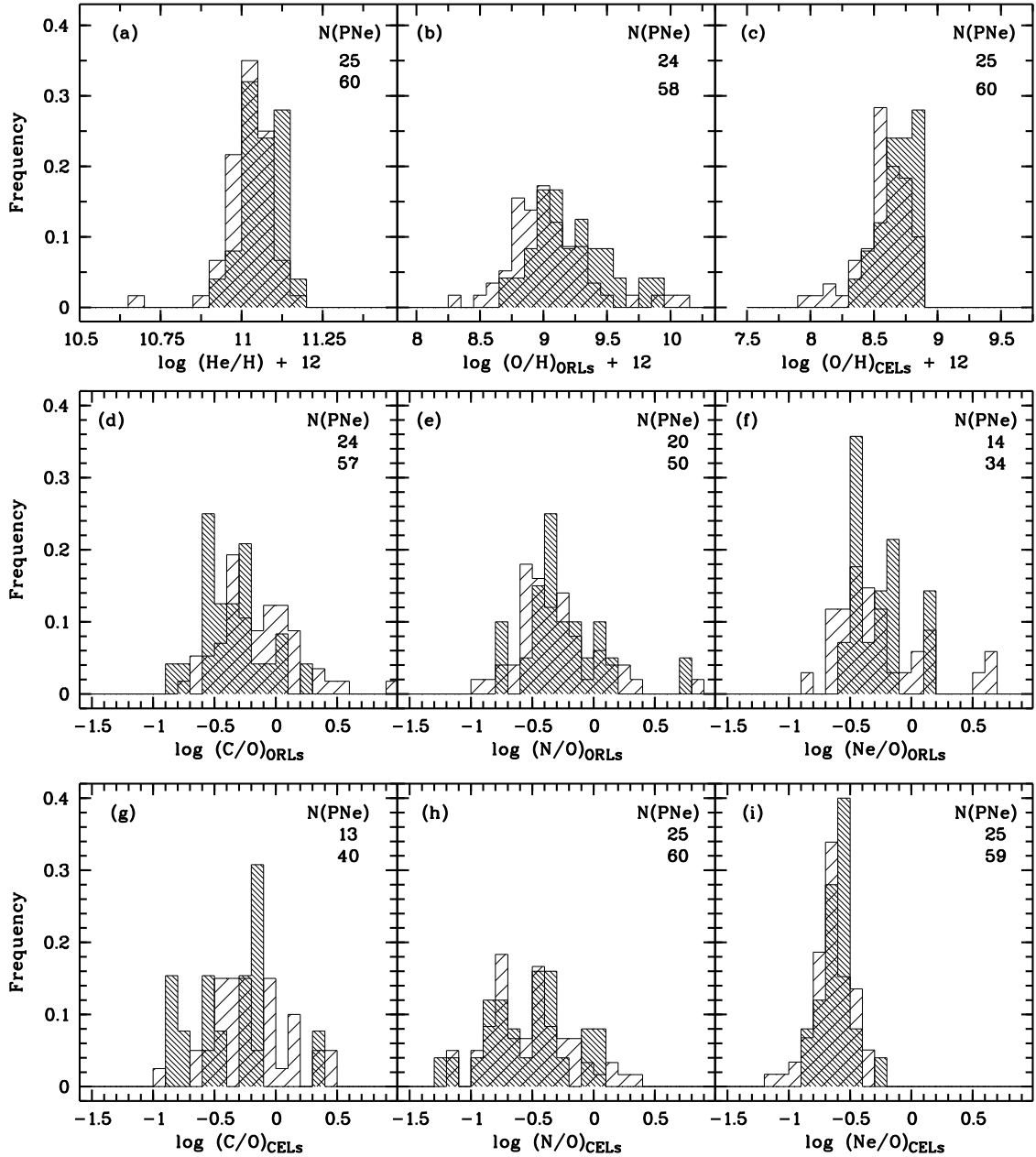, width=15.cm, bbllx=34,
bblly=34, bburx=541, bbury=604, clip=, angle=0}
\caption{Distributions, normalized to unit area, of elemental abundances 
relative to hydrogen, and of heavy element abundance ratios for GDPNe (lightly 
shaded histograms) and GBPNe (heavily shaded histograms). The ordinate gives 
the fractional number of nebulae in each bin. The bin size is 0.05 for panel 
(a) and 0.10 for all the other panels. In each panel, numbers of bulge and disk 
PNe for which measurements are available are marked from top to bottom.}
\label{b2dhist}
\end{figure*}

Table~\ref{totabun} shows that compared to the disk PN sample TLW, the average
elemental abundances of the bulge PN sample WL07, are consistently higher,
except for C/H derived from ORLs and Cl/H deduced from CELs. If we define
$\Delta M({\rm X})$ as the difference between the average abundances of element
X relative to H derived from disk bulge sample WL07 and from the disk sample
TLW, $\Delta M({\rm X}) \equiv  M({\rm X},{\rm GBPNe}) - M({\rm X},{\rm
GDPNe})$, then $\Delta M({\rm X})$ ranges from 0.04\,dex for He/H from ORLs,
0.10\,dex for O/H from both CELs and ORLs, $\sim 0.15$\,dex for Ne/H and Ar/H
from CELs, 0.15\,dex for Mg/H from ORLs, 0.17--0.20 for N/H from CELs and
from ORLs, and 0.21\,dex for S/H from CELs.

Fig.~\ref{b2dhist} compares distributions, normalized to unit area, of
elemental abundances relative to hydrogen and of heavy element abundance ratios
for the WL07 bulge sample (heavily shaded histograms) and the TLW disk sample
(lightly shaded histograms). Panels (a) -- (c) of Fig.~\ref{b2dhist} show that
abundances of He and O of bulge PNe peak at higher values compared to disk PNe.
Panels (d) and (g) show that, as already pointed out in \S\ref{orl2cel:adf}
(c.f. Fig.\,\ref{C2O_N2O_Ne2O}), bulge PNe have lower C/O ratios than disk PNe,
by approximately 0.2~dex. On the other hand, bulge and disk PNe seem to have
similar distributions of N/O and Ne/O abundance ratios (as discussed in
\S\ref{orl2cel:adf}, nitrogen and neon abundances deduced from ORLs and
consequently N/O and Ne/O abundance ratios based on ORLs are less reliable
compared to those of carbon and oxygen).

It is generally accepted that abundances of oxygen, neon, sulfur and argon are
hardly affected by nucleosynthesis and dredge-up processes that occur in
progenitor stars of PNe, thus their abundances should reflect values of
the ISM at the time those LIMS were formed. For these four elements, their
average abundances deduced from CELs for the bulge sample WL07 are consistently
higher by $\sim 0.1$--0.2\,dex than for the disk sample TLW. Corroborating evidence
is provided by the average O/H abundances derived from ORLs for which good
atomic data exist and line measurements are reasonably reliable. The results
show that the bulge is slightly overabundant with respect to the disk, by approximately
0.1\,dex.  Although the inferred abundance differences between the bulge and
disk samples are small, the consistency of results for all $\alpha$ elements
measured as well as from different tracers (CELs versus ORLs) strongly suggests
that differences are probably meaningful.

Supporting evidence that the Galactic bulge is more metal-rich than the disk
has also been found from other independent abundance studies of bulge PNe.
Spectrophotometric abundance analyses of 30 GBPNe by CMKAS00 show that while
the average abundances of O, S and Ar of GBPNe are comparable to those of
GDPNe, there is some evidence that higher elemental abundances are slightly
more frequently found amongst the bulge PNe. More recent work on 44 GBPNe by
\citet{gorny2004} also came to the same conclusion. They compared the average
oxygen abundance of their sample b (bulge PNe) with sample d (disk PNe within
the solar circle compiled from the literature) and found that the former is
about 0.13~dex (or 0.2~dex) more metal abundant than the latter. Their findings
are roughly in agreement with our results. Note that there are only a handful
of common objects between the three samples of bulge PNe and independent
analyses all indicate that GBPNe are slightly more metal-rich than GDPNe.

The lower C/O ratios observed in GBPNe compared to GDPNe may suggest that GBPNe
evolve from more massive progenitor stars. As a consequence, a larger
fraction of carbon has been converted to nitrogen, although it is difficult to
rule out selection effects as PNe of less massive cores are also less bright
and therefore harder to observe, particularly those in the distant, heavily
extincted region of the Galactic bulge. Alternatively, it may indicate that the
3rd dredge-up process is less efficient for metal-rich GBPNe than for GDPNe.
In \S~\ref{abundis:cn}, we will discuss in detail the lower C/O ratios
measured for GBPNe and compare the results with some of the recent theoretical
predictions.

The distribution of He/H, O/H and Ne/O abundance ratios plotted in
Fig.~\ref{b2dhist} can be well fitted using a single Gaussian profile, while
those of C/O and N/O appear to be more irregular and double-peaked. While
observational errors may play a role, the results may indicate that there are
two populations in each of the two samples.  A scrutiny of N/H abundances
obtained by EBW04 for a sample of GBPNe (see the solid histograms in their
Fig.~4) also reveals a double-peak distribution, peaking at 8.1 and 8.7\,dex,
respectively, similar to what we find for the current sample of GBPNe.
Similarly, N/H abundances of the ECM04 sample also exhibit two local maxima, at
7.7 and 8.4\,dex, respectively. It is interesting to note for each of the three
samples of GBPNe, the two local maxima observed in the N/H abundance
distribution are consistent with the average N/H abundance ratios for non-Type
I and Type-I PNe in the corresponding sample, respectively.  The fraction of
Type-I PNe in the WL07 bulge PN sample is 19\,per~cent, very similar to the
fraction of 21\,per~cent for the TLW disk sample.  It seems unlikely that the
slightly higher metallicity observed for GBPNe compared to GDPNe is caused by a
mixture of different populations in the two samples.

In a recent theoretical study, \citet{nakasato2003} have modelled the Galaxy
formation using three-dimensional, hydrodynamic N-body simulations, paying
particular attention to the formation of the bulge component. Their simulations
show that the bulge consists of two chemically different components -- one
resulted from a fast merging of subgalactic clumps in the proto-Galaxy while
the other formed gradually in the inner disk. Their results predict a
double-peak [Fe/H] abundance distribution -- the iron content in the non-merger
component is higher than the merger component as a result of pollution by
Type-Ia supernovae. It is unclear however how the double-peak [Fe/H]
distribution predicted for the bulge is related to the double-peak distribution
of N/H abundances observed in GBPNe.

To summarize, a comparison of elemental abundances of disk and bulge PNe shows
that GBPNe appear to be more metal-rich than GDPNe, by approximately 0.1 --
0.2\,dex.  In addition, the mean C/O ratio of GBPNe are lower than that of
GDPNe, by 0.16 and 0.27\,dex as deduced from CELs and ORLs, respectively. 

\subsection{Carbon and nitrogen}
\label{abundis:cn}

At the end of the main-sequence evolution when the central hydrogen is
exhausted, the star evolves to the red giant branch (RGB). A convection zone 
develops and dredge-ups partially CN-cycle processed material to the surface.
This first dredge-up process enhances the surface nitrogen abundance and reduces
the surface abundance of carbon as well as the $^{12}$C/$^{13}$C isotope ratio.
Following the exhaustion of central helium and the formation of an electron
degenerate CO core, the second dredge-up starts in early Asymptotic Giant
Branch (AGB) stars but only in those more massive ones with initial masses in
the (uncertain) range from 2.3 to 8\,$M_{\sun}$. The process increases the
surface abundances of both helium and nitrogen and reduces slightly those of
carbon and oxygen. The third dredge-up takes place in AGB stars during
the thermal pulses (helium shell flashes). Helium-burning products, mostly
$^{12}$C and {\it s}-process elements, are brought to the surface. On the way
to the surface, the material is subject to limited CN-cycle burning, allowing
for some conversion from C to N and decreasing the $^{12}$C/$^{13}$C ratio. In
all three episodes of the dredge-up process, the surface abundances of He
and N are increased, while that of $^{12}$C abundance and the $^{12}$C/$^{13}$C
isotope ratio first decrease in the first (and second, when it occurs) dredge-up
episode and then increase in the third. The amount of abundance changes depends
mainly on the initial mass but also on metallicity of the progenitor star.  The
abundances of C and N relative to other light elements thus provide tools to
probe the nucleosynthesis and mixing processes as well as properties of PN
progenitor stars, such as their initial masses and metallicities.

Since the pioneering work by \citet{renzini1981}, the roles of LIMS in
enhancing the chemical composition of the ISM have been studied extensively,
both observationally and theoretically. In recent work, Marigo (2001, Ma01
hereafter) presented quantitative predictions of the surface abundance patterns
of He, C, N and O for different values of initial stellar mass and metallicity
and mixing-length parameter $\alpha$.

\begin{figure}
\centering
\epsfig{file=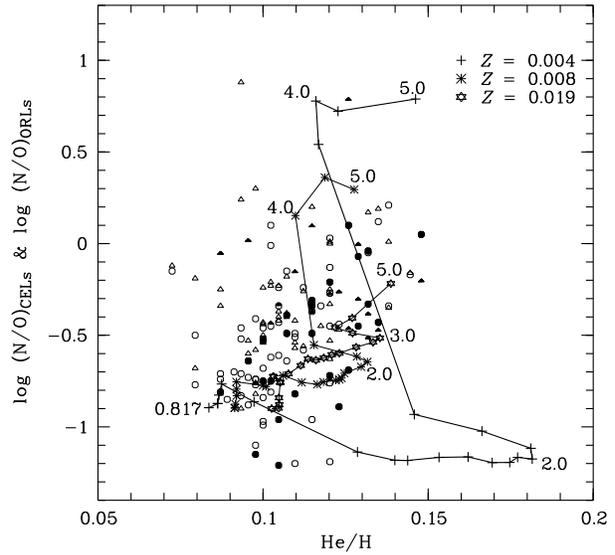, width=8.0cm, bbllx=36,
bblly=295, bburx=554, bbury=772, clip=, angle=0}
\caption{N/O abundance ratios derived from CELs (circles) and from ORLs
(triangles) for the WL07 sample of GBPNe (filled symbols) and the TLW sample
of GDPNe (open symbols), plotted against helium abundance He/H. Overplotted
are loci of initial mass predicted by Ma01 for initial metallicities $Z = 
0.004$, 0.008, 0.019 and mixing length parameter $\alpha = 1.68$. The tick 
marks on the tracks indicate initial stellar mass, ranging from $\sim 0.8$ to 
5 $M_\odot$. Only a few values are labelled.}
\label{NOvsHe_Ma01}
\end{figure}

\begin{figure}
\centering
\epsfig{file=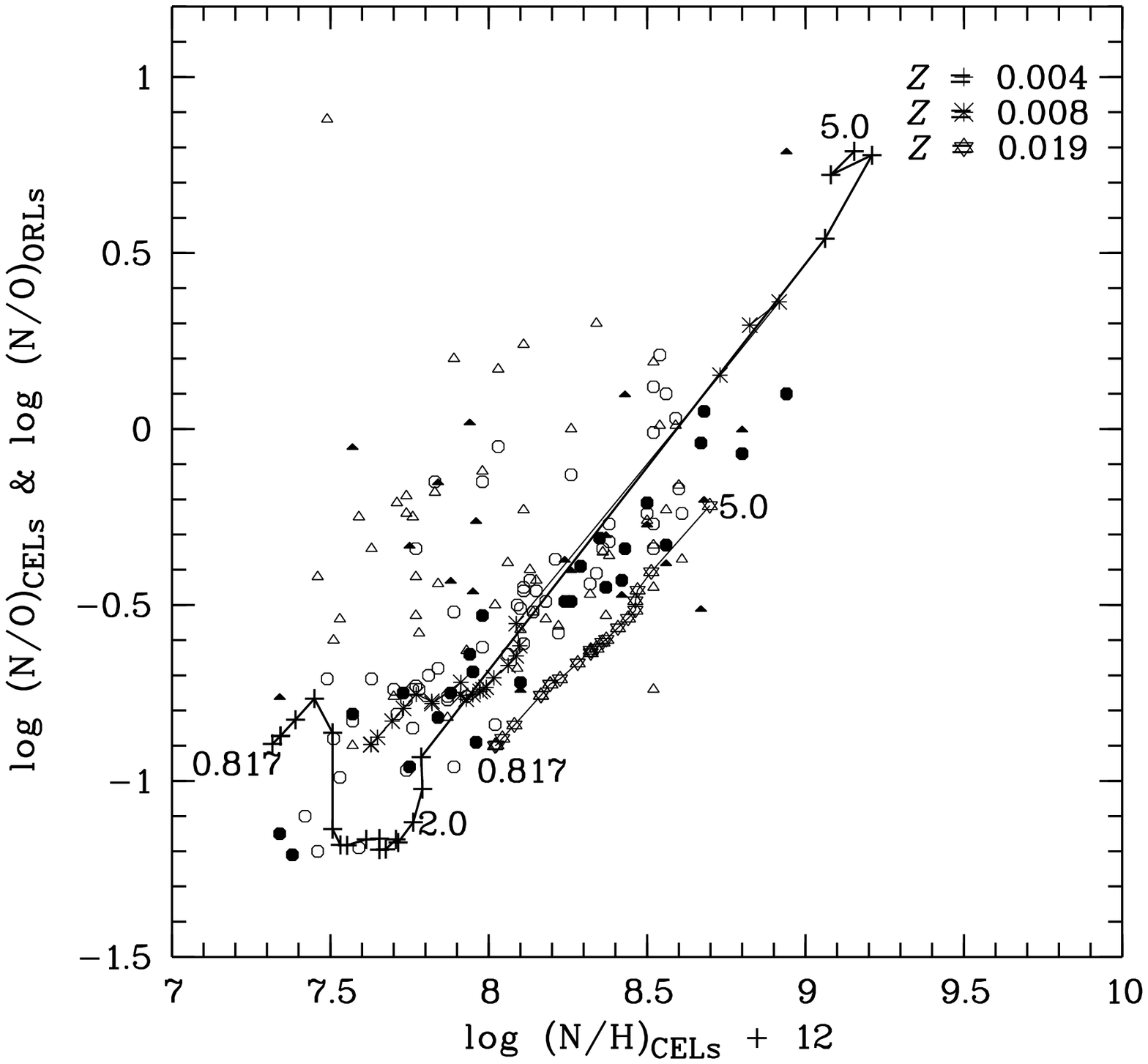, width=8.0cm, bbllx=36,
bblly=294, bburx=551, bbury=772, clip=, angle=0}
\caption{N/O abundance ratios plotted against N/H abundance. For an explanation
of the different symbols and lines, c.f. caption to Fig.\,\ref{NOvsHe_Ma01}.}
\label{NOvsN_Ma01}
\end{figure}

\begin{figure}
\centering
\epsfig{file=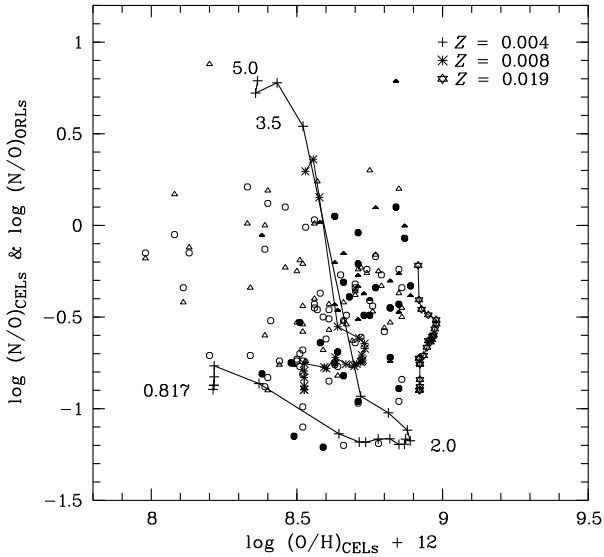, width=8.0cm, bbllx=36,
bblly=294, bburx=554, bbury=772, clip=, angle=0}
\caption{N/O abundance ratios plotted against O/H abundance derived from CELs. 
For an explanation of the different symbols and lines, c.f. caption to 
Fig.\,\ref{NOvsHe_Ma01}.}
\label{NOvsO_Ma01}
\end{figure}

In Figs.\,\ref{NOvsHe_Ma01}, \ref{NOvsN_Ma01} and \ref{NOvsO_Ma01}, we plot N/O
ratios measured for the WL07 sample of GBPNe and for the TLW sample of GDPNe,
deduced from CELs and from ORLs against He/H derived from ORL analysis and
against N/H and O/H obtained from CEL analysis. The data are compared with
theoretical tracks calculated by Ma01 as a function of initial mass for initial
metallicities $Z = 0.04$, 0.008 and 0.019, assuming 
$\alpha = 1.68$.

Fig.~\ref{NOvsHe_Ma01} shows that the observed distribution of N/O versus He/H
does not fit well with what is expected, with many data points falling to the
upper left of the theoretical tracks. Either the theoretical calculation
underestimates the N/O ratio or overestimates the He/H enhancement, assuming
that most PNe are descendants of LIMS, an assumption which is most likely true.
The effect is most obvious for GDPNe (open symbols).  Similarly,
Fig.~\ref{NOvsN_Ma01} shows that while most GBPNe (filled symbols) fall within
the band delineated by the tracks of various initial metallicities, more than
half the open symbols fall above and to the left of the track of even the
lowest metallicity $Z = 0.004$. The theoretical loci of N/O versus O/H become
almost orthogonal to the abscissa (Fig.\,\ref{NOvsO_Ma01}) for initial masses
higher than $\sim 2~M_\odot$. Again in this diagram, we find that bulge PNe are
mostly found between the two loci for initial $Z = 0.008$ and 0.016, while a
significant fraction of disk PNe fall to the left of the loci for $Z = 0.004$.
Figs.\,\ref{NOvsHe_Ma01}, \ref{NOvsN_Ma01} and \ref{NOvsO_Ma01} thus all seem
to indicate that bulge PNe originated from environments of higher metallicities
($Z \sim 0.013$ from a rough estimate) than disk PNe. In addition,
Fig.\,\ref{NOvsN_Ma01} shows that to reach a given value of the N/O ratio, a
much higher initial mass is required for a star of higher initial metallicity
than for one of lower initial metallicity.  Fig.\,\ref{NOvsN_Ma01} shows that
almost all disk PNe appear to evolve from stars of initial mass lower than
$\sim 4~M_\odot$ (assuming they have an initial metallicity close to or lower
than 0.008), while a significant fraction of bulge PNe have initial masses
higher than this. We thus conclude that PNe of massive progenitors are more
frequently found in the bulge than in the disk, at least for the limited sample
of bulge PNe analyzed in the current work.
 
\begin{figure}
\centering
\epsfig{file=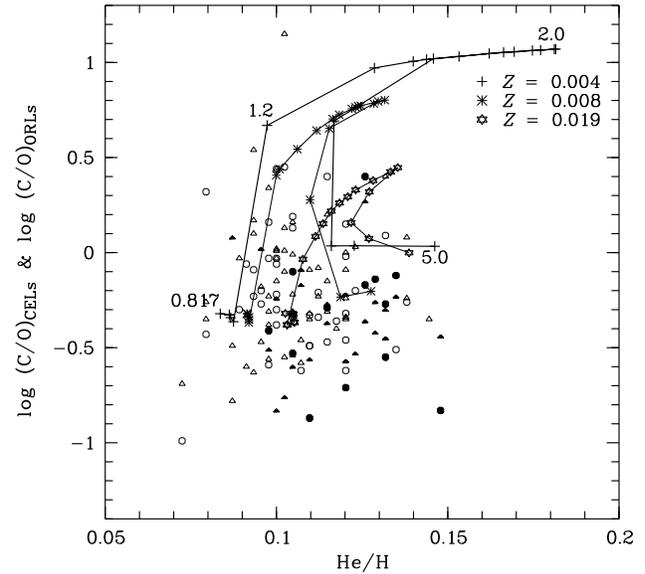, width=8.3cm, bbllx=36,
bblly=295, bburx=554, bbury=772, clip=, angle=0}
\caption{C/O abundance ratios derived from CELs and from ORLs plotted
against helium abundances for bulge and disk PNe. Overplotted are 
loci calculated by Ma01. See also the caption to Fig.\,\ref{NOvsHe_Ma01} for 
more description.}
\label{COvsHe_Ma01}
\end{figure}

\begin{figure}
\centering
\epsfig{file=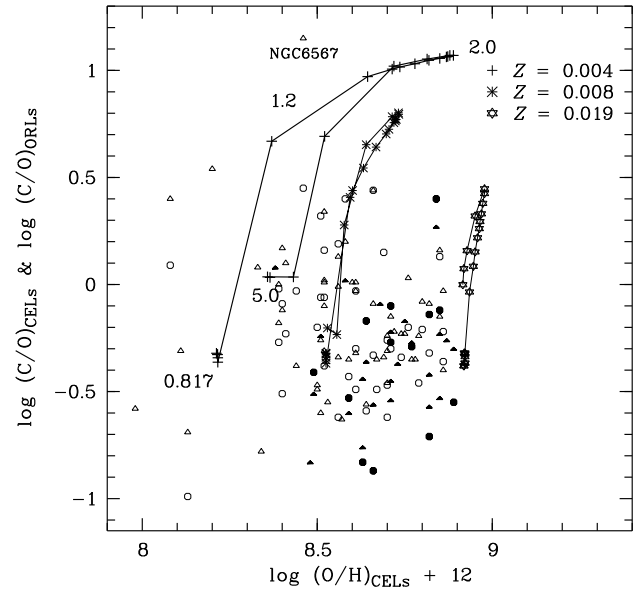, width=8.3cm, bbllx=36,
bblly=294, bburx=543, bbury=772, clip=, angle=0}
\caption{C/O ratios derived from CELs and from ORLs plotted against oxygen 
abundances derived from CELs for GBPNe and GDPNe. See also the caption to 
Fig.\,\ref{NOvsHe_Ma01} for more description.}
\label{COvsO_Ma01}
\end{figure}

Loci of the C/O ratio versus He/H and O/H abundances calculated by Ma01 are plotted
in Figs.~\ref{COvsHe_Ma01} and \ref{COvsO_Ma01}, respectively, and compared to
observations. For low-mass progenitor stars, in the absence of
Hot-Bottom-Burning (HBB) which destroys carbon, the final surface abundance of
carbon as well as the C/O ratio increase with increasing initial mass, as a result
of the increasing number of dredge-up episodes experienced during the
Thermal-Pulse-AGB phase.  The quantities reach maximum values around
2--3$M_\odot$. Beyond that, they  start to decline as initial mass increases as
a consequence of less dredge-up events. For even higher initial masses, HBB
kicks in and eventually prevails.

As noted earlier (\S\ref{orl2cel:adf}, Fig.\,\ref{C2O_N2O_Ne2O}), GBPNe appear
to have lower C/O ratios than GDPNe.  Fig.~\ref{COvsHe_Ma01} shows that the
majority of GBPNe have C/O ratios lower than the predicted values of Ma01 for
initial values ranging from $\sim 0.8$--5\,$M_\odot$. Even for GDPNe, over half
of them have C/O ratios lower than the lowest predicted values. Similar trends
are also found in Fig.~\ref{COvsO_Ma01}. Thus, it sees that the theoretical
calculations of Ma01 have overestimated the yield of carbon, by as much as
about 0.4\,dex.  The predicted C/O ratios can be lower for initial masses
higher than 5\,$M_\odot$. However, it seems highly unlikely that over half the
PNe in our sample descend from such massive stars.

Fig.~\ref{COvsO_Ma01} can be used to obtain better estimates of the average
initial metallicity for our disk and bulge PN samples than possible with
Fig.~\ref{NOvsO_Ma01}, because the C/O ratio is more sensitive to the initial
metallicity than N/O is. Fig.~\ref{COvsO_Ma01} shows that essentially all PNe
in our sample of GBPNe fall between the two loci of $Z = 0.008$ and 0.019,
while most PNe in the current sample of GDPNe fall close to the locus of $Z =
0.008$.

In summary, both from the actually observed distributions of oxygen abundances
(Fig.\,\ref{b2dhist}) for bulge and disk PNe as well as from the measured N/O
and C/O abundance ratios (Figs.\,~\ref{NOvsN_Ma01}, \ref{NOvsO_Ma01} and
\ref{COvsO_Ma01}), there is strong evidence indicating that GBPNe were formed
in environments of higher metallicities ($Z\sim 0.013$) compared to disk PNe.
The latter seem to have initial metallicities $\la 0.008$. We also find that
the current sample of GBPNe may originate from more massive stars than the disk
PNe.  Finally, we point out that Ma01 may have overestimated carbon yield by
over a factor of two.

\subsection{Oxygen, neon and argon}

\begin{figure}
\centering
\epsfig{file=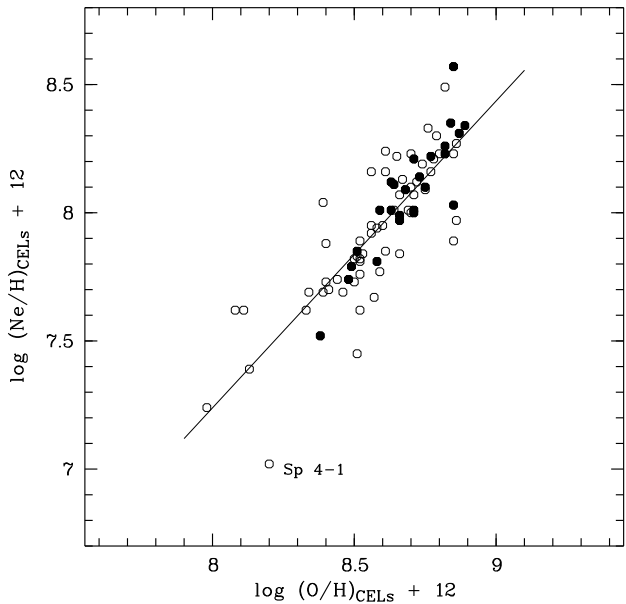, width=8.0cm, bbllx=52,
bblly=48, bburx=544, bbury=557, clip=, angle=-90}
\caption{Neon abundances plotted against those of oxygen for GBPNe (filled 
circles) and GDPNe (open circles). For both elements, abundances deduced from
CELs are adopted. The straight line represents a linear regression for both
bulge and disk PNe.} 
\label{NevsO}
\end{figure}

There is a tight linear relation between neon and oxygen abundances as
illustrated in Fig.~\ref{NevsO}. The only exception is Sp~4-1 (WLB05). For this
peculiar object, with the exception of oxygen, its abundances of all other
heavy elements are lower than the average of disk PNe by 0.5 to 0.8~dex.
Similar correlations between the neon and oxygen abundances have previously
been observed (e.g. \citealt{henry1989}, \citealt{henry2004}) and interpreted
as a direct consequence of the fact that both neon and oxygen originate from
primary nucleosynthesis in massive stars ($\ga 10$\,$M_\odot$), and is
therefore nearly independent of the evolution of LIMS, the progenitor stars of
PNe.  A linear regression of the data points for disk and bulge PNe yields
${\rm Ne/H} = (-2.42 \pm 0.70) + (1.21 \pm 0.08){\rm O/H}$, with a correlation
coefficient of 0.85. The result is compatible to what \citet{henry1989}
obtained for a sample of Galactic, Magellanic and M\,31 PNe: ${\rm Ne/H} =
(-2.14 \pm 0.33) + (1.16 \pm 0.04){\rm O/H}$. Similar results are also obtained
by \citet{henry2004} for a sample of  85 Galactic PNe and by
\citet{stanghellini2006} for a sample of 79 northern hemispheric PNe.  We note
that from both our linear regression and that of \citet{henry1989} for PNe
spanning a wide range of metallicities, there is marginal evidence that the
slope of linear relation is larger than unity. To further explore this
important issue, we have calculated the average Ne/O abundances for SMC and LMC
PNe, using the data published by \citet{LD06}. For the 30 SMC and 120 LMC PNe
in their sample, we obtain an average Ne/O ratio of 0.154 (or $-0.81$\,dex) and
0.183 ($-0.73$\,dex), respectively, significantly lower than the average values
of 0.24 ($-0.61$\,dex) and 0.25 ($-0.57$\,dex) for the Galactic disk and bulge
samples. The average oxygen abundances for the SMC and LMC samples are 8.09 and
8.38, respectively. The results thus strongly suggest that there is a time
delay of neon enrichment relative to oxygen and neon abundance is lower at low
metallicities -- a result that may bear some important implications for our
understanding of the enrichment of the ISM by massive stars and/or the possible
evolution of the initial mass function (IMF) with metallicity. 

Similar but looser linear correlations have also been found between abundances
of oxygen and those of the other two $\alpha$-elements, sulfur and argon.
Linear regression yields ${\rm S/H} = (-2.64\pm1.32) + (1.10\pm0.15){\rm O/H}$
and ${\rm Ar/H} = (-2.34\pm1.21) + (0.99\pm0.14){\rm O/H}$, with a correlation
coefficient of about 0.6 in both cases.

\citet{stanghellini2006} found an average Ne/O ratio of $0.27\pm0.10$ for their
sample of Galactic PNe. They argued that the ratio is independent of the
nebular morphology and thus on stellar population. For our samples of bulge and
disk PNe, we obtain an average Ne/O abundance ratio of $0.25\pm0.08$ and
$0.24\pm0.08$, respectively, in good agreement with the result of
\citet{stanghellini2006}. It is interesting that the Ne/O abundance ratios
obtained for the Orion Nebula by \citet{simpson1998} and for two H~{\sc ii}
regions, G333.6-0.2 and W43, by \citet{simpson2004} are also 0.25.  The Ne/O
abundance ratios derived from PN and H~{\sc ii} region observations are
therefore consistently higher than the solar value of 0.15
(\citealt{lodders2003}; \citealt{asplund2005}).  If we accept the average Ne/O
ratio of 0.25 deduced for Galactic PNe, and combine the value with the solar
oxygen abundance of 8.69 from \citet{lodders2003}, we obtain a neon abundance
of 8.09, which is significantly higher than the currently recommended value of
7.87 for the Sun but is remarkably consistent with the value 8.11 deduced from
solar coronal line measurements (\citealt{feldman2003}). 

The relatively high neon abundance of PNe has also been observed and discussed
previously. \citet{pottasch1999} suggested that neon is created in NGC~6302, a
bipolar PN ionized by a hidden ionizing source of probably the highest
temperature known for a star ($\ga 200,000$~K).  Using multi-waveband data,
they deduced ionic abundances from Ne$^+$ up to Ne$^{5+}$ and obtained a total
Ne elemental abundance of 8.34, which is 0.5 and 0.3\,dex higher than the solar
value and the average abundance of B-stars, respectively.  Theoretically,
\citet{KL2003} suggest that some PNe may have experienced moderate neon
enrichment. The mechanism occurs for stars of initial metallicities less than
0.008 and initial masses in the range of 2--4\,$M_\odot$, where $^{22}$Ne is
produced in sufficient quantities to affect the total neon abundance by more
than 20 pet cent, about 0.08~dex. However, if the mechanism only operates in
metal-poor stars in a narrow range of initial mass, then it is unlikely it will
affect the average neon abundance of PNe by any significant amount.
\citet{pottasch2006} argued that the higher Ne/O ratio in PNe compared to the
Sun is more likely caused by the destruction of oxygen in progenitor stars of
PNe. Model calculations by \citet{karakas2003} show that some oxygen is indeed
destroyed in stars more massive than 5\,$M_\odot$ of initial metallicity $Z =
0.004$ and 0.008. The process however also produce copious nitrogen, which is
inconsistent with observations. On the other hand, her calculations for stars
of initial metallicity $Z = 0.02$ (i.e. solar composition) and of initial
masses of 6 and 6.5~$M_\odot$, a moderate N/O ratio of about 1.2 is found,
which is more compatible with observations. To explain the above the average
neon abundance of PNe requires a significant fraction of PNe evolve from
progenitor stars in such a restrictive range of initial mass and metallicity,
an assumption that is obviously unrealistic.  The average oxygen abundance
deduced from CELs for the TLW disk sample of PNe is 8.60, whereas that of the
WL07 bulge sample is 8.70, close to the solar value of 8.69 (c.f.
Table~\ref{totabun}). Thus depletion of oxygen in PNe is likely to be small if
any, and it is unlikely that the high Ne/O ratios observed in PNe is caused by
the destruction of oxygen in their progenitor stars. From the above discussion,
we conclude that abundances deduced from PNe provide strong, albeit indirect,
evidence that the solar neon abundance recommended by \citet{asplund2005} may
have been underestimated.

Helioseismological studies have indeed pointed to a higher solar neon
abundance.  As a consequence of the application of a time-dependent 3D
hydrodynamical model of the solar atmosphere instead of the earlier 1D
hydrodynamical model, the metal content in the solar convection zone has
decreased by almost a factor of two. Amongst them, abundances of key elements
such as, C, N and O have been lowered by approximately 0.2--0.3~dex from the
earlier widely adopted values (see \citealt{asplund2005} for a review). Those
large downward revisions however have posed serious challenges, most notably
for helioseismology. Solar models which adopt the new solar abundances as input
can no longer satisfactorily match helioseismological observations. On the
other hand, \citet{bahcall2005} show that raising the solar neon abundance to
8.29 (or ${\rm Ne/O}=0.42$) can in principle bring the solar model back into
agreement with the helioseismological measurements. The suggestion is supported
by Chandra measurements of Ne/O ratios for a sample of 21 nearby solar-like
stars, yielding an error-weighted mean Ne/O ratio of 0.41 (\citealt{drake05}).
However, a reanalysis of the solar corona data by \cite{schmelz05} recovered
the old lower solar Ne/O ratio of 0.15 (\citealt{GS98}). In a more recent
study, \citet{LS06} derived a Ne/O ratio of $\sim 0.28$, an intermediate value
for the inactive, solar-like star $\alpha$\,Cen (primarily $\alpha$\,Cen~B,
which is the dominant component in X-rays). It seems that the problem of the
solar Ne/O abundance ratio and consequently the absolute abundance of Ne
remains to be sorted out. 

\subsection{Magnesium}     
\label{abundis:mg}

Table~\ref{totabun} shows that GBPNe have an average CEL abundance of the
$\alpha$-element oxygen that is almost identical to the solar value. While for
the third-row $\alpha$-element magnesium, its mean abundance deduced from ORLs,
7.71$\pm0.06$, is about 0.15~dex higher than the solar value 7.55. The latter
is almost identical to the average magnesium abundance 7.56, also derived from
ORLs, for the TLW disk sample of PNe. As previously argued by \citet{BPL2003},
depletion of magnesium onto dust grains in PNe is likely to be small, probably
less than 30 per~cent.

The result confirms previous studies which find that that magnesium is enhanced
in the Galactic bulge \citep{MR1994}. Large enhancements of $\alpha$-elements
neon and magnesium are also in the peculiar bulge PN M\,2-24 \citep{zhang02}.
In a recent abundance study of 27 RGB stars towards the Galactic bulge in
Baade's Window based on Keck/HIRES echelle observations, \citet{fulbright2005,
fulbright2006} found enhanced [$\alpha$/Fe] ratios for all bulge giants,
including those at super-solar metallicities.  The enhancement of magnesium in
GBPNe and the large [$\alpha$/Fe] ratios in bulge giants, combined with the old
age for the bulge, point to Type\,II supernovae as the primary enrichment
process in the bulge.

\section{Conclusions} 
\label{concl}

In the current work, we present careful abundance analyses for a sample of 25
GBPNe plus 6 GDPNe, using both CELs and ORLs. The results are compared to those
of disk PNe published in the recent literature and to the theoretical
predictions of Ma01. We find,    

\begin{enumerate}

\item CEL abundances deduced in the current work are consistent, within
0.2\,dex, with those published in the literature for samples of GBPNe,
including W88~\citep{webster1988}, RPDM97~\citep{ratag1997},
SRM98~\citep{stasinska1998}, CMKAS00~\citep{cuisinier2000} ,
ECM04~\citep{escudero04} and EBW04~\citep{exter2004}. An exception is
EC01~\citep{escudero01}, which yields average abundances that are at odds with
all other samples of GBPNe.

\item GBPNe are found to be slightly more metal rich on average than GDPNe, 
by approximately 0.1 to 0.2~dex. GBPNe also have C/O ratios systematically lower
than GDPNe, by approximately 0.2\,dex.

\item Bulge PNe, similar to disk PNe, also exhibit ubiquitous discrepancies
between \Te\, and heavy element abundances derived from CELs
and those deduced from CELs. The distribution of the abundance discrepancy
factor, adf, for the bulge PNe are similar to that of disk PNe. Analyses also
confirm many of the trends previously found for disk PNe, such as the adf is positively
correlated with the difference between the [O~{\sc iii}] forbidden line temperatures
and the H~{\sc i} Balmer jump temperatures, and that for a given nebula,
values of adf deduced for all abundant second-row elements are comparable.

\item Based on ORL abundance analyses, we have obtained reliable C/O abundance
ratios for a substantial sample of GBPNe.  By comparing the observed C/O ratios
with the recent theoretical predictions of Ma01, we conclude that GBPNe were
probably formed in more enriched environments than disk PNe, and that Ma01 may
have overestimated the carbon yield by as much as 0.4\,dex.  In addition, we
find that GBPNe tend to have more massive progenitor stars than GDPNe do. 

\item The average Ne/O ratio deduced from PNe in the WL07 and TLW samples is
about 0.25, much higher than the current recommended solar value. This may
points to a solar neon abundance that has been underestimated by 0.2~dex. 

\item For the first time, abundances of the important $\alpha$-element magnesium
have been determined using ORLs for a handful of bulge and disk PNe.  We find
that GBPNe have an average abundance about 0.13~dex higher than GDPNe. The
latter have an average abundance almost identical to the solar value.  The
enhancement of magnesium in GBPNe and [$\alpha$/Fe] in bulge giants suggests
that the primary enrichment process of the bulge was Type-II SNe.

\end{enumerate} 

\section{Acknowledgments}
The work is partially supported by Grant \#10325312 of the National Natural
Science Foundation of China. We would also like to thank Dr. R. Rubin for a
critical reading of the manuscript prior to its submission.

\appendix \section{Line fluxes}

\clearpage
\begin{table}
\caption{Ultraviolet and optical line fluxes on a scale where ${\rm H}\beta = 
100$. A superscript $^s$ denotes the emission is mainly from the central star 
and a $^{ps}$ indicates that stellar emission contributes at least partially 
the observed feature. (This is only a sample of this table: the full version is avaiable online 
as supplementary meterial.)}
\label{allline}
\begin{tabular}{c c c c c c}
\hline
\multicolumn{6}{l}{Cn~1-5}\\ 
$\lambda_{\rm obs}$ & $F(\lambda)$ & $I(\lambda)$ & Ion & $\lambda _0$  & Mult \\
\hline
 1246.53$^s$  &     11.480  &     72.136  &  N V       &   1240     &             \\
 1345.60$^s$  &      4.309  &     20.958  &  O VI      &   1342     &             \\
 1375.11$^s$  &      8.256  &     37.868  &  O V       &   1371.29  &             \\
 1553.86$^{ps}$  &     56.230  &    213.618  &  C IV      &   1551     &             \\
 1641.53$^{ps}$  &     16.560  &     59.797  &  He II     &   1640     &             \\
 1720.70$^s$  &      6.039  &     21.368  &  N IV      &   1718.55  &             \\
 1908.64  &     36.220  &    145.257  &  C III]    &   1906.68  &             \\
 3189.71  &      3.944  &      6.356  &  He I      &   3188.67  &  V3         \\
 3205.03$^s$  &      2.797  &      4.472  &  He II     &   3203.10  &  3.5       \\
 3704.33  &      1.703  &      2.284  &  H 16      &   3703.86  &  H16        \\
 3712.44  &      1.338  &      1.792  &  H 15      &   3711.97  &  H15        \\
 3722.31  &     20.510  &     27.409  &  H 14      &   3721.94  &  H14        \\
 3727.06  &     58.640  &     78.365  &  [O II]    &   3726.03  &  F1         \\
 3734.84  &      1.747  &      2.329  &  H 13      &   3734.37  &  H13        \\
 3750.62  &      1.950  &      2.594  &  H 12      &   3750.15  &  H12        \\
 3770.04  &      2.852  &      3.777  &  H 11      &   3770.63  &  H11        \\
 3799.11  &      5.574  &      7.341  &  H 10      &   3797.90  &  H10        \\
 3817.55  &      4.134  &      5.420  &  He I      &   3819.62  &  V22        \\
 3833.94  &      8.317  &     10.867  &  H 9       &   3835.39  &  H9         \\
 3867.80  &     70.260  &     91.077  &  [Ne III]  &   3868.75  &  F1         \\
 3887.79  &     18.000  &     23.228  &  H 8       &   3889.05  &  H8         \\
 3966.49  &     23.230  &     29.474  &  [Ne III]  &   3967.46  &  F1         \\
 3968.79  &     11.940  &     15.132  &  H 7       &   3970.07  &  H7         \\
 4008.82  &      0.236  &      0.296  &  He I      &   4009.26  &  V55        \\
 4068.16  &      3.894  &      4.820  &  [S II]    &   4068.60  &  F1         \\
 4071.56  &      0.204  &      0.252  &  O II      &   4072.16  &  V10        \\
 4075.82  &      1.383  &      1.708  &  [S II]    &   4076.35  &  F1         \\
 4088.94  &      0.055  &      0.068  &  O II      &   4089.29  &  V48a       \\
 4096.90  &      0.559  &      0.687  &  N III     &   4097.33  &  V1         \\
 4101.27  &     20.530  &     25.210  &  H 6       &   4101.74  &  H6         \\
 4120.15  &      0.292  &      0.357  &  O II      &   4121.46  &  V19        \\
 4132.42  &      0.087  &      0.106  &  O II      &   4132.80  &  V19        \\
 4143.32  &      0.449  &      0.546  &  He I      &   4143.76  &  V53        \\
 4152.80  &      0.100  &      0.121  &  O II      &   4153.30  &  V19        \\
 4155.96  &      0.068  &      0.082  &  O II      &   4156.53  &  V19        \\
 4236.61  &      0.049  &      0.058  &  N II      &   4236.91  &  V48a       \\
 4241.25  &      0.113  &      0.133  &  N II      &   4241.24  &  V48a       \\
 4266.66  &      1.109  &      1.305  &  C II      &   4267.15  &  V6         \\
 4275.85  &      0.092  &      0.108  &  O II      &   4275.55  &  V67a       \\
 4283.94  &      0.049  &      0.058  &  O II      &   4283.73  &  V67c       \\
 4339.93  &     40.520  &     46.763  &  H 5       &   4340.47  &  H5         \\
 4349.05  &      0.123  &      0.142  &  O II      &   4349.43  &  V2         \\
 4362.69  &      3.061  &      3.509  &  [O III]   &   4363.21  &  F2         \\
 4387.39  &      0.719  &      0.819  &  He I      &   4387.93  &  V51        \\
 4391.59  &      0.043  &      0.049  &  Ne II     &   4391.99  &  V55e       \\
 4437.13  &      0.066  &      0.074  &  He I      &   4437.55  &  V50        \\
 4470.96  &      5.974  &      6.650  &  He I      &   4471.50  &  V14        \\
 4480.86  &      0.021  &      0.023  &  Mg II     &   4481.21  &  V4         \\
 4530.00  &      0.053  &      0.058  &  N III     &   4530.86  &  V3         \\
 4561.97  &      0.076  &      0.082  &  Mg I]     &   4562.60  &             \\
 4570.59  &      0.221  &      0.240  &  Mg I]     &   4571.10  &             \\
 4630.08  &      0.134  &      0.143  &  N II      &   4630.54  &  V5         \\
 4633.66  &      0.294  &      0.313  &  N III     &   4634.14  &  V2         \\
 4639.80  &      0.457  &      0.486  &  N III     &   4640.64  &  V2         \\
 4641.01  &      0.540  &      0.574  &  N III     &   4641.84  &  V2         \\
 4648.86  &      0.378  &      0.401  &  O II      &   4649.13  &  V1         \\
 4657.71  &      0.941  &      0.995  &  [Fe III]  &   4658.10  &  F3         \\
 4661.39  &      0.081  &      0.086  &  O II      &   4661.63  &  V1         \\
\hline
\end{tabular}
\end{table}
  
 \begin{table}
\caption{Infrared line fluxes of PN\,Cn\,1-5, M\,2-23 and NGC~6567.}
\centering
\label{isoline}
\begin{tabular}{l c c }
\hline
 Lines &\multicolumn{1}{c}{$F(\lambda)$} &\multicolumn{1}{c}{$I(\lambda)$} \\
       & ($10^{-12}$\,ergs\,cm$^{-2}$\,s$^{-1}$) & [$I({\rm H}\beta) = 100$] \\
\hline
\noalign{\smallskip}       
\multicolumn{3}{l}{Cn~1-5}\\
\noalign{\smallskip}
 {[O\,{\sc iii}]}\,$52 \mu$m & 21.16 & 343\\ 
 {[O\,{\sc iii}]}\,$88 \mu$m & 8.39  & 136\\
  {[N\,{\sc iii}]}\,$57\mu$m & 9.66  & 157\\                   
   {[C\,{\sc ii}]}\,$158\mu$m & 0.75  & 12\\
\hline
\noalign{\smallskip}
\multicolumn{3}{l}{M~2-23}\\
\noalign{\smallskip}
H\,{\sc i} 6-4 $2.63 \mu$m & 0.75& 285 \\  
H\,{\sc i} 5-4 $4.05 \mu$m & 1.37& 520 \\
H\,{\sc i} 6-5 $7.46 \mu$m & 0.45& 171 \\
{[Ar\,{\sc iii}]}\,$8.99\mu$m & 1.06& 401 \\
 {[S\,{\sc iv}]}\,$10.5\mu$m & 2.43 & 923 \\
 {[Ne\,{\sc ii}]}\,$12.8\mu$m & 0.79 & 299 \\
{[Ne\,{\sc iii}]}\,$15.6\mu$m & 13.8 &5241 \\
 {[S\,{\sc iii}]}\,$18.7\mu$m & 0.85 &324 \\
{[Ne\,{\sc iii}]}\,$36.0\mu$m & 0.64 &243 \\
\hline
\noalign{\smallskip}
\multicolumn{3}{l}{NGC~6567}\\
\noalign{\smallskip}
H\,{\sc i} 5-4 $4.1 \mu$m & 0.81& 7.07    \\
{[Ar\,{\sc iii}]}\,$8.99\mu$m & 1.08& 9.39    \\
 {[S\,{\sc iv}]}\,$10.5\mu$m & 10.1 &87.7     \\
{[Ne\,{\sc iii}]}\,$15.6\mu$m & 28.9 &252   \\
 {[S\,{\sc iii}]}\,$18.7\mu$m & 1.25 &10.85 \\
{[Ne\,{\sc iii}]}\,$36.0\mu$m & 2.25 &19.6 \\
 {[O\,{\sc iii}]}\,$52 \mu$m & 51.1  & 455\\ 
 {[O\,{\sc iii}]}\,$88 \mu$m & 29.7  & 264\\
\hline
\end{tabular}
\end{table}

\end{document}